\documentclass[fleqn,usenatbib]{mnras}
\usepackage[T1]{fontenc}
\usepackage{ae,aecompl}

\DeclareRobustCommand{\VAN}[3]{#2}
\let\VANthebibliography\thebibliography
\def\thebibliography{\DeclareRobustCommand{\VAN}[3]{##3}\VANthebibliography}

\usepackage{soul}     % for: highlighting \hl{}
\usepackage{graphicx} % for: including figure files
\usepackage{float}    % for: figure placement, i.e., [H]
\usepackage{soul}     % for: highlighting \hl{}
\usepackage{multicol} % for: multi-column entries in tables
\usepackage[para]{threeparttable} % for: table notes
\usepackage[most]{tcolorbox} % for: tcolorboxes that envelope text
\usepackage{tikz}     % for: vector art

\usepackage{amsmath}   % for: advanced maths commands
\usepackage{amssymb}   % for: extra maths symbols
\usepackage{dsfont}    % for: greek integer symbol
\usepackage{esint}     % for: fancy integral symbols
\usepackage{abraces}   % for: notation of math expressions \aoverbrace
\usepackage{cleveref}  % for: referencing a range of eqns
\usepackage{bm}		   % for: bold maths symbols, including upright Greek
\usepackage{cancel}    % for: \cancelto{to-thing}{expression}
\usepackage{xfrac}     % for: sideways fraction: xfrac
\usepackage[a]{esvect} % for: longer vector accent: \vv. from: https://www.physicsread.com/latex-vector-arrow/
\usepackage{tikz}
\usepackage{circledsteps}
\usepackage{pstricks}

\usepackage{tikz}
\usepackage{scalerel}
\usetikzlibrary{svg.path}
\definecolor{orcidlogocol}{HTML}{A6CE39}
\tikzset{orcidlogo/.pic={\fill[orcidlogocol] svg{M256,128c0,70.7-57.3,128-128,128C57.3,256,0,198.7,0,128C0,57.3,57.3,0,128,0C198.7,0,256,57.3,256,128z}; \fill[white] svg{M86.3,186.2H70.9V79.1h15.4v48.4V186.2z} svg{M108.9,79.1h41.6c39.6,0,57,28.3,57,53.6c0,27.5-21.5,53.6-56.8,53.6h-41.8V79.1z M124.3,172.4h24.5c34.9,0,42.9-26.5,42.9-39.7c0-21.5-13.7-39.7-43.7-39.7h-23.7V172.4z} svg{M88.7,56.8c0,5.5-4.5,10.1-10.1,10.1c-5.6,0-10.1-4.6-10.1-10.1c0-5.6,4.5-10.1,10.1-10.1C84.2,46.7,88.7,51.3,88.7,56.8z};}}
\newcommand\orcidicon[1]{\href{https://orcid.org/#1}{\mbox{\scalerel*{
\begin{tikzpicture}[yscale=-1,transform shape]\pic{orcidlogo};
\end{tikzpicture}}{|}}}}
\defcitealias{burgers1948_turbulence_model}{B48}
\defcitealias{kolmogorov1941dissipation}{K41}
\newcommand{\tensor}[1]{\mathbfss{#1}}
\newcommand{\imatrix}{\mathbfss{I}} % identity matrix
\newcommand{\vect}[1]{\mathrm{{\bmath{\mathit{#1}}}}} % vector
\newcommand{\nvect}[1]{\widehat{\vect{#1}}} % normalised vector
\newcommand{\dd}[1]{\mathrm{d} #1} % derivative d
\newcommand{\pp}[1]{\partial #1} % partial derivative p
\newcommand{\driv}[2]{\frac{\dd{#1}}{\dd{#2}}} % derivative: d{} / d{}
\newcommand{\priv}[2]{\frac{\pp{#1}}{\pp{#2}}} %  partial derivative: p{} / p{}
\newcommand{\lfrac}[2]{\ensuremath{\frac{\displaystyle #1}{\displaystyle #2}}} % fraction with large text
\newcommand{\ave}[1]{\left\langle #1 \right\rangle}
\newcommand{\rbrac}[1]{\left( #1 \right)}
\newcommand{\sbrac}[1]{\left[ #1 \right]}
\newcommand{\cbrac}[1]{\left\{ #1 \right\}}
\newcommand{\nbrac}[1]{\left\| #1 \right\|}
\newcommand{\rfrac}[2]{\left( \frac{#1}{#2} \right)}

\newcommand{\ie}{\textit{i.e.},\,}
\newcommand{\eg}{\textit{e.g.},\,}

\newcommand{\viz}{\textit{viz.}\,}
\newcommand{\ts}{\mathrm{s}}
\newcommand{\tp}{\mathrm{p}}

\newcommand{\trms}{\mathrm{rms}}
\newcommand{\tturb}{\mathrm{turb}}
\newcommand{\tbox}{\mathrm{box}}

\newcommand{\tkin}{\mathrm{kin}}
\newcommand{\tmag}{\mathrm{mag}}
\newcommand{\tcur}{\mathrm{cur}}
\newcommand{\tratio}{\mathrm{ratio}}
\newcommand{\tsat}{\mathrm{sat}}
\newcommand{\tend}{\mathrm{end}}
\newcommand{\tcrit}{\mathrm{crit}}
\newcommand{\tres}{\mathrm{res}}
\newcommand{\tcor}{\mathrm{cor}}
\newcommand{\tshock}{\mathrm{shock}}
\newcommand{\tconst}{\mathrm{const}}
\newcommand{\Reyh}{\mbox{Re}}
\newcommand{\Reym}{\mbox{Rm}}
\newcommand{\Pranm}{\mbox{Pm}}

\newcommand{\Mach}{\mathcal{M}}

\newcommand{\vol}{\mathcal{V}}

\newcommand{\tbasis}{\nvect{b}}
\newcommand{\nbasis}{\nvect{\kappa}}
\newcommand{\ug}{\mathrm{G}}

\newcommand{\upc}{\mbox{pc}}
\newcommand{\ukpc}{\mbox{kpc}}
\newcommand{\umpc}{\mbox{Mpc}}

\newcommand{\umyr}{\mbox{Myr}}

\newcommand{\SimName}[3]{$\Mach${#1}$\Reyh${#2}$\Pranm${#3}}

\newcommand{\nquad}[1][1]{\hspace*{#1em}\ignorespaces}
\newcommand{\aref}[1]{\hyperref[#1]{Appendix~\ref{#1}}}

\defcitealias{kriel2022fundamental}{Fundamental Scales I}

\title[Supersonic small-scale dynamos]{Fundamental MHD scales -- II: the kinematic phase of the supersonic small-scale dynamo}

\author[Kriel, et al., 2024]{\centering
Neco Kriel$^{\orcidicon{0000-0002-3558-3926}\,1}$\thanks{E-mail: neco.kriel@anu.edu.au},
James R. Beattie$^{\orcidicon{0000-0001-9199-7771}\,2,3,1}$,
Christoph Federrath$^{\orcidicon{0000-0002-0706-2306}\,1,4}$,
Mark R. Krumholz$^{\orcidicon{0000-0003-3893-854X}\,1,4}$,
\newauthor
and Justin Kin Jun Hew$^{\orcidicon{0000-0002-5238-6115}\,1,5,6}$ \\
$^{1}$Research School of Astronomy \& Astrophysics, Australian National University, Canberra, ACT 2611, Australia \\
$^{2}$Department of Astrophysical Sciences, Nassau Street, Princeton University, Princeton, NJ 08544, USA \\
$^{3}$Canadian Institute for Theoretical Astrophysics, University of Toronto, 60 St. George Street, Toronto, ON M5S 3H8, Canada \\
$^{4}$Australian Research Council Centre of Excellence in All Sky Astrophysics (ASTRO3D), Canberra, ACT 2611, Australia \\
$^{5}$Space Plasma Power and Propulsion Laboratory, Department of Nuclear Physics and Accelerator Applications, Research School of \\ \;Physics, Australian National University, ACT 2601, Canberra, Australia \\
$^{6}$Mathematical Sciences Institute, Australian National University, Canberra, ACT 2601, Australia \\
}

\date{Accepted XXX. Received YYY; in original form ZZZ}

\pubyear{2024}
\begin{document}
\label{firstpage}
\pagerange{\pageref{firstpage}--\pageref{lastpage}}
\maketitle

\begin{abstract}
    Many astrophysical small-scale dynamos (SSDs) amplify weak magnetic fields via highly compressible, supersonic turbulence, but most established SSD theories have only considered incompressible flows. To address this gap, we perform visco-resistive SSD simulations across a range of sonic Mach numbers ($\mathcal{M}$), hydrodynamic Reynolds numbers ($\mathrm{Re}$), and magnetic Prandtl numbers ($\mathrm{Pm}$), focusing on the exponential growth phase. From these simulations, we develop robust measurements of the kinetic and magnetic energy dissipation scales ($\ell_\nu$ and $\ell_\eta$, respectively), and show that $\ell_\nu/\ell_\eta \sim \mathrm{Pm}^{1/2}$ is a universal feature of turbulent ($\mathrm{Re} \geq \mathrm{Re}_\mathrm{crit} \approx 100$), $\mathrm{Pm} \geq 1$ SSDs, regardless of $\mathcal{M}$. We also measure the scale of maximum magnetic field strength ($\ell_\mathrm{p}$), where we confirm that incompressible SSDs (where either $\mathcal{M} \leq 1$ or $\mathrm{Re} < \mathrm{Re}_\mathrm{crit}$) concentrate magnetic energy at $\ell_\mathrm{p} \sim \ell_\eta$ with inversely correlated field strength and curvature. By contrast, for compressible SSDs (where $\mathcal{M} > 1$ and $\mathrm{Re} \geq \mathrm{Re}_\mathrm{crit}$), shocks concentrate magnetic energy in large, over-dense, coherent structures with $\ell_\mathrm{p} \sim (\ell_\mathrm{turb} / \ell_\mathrm{shock})^{1/3} \ell_\eta \gg \ell_\eta$, where $\ell_\mathrm{shock}$ is the characteristic shock width, and $\ell_\mathrm{turb}$ is the outer scale of the turbulent field. When $\Pranm < \Reyh^{2/3}$, the shift of $\ell_\tp$ (from the incompressible to compressible flow regime) is large enough to move the peak magnetic energy scale out of the subviscous range, and the plasma converges on a hierarchy of scales: $\ell_\tturb > \ell_\tp > \ell_\tshock > \ell_\nu > \ell_\eta$. In the compressible flow regime, more broadly, we also find that magnetic field-line curvature becomes nearly independent of the field strength, not because the field geometry has changed, but instead the field becomes locally amplified through flux-frozen compression by shocks. These results have implications for various astrophysical plasma environments in the early Universe, and cosmic ray transport models in the interstellar medium.
\end{abstract}


\section{Introduction}

    \subsection{Setting the Stage for Present-Day Galactic Magnetism}
    
    As far as we understand, the Universe was born without magnetic fields. This, however, stands in stark contrast with the present-day Universe, where dynamically important magnetic fields are observed to be ubiquitous \citep[see][for recent reviews on magnetic fields in galaxies and their impact on star formation]{krumholz2019role, brandenburg2023galactic}. While the origin of these fields remains uncertain, two main candidates exist: phase transitions during inflation, which could have generated fields with strengths ranging from $10^{-36}$ to $10^{-8} \,\ug$, varying on $\sim \umpc$ scales \citep{quashnock1989magnetic, sigl1997primordial, kahniashvili2013evolution}, and battery processes during the epoch of re-ionisation ($z \sim 35$--$6$) \citep{biermann1950ursprung, naoz2013generation}, which could have produced fields of around $10^{-24} \,\ug$ on $\sim 10 \,\ukpc$ scales.
    
    Regardless of which mechanism seeded the first magnetic fields, primordial fields are believed to have decayed until the structure formation era at $z \sim 2$ \citep[\eg][]{brandenburg2017evolution, vachaspati2021progress, hosking2022cosmic, mtchedlidze2022evolution, mtchedlidze2023inflationary, Hosking2023_cosmic_voids_nature}, by which time they would have been more than $15$ orders of magnitude weaker than the $\sim \mu\ug$ fields observed on $\sim \ukpc$ scales in the Milky Way and in other nearby galaxies \citep[\eg][]{kulsrud1997protogalactic, beck2019synthesizing, shah2021magnetic, lopez2022extragalactic}. Present-day magnetic fields, therefore, cannot simply be relics from electroweak phase transitions or battery processes in the early Universe, and instead some mechanism must have amplified them dramatically.

    \subsection{Flavours of Dynamo}
    
    Dynamo action is believed to be the most plausible mechanism for amplifying primordially-produced fields to the levels we observe in the present-day \citep[\eg][]{latif2013small, mtchedlidze2022evolution, mtchedlidze2023inflationary}, and broadly describes the process by which initially weak magnetic fields are amplified and subsequently maintained through the conversion of kinetic into magnetic energy (see \citealt{rincon2019dynamo, brandenburg2023galactic} for recent reviews, and \citealt{tzeferacos2018laboratory, bott2021time, bott2022insensitivity} for recent laboratory experiments). These mechanisms are categorised either as small-scale dynamos (SSDs) or large-scale dynamos (LSDs), determined by the scale on which magnetic fields are grown relative to the fields that amplify them. In the current paradigm for the origin of galactic magnetic fields, both SSDs and LSDs are believed to be important, but are understood to play very different roles \citep[see, \eg][for recent works]{bhat2016unified, pakmor2017magnetic, rieder2017dynamo2, rieder2017dynamo3, steinwandel2023towards, gent2023transition}.

    \subsubsection{Small-Scale Dynamos}
    
    SSDs in the interstellar medium (ISM) of galaxies are primarily driven by supersonic turbulence, in turn supported by a combination of supernova feedback and gravitational instabilities \citep{krumholz2016turbulence, bacchini2020evidence}. Starting with initially weak (primordial) seed fields, these SSDs follow roughly a three-stage process: (1) random kinetic motions (or more precisely, random velocity gradients) \citep[\eg][]{vainshtein1972origin, zel1984kinematic, archontis2003numerical} amplify magnetic energy exponentially fast-in-time (termed the kinematic phase); (2), once magnetic fields are strong enough to impart a backreaction on the flow via the Lorentz force, growth slows down to polynomial-in-time \citep[termed the non-linear phase; \eg][see also \autoref{sec:discussion:saturation} for a discussion]{schekochihin2004simulations, beresnyak2012universal, Schleicher2013_quadratic_growth, Xu2016_dynamo, seta2020seed}; finally (3), once the magnetic field is in close equipartition with kinetic energy, the field strength saturates and continues to be maintained at this level via the velocity field \citep[\eg][]{schekochihin2002spectra, cho2009growth, seta2021saturation, beattie2022growth}.

    \subsubsection{Large-Scale Dynamos}
    
    Galactic LSDs are also capable of exponential magnetic growth over $\sim {\rm Gyr}$ timescales, which could be driven by a mix of helical turbulence \citep[\eg][]{bhat2016unified, rincon2021helical}, potentially due in part to galactic differential rotation/shear \citep[\eg][]{vishniac1997incoherent, kapyla2008large, squire2015generation, gent2023transition}, or magnetic instabilities \citep[\eg][]{johansen2008high, qazi2023nonlinear}. While catastrophic quenching at high magnetic resistivity has been a concern, particularly in domains with either closed or periodic boundaries, where magnetic helicity cannot escape \citep[see][]{vishniac2001magnetic}, this effect is believed to be less severe than previously thought, especially in shearing systems \citep{hubbard2012catastrophic}. Moreover, quenching can be alleviated by factors like non-conservation of helicity flux \citep{verma2003energy, sur2007galactic, del2013turbulent}, as would be the case in the presence of outflows (\eg galactic winds) and accreting flows, or with the additional influence of cosmic ray pressure \citep{parker1992fast, hanasz2004amplification}. Nonetheless, LSDs alone are not believed to be solely responsible for amplifying primordial fields to the magnitudes we observe today, as they operate much slower than the kinematic phase of SSDs \citep{schober2012magnetic, gent2023transition} and are less efficient in the presence of low resistivity SSDs \cite[\eg][]{gent2023transition}. However, some LSDs, like the shear-current dynamo \citep{squire2015generation}, can directly feed on the turbulent magnetic fluctuations produced by SSDs \citep[see][for a recent, thorough review of LSDs]{brandenburg2023galactic}.

    \subsubsection{Dynamos in Young Galaxies}
    
    Supernova-driven SSDs operating in the ISM of young, isolated Milky Way-like galaxies, have been numerically shown to be capable of amplifying primordial magnetic fields with an e-folding time of $\sim 100\,\umyr$, reaching a saturated field strength of $10$--$50 \,\umu\ug$ (corresponding to $\sim 10\%$ of equipartition with the kinetic energy) by $z \sim 3$--$2$ \citep{pakmor2017magnetic, rieder2016dynamo1, rieder2017dynamo2, steinwandel2023towards, Muni2024_seed_field_to_dynamo}. As these galaxies evolve and become more quiescent, some yet unconstrained combination of the aforementioned galactic mechanisms creates a LSD that operates on the SSD-amplified fields, gradually contributing to the total magnetic field amplitude and efficiently developing large-scale, ordered fields observed in present-day galaxies \citep{rieder2017dynamo2, gent2023transition, brandenburg2023galactic}. This was demonstrated by \citet{gent2023transition}, where a rotation-driven LSDs was shown to be capable of amplifying the saturated SSD energy state by a factor $\lesssim 10$, over several ${\rm Gyr}$, and in doing so, efficiently reorganise SSD amplified magnetic fields to produce large-scale, coherent structures on scales larger than where the SSD operates.
    
    Galaxy mergers are a example of systems where both a SSD and LSD operates in concert \citep{berlok2019impact, pfrommer2022simulating}. While the exact interaction between these two dynamo processes largely remains an open problem, the interplay between the non-linear phase of the SSD and the LSD is important for re-establishing large-scale order after temporary merger-induced disruption, and for redistributing kinematic SSD-amplified fields from the inner regions, outward \citep{whittingham2021impact, whittingham2023impact}. Nonetheless, for most of this paper, we focus on the kinematic phase, returning briefly to the non-linear phase in \autoref{sec:discussion:saturation}.

    \subsection{The Question: Does The Small-Scale Dynamo Always Produce Small-Scale Magnetic Fields?}
    
    Our current paradigm for magnetogenesis has recently been challenged by the detection of \mbox{$\lesssim 500 \,\mu\ug$} magnetic fields ordered on $5 \,\ukpc$ scales in a distant ($z \sim 2.6$) gravitationally micro-lensed galaxy \texttt{9io9} \citep{geach2023polarized}, as well as $\sim \ukpc$ fields in a $z \sim 5.6$ galaxy \citep{chen2024kiloparsec}. While the field strength of \texttt{9io9} remains poorly constrained\footnote{\citet{geach2023polarized} arrive at an upper bound for the magnetic field strength based on the assumption of energy equipartition between magnetic and kinetic energy. However, even the most efficient dynamos do not reach perfect equipartition on all scales \citep[\eg][]{federrath2011mach, federrath2014turbulent, kriel2022fundamental, beattie2022growth}. For supersonic dynamos, which are expected to be important in the cold molecular gas traced in \texttt{9io9}, the final saturation is more likely $\sim 1\%$, which means \texttt{9io9} is more likely in a $\sim \,\mu\ug$ state, making it consistent with the field strengths of modern galaxies.}, the fact that it is ordered on $\sim \ukpc$ scales poses a problem for the model outlined above, since a LSD would not have had enough time to become established\footnote{\citet{geach2023polarized} estimate that the dynamo number for \texttt{9io9} is $\approx 3$, which is lower than the critical value $\approx 7$ which is required for a LSD to be estimated as dynamically important in Milky Way-like disk galaxies \citep{ruzmaikin1988magnetism}.} at $z \sim 2.6$. Moreover, SSDs have been traditionally thought to produce fields that are chaotic on large scales, only becoming ordered on the smallest scales allowed by magnetic dissipation \citep[\eg][]{schekochihin2004simulations, kriel2022fundamental, brandenburg2023dissipative}, which are expected to be $\ll \upc$ in size for ISM conditions \citep{marchand2016chemical}. A similar problem exists in galaxy mergers, where the rapid growth of magnetic energy seen in simulations points to SSD action, but the fields are correlated on $\sim \ukpc$ scales, which have been thought to be too large scale to be generated during the kinematic SSD phase \citep[\eg][]{rodenbeck2016magnetic, basu2017detection, brzycki2019parameter, whittingham2021impact}.

    Now, the expectation that SSDs only produce small-scale fields, stems from extensive explorations of SSDs in incompressible (\eg subsonic) flow regimes \citep[see seminal works by, \eg][]{Kazantsev1968, vainshtein1982theory, vincenzi2002kraichnan, schekochihin2004simulations, boldyrev2004magnetic}. This expectation was confirmed in Paper I of this series \citep[][herein \citetalias{kriel2022fundamental}]{kriel2022fundamental}, where we used direct numerical simulations (DNSs) to explore the magnetic fields produced during the kinematic phase, and showed that the bulk of magnetic energy becomes concentrated at the smallest possible scales in the incompressible problem, \ie the scale where magnetic fields dissipate. Given that these scales are much smaller than a galactic scale height, and fall within the regime where turbulence becomes independent of the anisotropy imposed by the galactic potential \citep{Kortgen2021_power_spectrum_of_isolated_galaxies}, the presence of large-scale shear, differential rotation or other similar organised galactic flows, does not alter this conclusion. However, this understanding may not fully capture the dynamics of more compressible environments.
    
    Both the first galaxies \citep{maio2011impact, mandelker2020instability} and later galaxy mergers \citep{geng2012magnetic, sparre2022gas}, are expected to host highly compressible (\ie supersonic) turbulence. This is because the great majority of their mass, and a substantial fraction of their volume, consists of dense, atomic and molecular gas \citep{cox2006kinematic, krumholz2009star, popping2014evolution, nandakumar2023large} where rapid cooling keeps the sound speed well below the characteristic flow velocity \citep{rees1977cooling, white1978core, birnboim2003virial, Krumholz2020_cr_transport_starburst, li2020impact}. While it has been shown that supersonic flow decreases the efficiency of SSDs \citep{federrath2011mach, federrath2014turbulent, seta2021saturation, seta2022turbulent, hew2023lagrangian}, there has to date been no systematic study of how compressibility changes magnetic field geometry, characteristic turbulence and dynamo scales, or structure of the magnetic field. Given the discrepancy between the observations and the predictions of incompressible SSD models, there is a clear need for such a study. Our goal in this paper is to carefully explore how magnetic fields become organised during the kinematic phase of SSDs driven by supersonic turbulence, and in turn to explore its implications for galaxy mergers and other phenomena that depend on magnetic field structure, most notably new cosmic ray transport models relying on curvature and gyro-radii resonances \citep[\eg][]{kempski2023cosmic, Lemoine2023_field_reversal_CR_transport}.

    \subsection{Structure of the Paper}
    
    The remainder of this paper is structured as follows. In \autoref{sec:numerics} we describe the numerical simulation suite we use to develop our theoretical model for compressible SSDs. In \autoref{sec:results} we present and interpret the simulation results, and in \autoref{sec:discussion} we discuss the implications of these results for a variety of astrophysical systems. We summarise our results and conclusions in \autoref{sec:conclusion}.

\section{Simulating the Problem}
\label{sec:numerics}

    In this study we use DNSs to explore the kinematic phase of SSDs in plasma flows ranging from viscous to turbulent, and subsonic to supersonic. In \autoref{sec:mhd_model} we introduce the basic equations that we solve, the numerical method by which we do so, and the initial conditions for our simulations. In \autoref{sec:ICs} we introduce the key dimensionless parameters that describe different plasma flow regimes, and how we vary these parameters to explore different flow properties. In \autoref{sec:ICs:nres}, we then discuss issues of convergence. 
    
    \subsection{Numerical Approach}
    \label{sec:mhd_model}

    \subsubsection{Solving the MHD Equations}
    
    For all the simulations in this study, we solve the compressible set of non-ideal (visco-resistive) magnetohydrodynamical (MHD) fluid equations, which in conservative form are
    \begin{align}
        \priv{\rho}{t}
            + \nabla\cdot(\rho \vect{u})
            &= 0
            , \label{mhd:continuity} \\
        \priv{\rho\vect{u}}{t}
            + \nabla\cdot\bigg[
                \rho\vect{u}\otimes\vect{u}
                - \frac{1}{4\pi}\vect{b}\otimes\vect{b}
                \nquad[7]\nonumber \\
                + \rbrac{c_\ts^2\rho + \frac{b^2}{8\pi}}\tensor{\imatrix}
                - 2\nu\rho\tensor{S}
            \bigg] &= \rho\vect{f}
            , \label{mhd:momentum} \\
        \priv{\vect{b}}{t}
            - \nabla\times\rbrac{\vect{u}\times\vect{b} - \eta\vect{j}}
            &= 0
            , \label{mhd:induction} \\
        \nabla\cdot\vect{b}
            &= 0
            , \label{mhd:divb_free}
    \end{align}
    for an isothermal plasma evolving over a uniformly discretised, cubic-domain $\ell_{(x, y, z)} \in [0, \ell_\tbox]$, with triply periodic boundary conditions. Here we use constant (in both space and time) kinematic shear viscosity and Ohmic resistivity, parameterised by the coefficients $\nu$ and $\eta$, respectively, in combination with an external forcing field $\vect{f}$, to achieve flows with desired plasma numbers (see \autoref{sec:ICs} for details). The remaining quantities in the equations are the gas density $\rho$, the gas velocity $\vect{u}$, the sound speed $c_\ts$, the current density $\vect{j} = \nabla\times\vect{b} /(4\pi)$, and the magnetic field $\vect{b} = \vect{b}_0 + \vect{\delta b}$, which has mean field $\vect{b}_0$, and fluctuating (turbulent) field $\vect{\delta b}$ components. Finally, our viscosity model is based on the traceless strain rate tensor, $\tensor{S}$, where
    \begin{equation}
        \tensor{S}
            = \frac{1}{2} \rbrac{\nabla\otimes\vect{u} + \rbrac{\nabla\otimes\vect{u}}^T}
            - \frac{1}{3} \rbrac{\nabla\cdot\vect{u}}\tensor{\imatrix}
            , \label{eqn:strain_rate_tensor}
    \end{equation}
    and $\otimes$ is the tensor product $\nabla\otimes\vect{u} \equiv \pp_i u_j$.
    
    We solve \autoref{mhd:continuity}-\ref{mhd:divb_free} with a modified version of the finite-volume \textsc{flash} code \citep{fryxell2000flash, dubey2008introduction}, employing a second-order conservative MUSCL-Hancock 5-wave approximate Riemann solver, described in \citet{bouchut2007multiwave, bouchut2010multiwave}, and implemented into \textsc{flash} by \citet{waagan2011robust}, who showed that it possesses excellent stability properties for highly supersonic MHD flows, with improved efficiency and stability compared with Roe-type solvers. Since our primary interest lies in studying the effects of shocks (a hallmark of supersonic flows), this solver proves highly suitable. Moreover, we utilise the parabolic divergence-cleaning method described by \citet{Marder1987_fluxcleaning} to enforce that $\nabla\cdot\vect{b} = 0$ modes are diffused away.

    \subsubsection{Initial Conditions}
    
    All our simulations use a dimensionless unit system where the simulation box size $\ell_\tbox = 1$, mean density $\rho_0 = 1$, the sound speed $c_\ts = 1$, and magnetic fields are measured in units of $\rho_0^{1/2} c_\ts = 1$. We initialise every simulation with uniform density $\rho = \rho_0 = 1$, zero velocity $\vect{u} = \vect{0}$, and zero mean magnetic field $\vect{b}_0 = \vect{0}$. Since there is no mean magnetic field, and magnetic flux through the simulation volume is conserved, only a fluctuating component can exist, $\vect{b} = \vect{\delta b}$. We initialise this fluctuating component with a spectral distribution \citep[using \mbox{\textsc{TurbGen}};][]{Federrath2022_TurbGen} that is non-zero only over the wavenumber range $1 \leq k \ell_\tbox / 2\pi \leq 3$, with a parabolic profile that peaks at $k \ell_\tbox/2\pi = 2$, and goes to zero at $k \ell_\tbox/2\pi = 1$ and $k \ell_\tbox/2\pi = 3$. Here, the isotropic wavenumber $k$ is defined as per usual: $k \equiv 2\pi / \ell$. We choose the amplitude of this initial parabolic $\vect{\delta b}$ profile such that the plasma--$\beta \equiv p_{\rm th}/p_\tmag = 8\pi c_\ts^2 \rho_0 / b^2 = 10^{10}$. We note that the exact configuration of the initial seed $\vect{\delta b}$ field is not important, because it is quickly forgotten by the Markovian-like flow dynamics, and has been shown to not affect the amplification nor the final saturation of the dynamo \citep{seta2020seed, bott2022insensitivity, beattie2022growth}.

    \subsection{Important Dimensionless Numbers and Flow Regimes}
    \label{sec:ICs}

    In this study we explore 34 different simulation configurations, each parameterised by a set of dimensionless numbers that characterise the flow and plasma regimes. Here we introduce each of these numbers, as well as the range of values over which we vary them, before summarising our full set of simulations in \autoref{table:summary}.
    
    \subsubsection{Sonic Mach Number}
    \label{sec:ICs:Mach}
    
    For all our simulations we produce an isotropic, smoothly varying (in time and space) acceleration field via the forcing term, $\vect{f}$, in \autoref{mhd:momentum}, which is modelled with a generalisation of the Ornstein-Uhlenbeck process in wavenumber-space \citep{eswaran1988examination, schmidt2006numerical, schmidt2009numerical, federrath2010comparing, Federrath2022_TurbGen} using \textsc{TurbGen}. We choose to drive the acceleration field with purely solenoidal modes, \ie $\nabla\cdot\vect{f} = 0$, because they produce motions that are the most efficient at amplifying magnetic energy \citep{federrath2011mach, federrath2014turbulent, martins2019kazantsev, chirakkara2021efficient}, and tune the amplitude of $\vect{f}$ in each of our simulations to achieve a root-mean-squared (rms) gas velocity dispersion, $\ave{u}_\vol^{1/2}$, on the driving (outer) scale, $\ell_\tturb$, that lies within $5\%$ of our desired value, $u_\tturb$; the notation $\ave{q}_{\vol}$ indicates the volume average of quantity $q$ over the entire simulation domain $\vol \equiv \ell_\tbox^3$. We choose $\ell_\tturb = \ell_\tbox/2$ for all of our simulations to maximise the scale separation between velocity fields and the small-scale magnetic fields they generate, which we achieve by driving $\vect{f}$ with a parabolic power spectrum that is non-zero only over the wavenumber range $1 \leq k \ell_\tbox/2\pi \leq 3$, peaking at $k \ell_\tbox / 2\pi = 2$, and falling to zero at $k \ell_\tbox/2\pi = 1$ and $k \ell_\tbox/2\pi = 3$. The corresponding autocorrelation time of $\vect{f}$, and hence the velocity field, is
    \begin{align}
        t_\tturb
            \sim \frac{\ell_\tturb}{u_\tturb}
            = \frac{2\pi}{k_\tturb u_\tturb} .
    \end{align}
    
    Even though $c_\ts = 1$ in our simulations, it is convenient to express the flow velocity relative to the sound speed, \ie the sonic Mach number
    \begin{align}
        \Mach
            &\equiv \frac{u_\tturb}{c_\ts}
            . \label{dfn:Mach}
    \end{align}
    We run simulations spanning a wide range of $\Mach$, with $\Mach = 0.3, 1, 5$, and $10$. To get a baseline for how the plasma behaves in the absence of compressible effects, we first run 11 simulations with $\Mach = 0.3$, where incompressibility in the probability density function (PDF) of $\Mach$ values in $\vol$ holds up to $3$-sigma fluctuations\footnote{From the density dispersion-Mach relation we expect $\Mach = 0.3$ fields driven by solenoidal forcing to have $\approx 10\%$ fluctuations in the density field \citep{padoan1997supersonic, passot2003correlation, federrath2008density, federrath2010comparing, price2010density, gerrard2023new}.} (assuming Gaussianity). This is the regime we previously explored in \citetalias{kriel2022fundamental}, and is relevant for studying \citet{kolmogorov1941dissipation}-like turbulence. In addition to these simulations, we also run a set of four transsonic simulations, $\Mach = 1$, but then turn most of our attention towards the supersonic flow regime, where we run 16 simulations with $\Mach = 5$ and three simulations with $\Mach = 10$. Here, in the highly-compressible ($\Mach \gtrsim 1$) flow regime, one expects to see \citet{burgers1948_turbulence_model}-like turbulence \citep[see for example][]{federrath2013universality, Federrath2021_sonic_scale}.

    \subsubsection{Hydrodynamic Reynolds Number}
    
    The second dimensionless parameter that characterises our simulations is the hydrodynamic Reynolds number 
    \begin{align}
        \Reyh
            &\equiv \frac{
                    \|\nabla\cdot(\rho\vect{u}\otimes\vect{u})\|
                }{
                    \|\nabla\cdot(2\nu\rho\tensor{S})\|
                }
            \sim \frac{u_\tturb \,\ell_\tturb}{\nu}
            , \label{dfn:Reyh}
    \end{align}
    which describes the relative importance of inertial compared to viscous forces in a flow; here $\|\dots\|$ is the 2-norm.
    
    In \citetalias{kriel2022fundamental}, we identified $\Reyh_\tcrit \approx 100$ as a critical threshold for $\Reyh$ during the kinematic phase, separating viscous ($\Reyh < \Reyh_\tcrit$) from turbulent ($\Reyh \geq \Reyh_\tcrit$) flows, with several flow properties changing across this boundary. Above this threshold, the flow becomes more turbulent, leading to intermittent velocity fluctuations\footnote{Velocity gradients reflect local deformation rates and energy dissipation, where larger, more intermittent gradients indicating turbulence, and conversely, smaller, more uniform gradients suggest viscous flow \citep{schumacher2014small}.} (super-Gaussian kurtosis), while flows below this value show sub-Gaussian kurtosis. Similarly, the kinetic energy\footnote{In turbulence theory, naming conventions originally inspired by \citet{kolmogorov1941dissipation}, primarily apply to incompressible (\mbox{$\Mach \ll 1$}) flows. In these flows, the time-derivative of the average kinetic energy $\pp_t\big[\ave{E_\tkin}_\vol\big] = 0.5\, \ave{\rho}_\vol \pp_t\sbrac{\ave{u^2}_\vol} \sim \pp_t\sbrac{\ave{u^2}_\vol}$, assuming that density fields are roughly constant in time, while velocity fields vary in time. However, for highly compressible ($\Mach \gg 1$) flows, these conventions need to be adjusted to account for the time variation in both $\rho$ and $u$, as well as their covariance. As a result, $\ell_\nu$ represents the characteristic dissipation scale of kinetic energy, not just the velocity field.\label{note:naming_convesions}} dissipation (viscous) scale\footnote{While $\ell_\nu$ is the characteristic viscous scale, where kinetic fluctuations transition from inertial dominated to viscous dominated, there are in fact a whole range of scales $\ell \lesssim \ell_\nu$ over which dissipation takes place \citep[\eg][]{frisch1991prediction, chen1993far}. These dissipation scales have also been shown to be directly affected by the degree of intermittency of velocity gradient fluctuations \citep[\eg][]{schumacher2007sub}.} follows the theoretically-expected scaling $\ell_\nu \sim \Reyh^{3/4}$ \citep{kolmogorov1941dissipation} for turbulent flows, and then scales as $\ell_\nu \sim \Reyh^{3/8}$ for viscous flows \citepalias{kriel2022fundamental}.

    Based on this, we explore $10 \leq \Reyh < 3000$, where we run most of our simulations with $\Reyh \geq \Reyh_\tcrit$ (in the turbulent regime), and dedicate a small portion of our simulations to $10 \leq \Reyh < \Reyh_\tcrit$, to explore the transitional regime towards viscous flows.

    \subsubsection{Magnetic Prandtl Number}
    \label{sec:ICs:Pm}
    
    Our final two dimensionless numbers\footnote{Instead of parameterising our flows with respect to the dimensionless Alfv\'enic Mach number ($\Mach_A$), we follow SSD conventions, and use the magnetic-to-kinetic energy ratio ($E_\tmag / E_\tkin$). In purely fluctuation plasmas, these two quantities are directly related: $\Mach_A = (4\pi \rho)^{1/2} \ave{u^2}_\vol^{1/2} / \ave{b^2}_\vol^{1/2} = (E_\tmag / E_\tkin)^{-1/2}$. Thus, we choose $E_\tmag / E_\tkin$ as our measure for the significance of Lorentz forces in the flow, where, during the kinematic phase, $E_\tmag / E_\tkin \ll 1$, meaning $\Mach_A \gg 1$ and Lorentz forces are negligible.} are the magnetic Reynolds number and the magnetic Prandtl number. The magnetic Reynolds number is defined as
    \begin{align}
        \Reym
            &\equiv \frac{
                    \|\nabla\times(\vect{u}\times\vect{b})\|
                }{
                    \|\eta\nabla\times\vect{j}\|
                }
            \sim \frac{u_\tturb \,\ell_\tturb}{\eta}
            , \label{dfn:Reym}
    \end{align}
    which, analogously to the hydrodynamic Reynolds number, characterises the relative importance of magnetic induction compared with magnetic (Ohmic) dissipation. The magnetic Prandtl number is then given by
    \begin{align}
        \Pranm
            \equiv \frac{\Reym}{\Reyh}
            \sim \frac{\nu}{\eta}
            , \label{dfn:Pranm}
    \end{align}
    and it characterises the relative strength of the magnetic and kinetic energy dissipation. In collisional plasmas, $\Pranm$ is an intrinsic property of the plasma itself, unlike $\Reyh$ and $\Reym$, which depend on the flow properties (\ie $\ell_\tturb$ and $u_\tturb$). In our simulations, we choose values of $1 \leq \Pranm \leq 300$ to explore plasma regimes relevant for most of the gas in the ISM, with a focus on $\Pranm > 1$ scenarios. However, resolution constraints prevent us from exploring turbulent flows with large $\Pranm$. Regardless, in all cases, the characteristic kinetic and magnetic energy ($\ell_\eta$) dissipation scales are organised such that $\ell_\nu \geq \ell_\eta$.
    
    Now, during the kinematic phase of subsonic, $\Pranm > 1$ SSDs, there exist well-established theoretical predictions for the relationships between key MHD length scales: $\ell_\tturb$, $\ell_\nu$, $\ell_\eta$, and the scale $\ell_\tp$ where magnetic fields are strongest. \citet{schekochihin2002spectra, schekochihin2004simulations} predicted -- and \citetalias{kriel2022fundamental} and \citet{brandenburg2023dissipative} confirmed numerically -- that magnetic energy becomes strongest on the smallest scales allowed by magnetic dissipation, yielding an ordering of scales $\ell_\tp \sim \ell_\eta \sim \ell_\nu \, \Pranm^{-1/2}$. One of the primary goals of our study is to test whether this hierarchy also holds in the supersonic regime.

    \subsubsection{Choice of Simulation Parameters}
    
    The discussion of dimensionless numbers above motivates our choice of simulation parameters. To determine the scaling behaviour of $\ell_\eta$ and $\ell_\tp$ in compressible flows, we first run a set of eight simulations with $\Reym = 3000$, where we vary $1 \leq \Pranm \leq 300$ while keeping $\Mach = 5$. To isolate the role of compressibility, we then also run a subset of these simulations with $\Mach = 0.3, 1$, and $10$. Next, we run three $\Reyh = 500$, $\Mach = 5$ simulations, with $\Pranm = 1, 2$, and $4$, and four $\Reyh = 10$, $\Mach = 5$ simulations with $25 \leq \Pranm \leq 250$. Then, to explore the transition from turbulent to viscous flows, we run a set of four $\Mach = 0.3$ simulations, where we fix $\Reym = 500$ and vary $1 \leq \Pranm \leq 50$. Finally, we run two simulations with $\Reyh = 2000$ and $\Pranm = 5$ (which gives $\Reym = 10000$), with $\Mach = 0.3$ and $5$, respectively, to confirm that our findings in both the subsonic and supersonic regimes hold in the high-$\Reym$ limit.

    We summarise our full set of simulations in \autoref{table:summary}, adopting a naming convention whereby each simulation is labeled \SimName{\texttt{MMM}}{\texttt{RRR}}{\texttt{PPP}}. Here, \texttt{MMM}, \texttt{RRR}, and \texttt{PPP} give the numerical values of the sonic Mach number, hydrodynamic Reynolds number, and magnetic Prandtl number, respectively. For example, \SimName{5}{600}{5} corresponds with a simulation where $\Mach = 5.0$, $\Reyh=600$, and $\Pranm=5$.
    
    \subsection{Numerical Convergence in Time and Resolution}
    \label{sec:ICs:nres}

    To ensure well-sampled statistics, we run all of our simulations for a duration of $t = 100 \,t_\tturb$ (\ie 100 autocorrelation times of the forcing field), which we show extends well beyond the kinematic phase and into the saturated state of the dynamo for all of our simulations. We also collect data every $t = 0.1 \,t_\tturb$ to ensure well-sampled temporal statistics.
    
    To ensure convergence with regard to spatial resolution, we systematically run each of our simulation setups at progressively higher resolution, until our measurements of key characteristic scales: $\ell_\nu$, $\ell_\eta$, and $\ell_\tp$ converge. All our simulations use a uniform, cubic grid of $N_\tres^3$ cells, where we test for convergence by carrying out simulations at resolutions $N_\tres = 18, 36, 72, 144$, and $288$, and then for a subset of our simulations, we also run at higher resolutions of $N_\tres = 576$, $1152$ as required (we indicate these simulations in column 13 of \autoref{table:summary}); we defer a discussion of how we assess convergence to \autoref{sec:results:nres}.
    
    \begin{table*}
    \renewcommand{\arraystretch}{1.05}
    \setlength{\tabcolsep}{2.1pt}
    \caption{Main simulation parameters and derived quantities.}
    \vspace{-0.5 cm}
    \label{table:summary}
    \begin{center}
    \begin{tabular}{l c c c c c c c c c c c c c c}
        \hline\hline
        Sim. ID
        & $\Reyh$ & $\Reym$ & $\Pranm$
        & $\sfrac{\nu t_\tturb}{\ell_\tturb^2}$ & $\sfrac{\eta t_\tturb}{\ell_\tturb^2}$
        & $\Mach$ & $\gamma \, t_\tturb$ & $\rbrac{\frac{E_\tmag}{E_\tkin}}_\tsat$
        & $k_\nu / k_\tbox$ & $k_\eta / k_\tbox$ & $k_\tp / k_\tbox$
        & Extra $N_\tres$ \\[0.4em]
        \multicolumn{1}{l}{(1)} & (2) & (3) & (4) & (5) & (6) & (7) & (8) & (9) & (10) & (11) & (12) & (13) \\
\hline
\hline
\multicolumn{13}{c}{$\Mach = 0.3$} \\
\hline
$\mathcal{M}$0.3Re500Pm1
	& $500$ & $500$ & $1$
	& $3.0 \times 10^{-4}$ & $3.0 \times 10^{-4}$ & $0.30 \pm 0.01$
	& $0.42 \pm 0.01$ & $0.14 \pm 0.03$
	& $23.9 \pm 0.5$ & $13 \pm 1$ & $4 \pm 1$
	& -- \\
$\mathcal{M}$0.3Re100Pm5    
	& $100$ & $500$ & $5$
	& $1.5 \times 10^{-3}$ & $3.0 \times 10^{-4}$ & $0.30 \pm 0.02$
	& $0.48 \pm 0.01$ & $0.36 \pm 0.06$
	& $8.4 \pm 0.2$ & $9 \pm 1$ & $4 \pm 1$
	& -- \\
$\mathcal{M}$0.3Re50Pm10
	& $50$ & $500$ & $10$
	& $3.0 \times 10^{-3}$ & $3.0 \times 10^{-4}$ & $0.30 \pm 0.02$
	& $0.46 \pm 0.01$ & $0.4 \pm 0.1$
	& $5.8 \pm 0.2$ & $9 \pm 1$ & $4 \pm 1$
	& -- \\
$\mathcal{M}$0.3Re10Pm50
	& $10$ & $500$ & $50$
	& $1.5 \times 10^{-2}$ & $3.0 \times 10^{-4}$ & $0.29 \pm 0.02$
	& $0.40 \pm 0.01$ & $0.05 \pm 0.06$
	& $3.0 \pm 0.1$ & $9 \pm 1$ & $4 \pm 1$
	& -- \\
$\mathcal{M}$0.3Re3000Pm1
	& $3000$ & $3000$ & $1$
	& $5.0 \times 10^{-5}$ & $5.0 \times 10^{-5}$ & $0.29 \pm 0.01$
	& $0.76 \pm 0.01$ & $0.32 \pm 0.04$
	& $75.5 \pm 1.3$ & $41 \pm 1$ & $16 \pm 1$
	& $576$ \\
$\mathcal{M}$0.3Re600Pm5
	& $600$ & $3000$ & $5$
	& $2.5 \times 10^{-4}$ & $5.0 \times 10^{-5}$ & $0.31 \pm 0.01$
	& $1.00 \pm 0.01$ & $0.43 \pm 0.05$
	& $27.7 \pm 0.6$ & $30 \pm 1$ & $10 \pm 1$
	& $576$ \\
$\mathcal{M}$0.3Re300Pm10
	& $300$ & $3000$ & $10$
	& $5.0 \times 10^{-4}$ & $5.0 \times 10^{-5}$ & $0.30 \pm 0.01$
	& $0.93 \pm 0.01$ & $0.7 \pm 0.1$
	& $16.7 \pm 0.3$ & $22 \pm 1$ & $8 \pm 1$
	& -- \\
$\mathcal{M}$0.3Re100Pm30
	& $100$ & $3000$ & $30$
	& $1.5 \times 10^{-3}$ & $5.0 \times 10^{-5}$ & $0.30 \pm 0.01$
	& $0.76 \pm 0.01$ & $0.9 \pm 0.2$
	& $8.4 \pm 0.2$ & $20 \pm 1$ & $8 \pm 1$
	& -- \\
$\mathcal{M}$0.3Re24Pm125
	& $24$ & $3000$ & $125$
	& $6.3 \times 10^{-3}$ & $5.0 \times 10^{-5}$ & $0.31 \pm 0.02$
	& $0.83 \pm 0.01$ & $1.6 \pm 0.3$
	& $4.2 \pm 0.1$ & $16 \pm 1$ & $7 \pm 1$
	& -- \\
$\mathcal{M}$0.3Re10Pm300
	& $10$ & $3000$ & $300$
	& $1.5 \times 10^{-2}$ & $5.0 \times 10^{-5}$ & $0.30 \pm 0.02$
	& $0.79 \pm 0.01$ & $2.4 \pm 0.4$
	& $3.0 \pm 0.1$ & $14 \pm 1$ & $7 \pm 1$
	& -- \\
$\mathcal{M}$0.3Re2000Pm5
	& $2000$ & $10000$ & $5$
	& $7.0 \times 10^{-5}$ & $1.0 \times 10^{-5}$ & $0.29 \pm 0.01$
	& $1.34 \pm 0.01$ & $0.36 \pm 0.04$
	& $70.6 \pm 1.0$ & $80 \pm 1$ & $26 \pm 1$
	& $576, 1152$ \\
\hline
\multicolumn{13}{c}{$\Mach = 1$} \\
\hline
$\mathcal{M}$1Re3000Pm1
	& $3000$ & $3000$ & $1$
	& $1.7 \times 10^{-4}$ & $1.7 \times 10^{-4}$ & $0.97 \pm 0.04$
	& $0.59 \pm 0.01$ & $0.18 \pm 0.03$
	& $78.0 \pm 1.3$ & $38 \pm 1$ & $13 \pm 1$
	& $576$ \\
$\mathcal{M}$1Re600Pm5
	& $600$ & $3000$ & $5$
	& $8.5 \times 10^{-4}$ & $1.7 \times 10^{-4}$ & $1.02 \pm 0.03$
	& $0.72 \pm 0.01$ & $0.40 \pm 0.05$
	& $30.7 \pm 0.9$ & $27 \pm 1$ & $8 \pm 1$
	& $576$ \\
$\mathcal{M}$1Re300Pm10
	& $300$ & $3000$ & $10$
	& $1.7 \times 10^{-3}$ & $1.7 \times 10^{-4}$ & $1.04 \pm 0.04$
	& $0.78 \pm 0.01$ & $0.45 \pm 0.08$
	& $18.9 \pm 0.8$ & $22 \pm 1$ & $8 \pm 1$
	& -- \\
$\mathcal{M}$1Re24Pm125
	& $24$ & $3000$ & $125$
	& $2.1 \times 10^{-2}$ & $1.7 \times 10^{-4}$ & $1.04 \pm 0.06$
	& $0.83 \pm 0.01$ & $1.1 \pm 0.2$
	& $4.4 \pm 0.1$ & $16 \pm 1$ & $8 \pm 1$
	& -- \\
\hline
\multicolumn{13}{c}{$\Mach = 5$} \\
\hline
$\mathcal{M}$5Re10Pm25
	& $10$ & $250$ & $25$
	& $2.5 \times 10^{-1}$ & $1.0 \times 10^{-2}$ & $5.1 \pm 0.3$
	& $0.43 \pm 0.01$ & $0.14 \pm 0.05$
	& $3.2 \pm 0.1$ & $7 \pm 1$ & $3 \pm 1$
	& $576$ \\
$\mathcal{M}$5Re10Pm50
	& $10$ & $500$ & $50$
	& $2.5 \times 10^{-1}$ & $5.0 \times 10^{-3}$ & $5.1 \pm 0.3$
	& $0.64 \pm 0.01$ & $0.26 \pm 0.06$
	& $3.2 \pm 0.1$ & $9 \pm 1$ & $4 \pm 1$
	& $576$ \\
$\mathcal{M}$5Re10Pm125
	& $10$ & $1250$ & $125$
	& $2.5 \times 10^{-1}$ & $2.0 \times 10^{-3}$ & $5.0 \pm 0.3$
	& $0.77 \pm 0.01$ & $0.29 \pm 0.04$
	& $3.2 \pm 0.1$ & $13 \pm 1$ & $6 \pm 1$
	& $576$ \\
$\mathcal{M}$5Re10Pm250
	& $10$ & $2500$ & $250$
	& $2.5 \times 10^{-1}$ & $1.0 \times 10^{-3}$ & $5.0 \pm 0.3$
	& $0.81 \pm 0.01$ & $0.40 \pm 0.07$
	& $3.2 \pm 0.1$ & $16 \pm 1$ & $9 \pm 1$
	& $576$ \\
$\mathcal{M}$5Re500Pm1
	& $500$ & $500$ & $1$
	& $5.0 \times 10^{-3}$ & $5.0 \times 10^{-3}$ & $5.1 \pm 0.3$
	& $0.19 \pm 0.01$ & $0.01 \pm 0.01$
	& $37.5 \pm 0.7$ & $16 \pm 4$ & $3 \pm 1$
	& -- \\
$\mathcal{M}$5Re500Pm2
	& $500$ & $1000$ & $2$
	& $5.0 \times 10^{-3}$ & $2.5 \times 10^{-3}$ & $5.2 \pm 0.3$
	& $0.34 \pm 0.01$ & $0.04 \pm 0.01$
	& $37.8 \pm 0.7$ & $21 \pm 3$ & $4 \pm 2$
	& -- \\
$\mathcal{M}$5Re500Pm4
	& $500$ & $2000$ & $4$
	& $5.0 \times 10^{-3}$ & $1.3 \times 10^{-3}$ & $5.1 \pm 0.3$
	& $0.42 \pm 0.01$ & $0.06 \pm 0.01$
	& $37.6 \pm 0.7$ & $25 \pm 2$ & $4 \pm 2$
	& -- \\
$\mathcal{M}$5Re3000Pm1
	& $3000$ & $3000$ & $1$
	& $8.3 \times 10^{-4}$ & $8.3 \times 10^{-4}$ & $5.1 \pm 0.2$
	& $0.35 \pm 0.01$ & $0.03 \pm 0.01$
	& $101.5 \pm 1.2$ & $46 \pm 5$ & $5 \pm 2$
	& $576$ \\
$\mathcal{M}$5Re1500Pm2
	& $1500$ & $3000$ & $2$
	& $1.7 \times 10^{-3}$ & $8.3 \times 10^{-4}$ & $5.1 \pm 0.3$
	& $0.37 \pm 0.01$ & $0.05 \pm 0.01$
	& $80.8 \pm 1.2$ & $41 \pm 3$ & $4 \pm 2$
	& $576$ \\
$\mathcal{M}$5Re600Pm5
	& $600$ & $3000$ & $5$
	& $4.2 \times 10^{-3}$ & $8.3 \times 10^{-4}$ & $5.0 \pm 0.2$
	& $0.44 \pm 0.01$ & $0.07 \pm 0.01$
	& $43.9 \pm 0.8$ & $34 \pm 3$ & $4 \pm 1$
	& $576$ \\
$\mathcal{M}$5Re300Pm10
	& $300$ & $3000$ & $10$
	& $8.3 \times 10^{-3}$ & $8.3 \times 10^{-4}$ & $5.1 \pm 0.3$
	& $0.50 \pm 0.01$ & $0.11 \pm 0.04$
	& $27.0 \pm 0.6$ & $25 \pm 2$ & $5 \pm 1$
	& -- \\
$\mathcal{M}$5Re120Pm25
	& $120$ & $3000$ & $25$
	& $2.1 \times 10^{-2}$ & $8.3 \times 10^{-4}$ & $5.0 \pm 0.3$
	& $0.57 \pm 0.01$ & $0.13 \pm 0.01$
	& $14.3 \pm 0.4$ & $25 \pm 1$ & $7 \pm 1$
	& $576$ \\
$\mathcal{M}$5Re60Pm50
	& $60$ & $3000$ & $50$
	& $4.2 \times 10^{-2}$ & $8.3 \times 10^{-4}$ & $4.9 \pm 0.3$
	& $0.72 \pm 0.01$ & $0.25 \pm 0.03$
	& $9.0 \pm 0.3$ & $23 \pm 1$ & $7 \pm 1$
	& $576$ \\
$\mathcal{M}$5Re24Pm125
	& $24$ & $3000$ & $125$
	& $1.0 \times 10^{-1}$ & $8.3 \times 10^{-4}$ & $4.9 \pm 0.3$
	& $0.76 \pm 0.01$ & $0.38 \pm 0.09$
	& $5.0 \pm 0.2$ & $19 \pm 1$ & $8 \pm 1$
	& $576$ \\
$\mathcal{M}$5Re12Pm250
	& $12$ & $3000$ & $250$
	& $2.1 \times 10^{-1}$ & $8.3 \times 10^{-4}$ & $4.8 \pm 0.3$
	& $0.72 \pm 0.01$ & $0.42 \pm 0.06$
	& $3.4 \pm 0.1$ & $17 \pm 1$ & $8 \pm 1$
	& $576$ \\
$\mathcal{M}$5Re2000Pm5
	& $2000$ & $10000$ & $5$
	& $1.3 \times 10^{-3}$ & $2.5 \times 10^{-4}$ & $4.9 \pm 0.3$
	& $0.49 \pm 0.01$ & $0.09 \pm 0.02$
	& $104.2 \pm 1.5$ & $71 \pm 3$ & $9 \pm 2$
	& $576, 1152$ \\
\hline
\multicolumn{13}{c}{$\Mach = 10$} \\
\hline
$\mathcal{M}$10Re3000Pm1
	& $3000$ & $3000$ & $1$
	& $1.7 \times 10^{-3}$ & $1.7 \times 10^{-3}$ & $9.6 \pm 0.5$
	& $0.44 \pm 0.01$ & $0.02 \pm 0.01$
	& $98.5 \pm 1.7$ & $53 \pm 4$ & $5 \pm 2$
	& $576$ \\
$\mathcal{M}$10Re600Pm5
	& $600$ & $3000$ & $5$
	& $8.3 \times 10^{-3}$ & $1.7 \times 10^{-3}$ & $9.8 \pm 0.5$
	& $0.55 \pm 0.01$ & $0.05 \pm 0.01$
	& $45.5 \pm 0.8$ & $37 \pm 2$ & $6 \pm 1$
	& $576$ \\
$\mathcal{M}$10Re300Pm10
	& $300$ & $3000$ & $10$
	& $1.7 \times 10^{-2}$ & $1.7 \times 10^{-3}$ & $10.1 \pm 0.5$
	& $0.65 \pm 0.01$ & $0.07 \pm 0.02$
	& $28.5 \pm 0.6$ & $26 \pm 2$ & $6 \pm 2$
	& -- \\
\hline
\hline
    \end{tabular}
    \end{center}
    \vspace{-0.25 cm}
    \begin{tablenotes}[para]
         \textbf{Column (1):} unique simulation ID. \textbf{Column (2):} the hydrodynamic Reynolds number (\autoref{dfn:Reyh}). \textbf{Column (3):} the magnetic Reynolds number (\autoref{dfn:Reym}). \textbf{Column (4):} the magnetic Prandtl number (\autoref{dfn:Pranm}). \textbf{Columns (5) and (6):} the kinematic viscosity (in \autoref{mhd:momentum}) and magnetic resistivity (in \autoref{mhd:induction}) expressed in units of the turbulent turnover-time ($t_\tturb$), and the driving scale ($\ell_\tturb$). \textbf{Column (7):} the turbulent sonic Mach number ($\Mach = u_\tturb / c_s$), where $c_s$ is the speed of sound. \textbf{Column (8):} the exponential growth rate of the volume-integrated magnetic energy ($E_{\tmag}$) during the exponential-growth (kinematic) phase of the small-scale dynamo (SSD), in units of $t_\tturb$. \textbf{Column (9):} the ratio of the volume-integrated magnetic to kinetic energy ($E_{\tkin}$) in the saturated state of the SSD. \textbf{Columns (10), (11), and (12):} the characteristic kinetic dissipation wavenumber ($k_{\nu}$; see \autoref{sec:tools:k_nu}), magnetic dissipation wavenumber ($k_{\eta}$; see \autoref{sec:tools:k_eta}), and peak scale of the magnetic energy power spectrum ($k_{\tp}$; see \autoref{sec:tools:k_p}) during the kinematic phase. Note, all scales are expressed in units of $k_{\tbox} = \ell_{\tbox}/(2\pi)$. \textbf{Column (13):} extra grid resolutions that were explored in addition to the default $N_\tres \in \{18, 36, 72, 144, 288\}$ (see \autoref{sec:ICs:nres} for details).
    \end{tablenotes}
\end{table*}

\section{Results}
\label{sec:results}

    Before we detail our methods for characterising field structures in our simulations, we first confirm that we measure dynamo growth for all our simulations in \autoref{sec:results:dynamo_phases}. We also discuss the effect of compressibility on the efficiency of the dynamo, and define how we isolate the time-range corresponding with the kinematic phase. Then, in \autoref{sec:results:field_slices}, we compare magnetic field morphologies in the subsonic and supersonic regimes, which motivates our methods for measuring characteristic MHD scales described in \autoref{sec:results:tools}. We evaluate convergence in \autoref{sec:results:nres}, and then analyse trends of (converged) characteristic scales in \autoref{sec:results:dissipation} and \ref{sec:results:peak}.
    
    \subsection{Simulation Phases}
    \label{sec:results:dynamo_phases}
    
    \begin{figure}
        \centering
        \includegraphics[width=\linewidth]{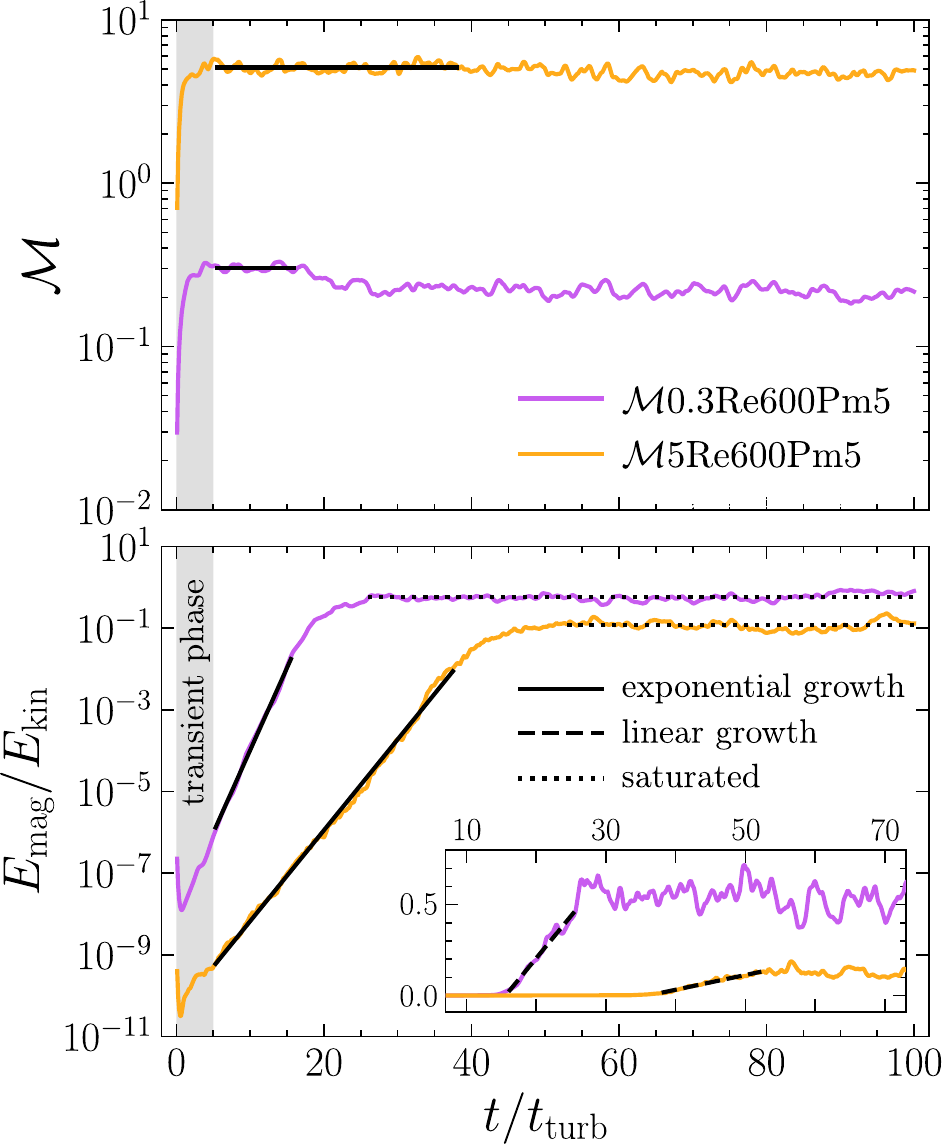}
        \caption{
        Time evolution of the root-mean-squared sonic Mach number ($\Mach$; top panel) and the ratio of the volume-integrated magnetic to kinetic energy ($E_\tmag/E_\tkin$; bottom panel) for \SimName{0.3}{600}{5} (purple) and \SimName{5}{600}{5} (yellow). Both plasmas have $\Reyh = 600 > \Reyh_\tcrit \approx 100$ and $\Pranm = 5$, but $\Mach = 0.3$ and $5$, respectively. We indicate four distinct phases in the simulations: (1) a transient phase where the turbulent velocity field becomes fully established (grey shaded region); (2) a kinematic phase when the magnetic field grows exponentially in time (black solid lines show fits of an exponential model); (3) a linear-growth phase that begins once magnetic fields are strong enough to suppress some of the kinetic motions (dashed black lines in the inset panel show linear fits); (4) a saturated phase that begins once the magnetic energy is close to equipartition with the kinetic energy (dotted horizontal lines). We report the measured exponential growth rate, and saturated energy ratio for each of our simulation setups in columns (8) and (9) of \autoref{table:summary}, respectively.
        }
        \label{fig:time_evolution}
    \end{figure}

    In \autoref{fig:time_evolution} we plot the time evolution of the rms $\Mach$ and volume-integrated ratio of magnetic to kinetic energy,
    \begin{align}
        E_\tratio
            \equiv \frac{E_\tmag}{E_\tkin}
            = \lfrac{
                    \int_\vol b^2 / (8\pi) \,\dd{\vol}
                }{
                    \int_\vol \rho u^2 / 2\, \dd{\vol}
                }
    \end{align}
    for two representative simulations, \SimName{0.3}{600}{5} in purple and \SimName{5}{600}{5} in yellow. These runs have identical plasma numbers, $\Reyh = 600$ with $\Pranm = 5$, but differ in $\Mach$. In both simulations shown, and in fact for all our simulations (see \autoref{table:summary}), we identify four distinct phases: a transient phase at the start of the simulation, immediately followed by the exponential-growth (kinematic), linear-growth, and finally, saturated dynamo phase. 
    
    The transient phase roughly spans $0 \leq t/t_\tturb \leq 5$, and may be a mixture of the time it takes for our imposed forcing field to accelerate the plasma into a fully developed (statistically stationary) turbulent state, or the so-called diffusion-free regime, where the magnetic field itself moves down to the resistive scale \citep{schekochihin2002spectra}. During this initial phase, the magnetic field reorganises itself out of its initial configuration, and in the case of subsonic turbulence, into a self-similar configuration that has most of its energy concentrated on the smallest scales (\eg \citetalias{kriel2022fundamental} and \citealt{beattie2022growth}). However, we do not focus on this initial, potentially diffusion-free phase, and instead move directly to the classical kinematic stage, which is the focus of this study.
    
    Once the magnetic field progresses past the initial transient state, we see the onset of the kinematic phase. Here, $E_\tmag$ grows exponentially fast, amplifying the magnetic energy by more than 7~orders of magnitude, until it reaches $\sim 10\%$ of the kinetic energy. We measure the growth rate, $\gamma$, of magnetic energy during this phase by fitting each simulation with an exponential model, $E_\tmag(t) \sim \exp(\gamma t)$, over the time range $5t_\tturb \leq t \leq t_\tend$, where $t_\tend$ is implicitly defined as the time when $E_\tratio(t_\tend) = 10^{-2}$. For the two simulations shown, namely \SimName{0.3}{600}{5} and \SimName{5}{600}{5}, we measure growth rates $\gamma = (1.00 \pm 0.01)t_{\tturb}^{-1}$ and $\gamma = (0.44 \pm 0.01)t_{\tturb}^{-1}$, and report the measured $\gamma$ for all of our simulations in column 8 of \autoref{table:summary}. Inspection of these values support the idea that at fixed $\Reyh$ and $\Pranm$, $\gamma$ is generally lower in supersonic compared with subsonic SSDs \citep[see, \eg][for more detailed analysis on this effect]{federrath2011mach, schober2012magnetic, federrath2014turbulent, chirakkara2021efficient}. During this (kinematic) phase we also confirm that $\Mach$ remains statistically stationary, and within $5\%$ of our desired value for all our simulations; $\Mach = 0.31 \pm 0.01$ for \SimName{0.3}{600}{5}, and $\Mach = 5.0 \pm 0.2$ for \SimName{5}{600}{5} (see column 7 of \autoref{table:summary} for all other simulations).
    
    Following the kinematic phase, the magnetic energy growth transitions from an exponential-in-time process to a linear-in-time (or secular) process for both $\Mach <1$ and $\Mach \geq 1$ SSDs. To illustrate this, we plot $E_\tratio$ for our two representative simulations on a linear-linear scale in the inset axis of the bottom panel in \autoref{fig:time_evolution}. Here, for our $\Mach \geq 1$ SSDs, we do not find a transition into quadratic growth, as has been suggested should be the case by \citet{Schleicher2013_quadratic_growth}. Moreover, as was the case in the kinematic phase, we find that the growth rate in the linear-growth phase is lower for the supersonic, \ie \SimName{5}{600}{5}, compared with the subsonic, \ie \SimName{0.3}{600}{5}, SSDs, which is not directly predicted by any linear growth model based on the energy flux of the hydrodynamical cascade \citep[e.g.,]{Xu2016_dynamo,StOnge2020_weakly_collisional_dynamo,beattie2023bulk}. Interestingly, for all of our simulations we find that the duration of the kinematic and linear-growth phases are roughly equal. While the duration of the kinematic phase decreases (as the exponential growth rate increases) with plasma numbers (\ie $\Reyh$ and $\Pranm$), the duration of the linear-growth phase also decreases by a proportional amount.
    
    Finally, once the magnetic energy approaches equipartition with the kinetic energy, the energy ratio saturates, and is maintained thereafter at a nearly constant value by the forcing field; this defines the saturated phase. We measure $(E_\tratio)_\tsat = 0.43 \pm 0.05$ and $0.07 \pm 0.01$ for \SimName{0.3}{600}{5} and \SimName{5}{600}{5}, respectively, and report this ratio for all simulations in column~9 of \autoref{table:summary}. Again, these values are generally smaller in the supersonic compared with subsonic regimes.
    
    \subsection{Magnetic Structures}
    \label{sec:results:field_slices}
    
    \begin{figure*}
        \centering
        \includegraphics[width=\linewidth]{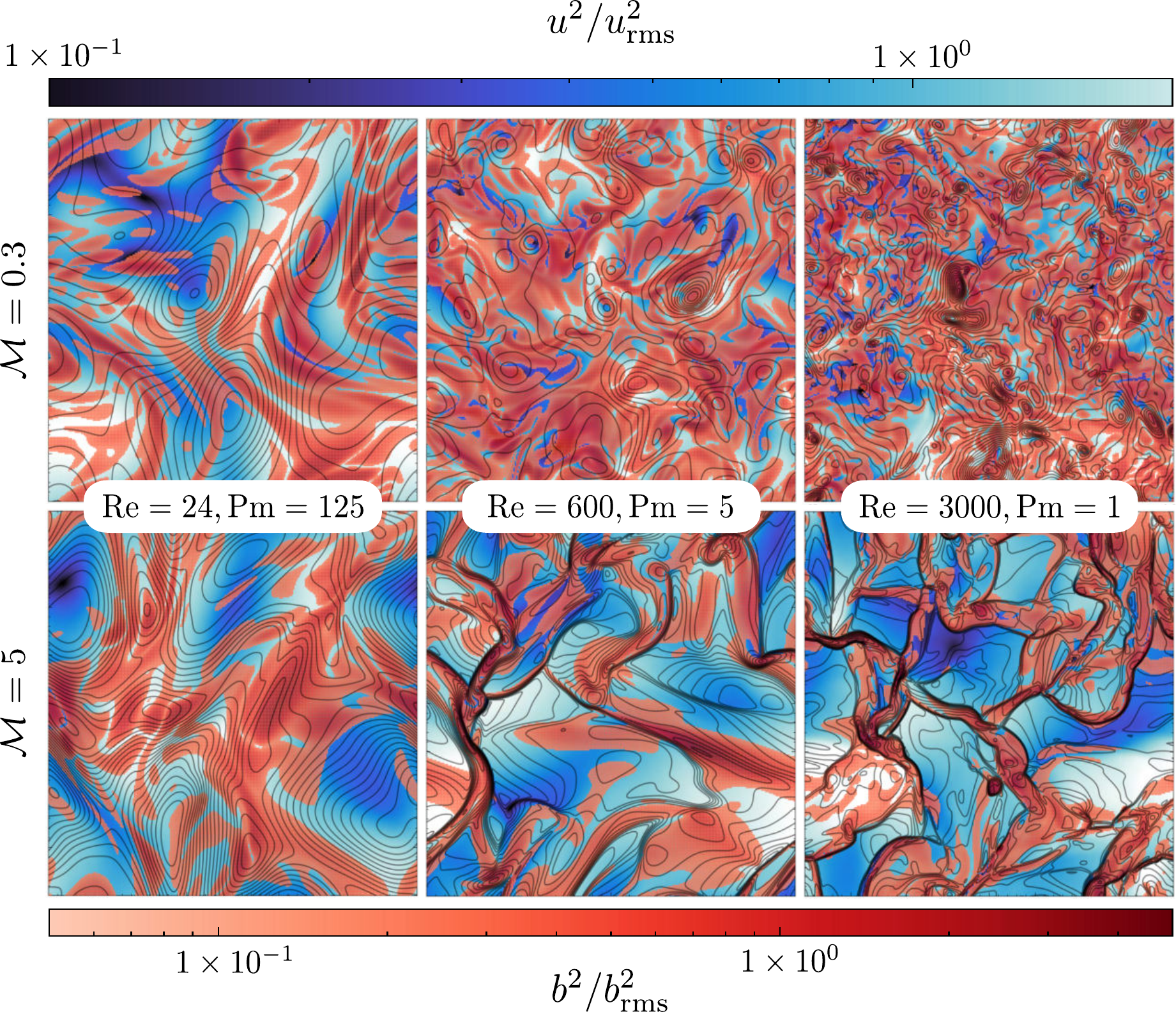}
        \caption{Two-dimensional slices of $u^2/u_\trms^2$ (blue), $b^2/b_\trms^2$ (red), and $\rho/\rho_\trms$ (black contours) fields for six different simulations at $N_\tres = 576$, spanning a range of plasma regimes. Note that the colourbar for $b^2/b_\trms^2$ is transparent for values $b^2/b_\trms^2 < 0.5$, so regions of low magnetic energy density are not visible. The six simulations shown are \SimName{0.3}{24}{125}, \SimName{0.3}{600}{5}, and \SimName{0.3}{3000}{1} in the top row, and \SimName{5}{24}{125}, \SimName{5}{600}{5}, and \SimName{5}{3000}{1} in the bottom row, respectively, where all simulations have $\Reym = 3000$, with $\Pranm$ (and therefore also $\Reyh$) changing, and velocity flows are subsonic ($\Mach = 0.3$) in the top row, and supersonic ($\Mach = 5$) in the bottom row. We see that in the viscous flow regime ($\Reyh < \Reyh_\tcrit \approx 100$; left column), both the subsonic and supersonic simulations produce similar, large-scale magnetic structures. Moving towards the turbulent regime ($\Reyh \gg \Reyh_\tcrit$; middle and right columns), structures that carry most of the magnetic field energy live on significantly smaller scales compared with the viscous regime. In the subsonic regime (top row), magnetic fields are more uniformly distributed and space-filling, occupying even smaller scales than in the corresponding supersonic simulations (bottom row). In the supersonic simulations, magnetic fields are concentrated in high-density shocked regions bounded by sharp jumps in the velocity magnitude, with an almost constant characteristic length of approximately $\ell_\tturb$, and shock width $\ell_\tshock$ decreasing with increasing $\Reyh$.}
        \label{fig:field_slices}
    \end{figure*}
    
    Now that we have confirmed we observe SSD growth in all of our simulations, we turn our attention to field morphologies that are produced during the kinematic phase. In \autoref{fig:field_slices} we plot 2D field slices for six simulations from \autoref{table:summary} (with the two simulations in \autoref{fig:time_evolution} plotted in the two middle column-panels). All slices are taken from the middle of the simulation domain, $(x, y, z=\ell_\tbox/2)$, at a time realisation midway through the kinematic phase, \viz $(5 + t_\tend)/2$, where $t_\tend$ is defined as in the previous section. The top and bottom row panels show slices for simulations in two different $\Mach$ regimes, where the top row shows subsonic simulations with $\Mach = 0.3$, and the bottom row shows supersonic simulations with $\Mach = 5$. For all simulations in this figure, we keep \mbox{$\Reym = 3000$} fixed, and vary $\Pranm$ (and thereby $\Reyh$, or more explicitly\footnote{Because $\ell_\tturb$ and $u_\tturb$ is fixed in each row.} $\nu$) between the columns.
    
    In the left column we show two simulations in the viscous-flow regime, where $\Reyh = 24 < \Reyh_\tcrit$, in the middle column we show mildly turbulent flows, $\Reyh = 600 \gtrsim \Reyh_\tcrit$, and the right column we show highly turbulent flows, $\Reyh = 3000 \gg \Reyh_\tcrit$. In each panel we plot $u^2 / u_\trms^2$ in blue, $b^2/b_\trms^2$ in red, and $\rho/\rho_\trms$ contours in black, where we have normalised both the velocity and magnetic fields by their rms values to reveal the underlying structure. Note that we truncate the magnetic energy colourbar to only show regions where $b^2/b_\trms^2 > 0.5$, so that the weak-field regions are not shown (\ie transparent).

    Qualitatively, the top row of \autoref{fig:field_slices} showcases how the distribution of magnetic energy in subsonic plasmas shifts from being primarily present in large-scale structures in viscous flows (top-left panel), to smaller scale structures in turbulent flows (top-right panel). This systematic transition is characteristic of subsonic SSDs during the kinematic phase, where magnetic fields become stretched, folded, and ultimately organised with most of their energy concentrated at the smallest available scales allowed by Ohmic dissipation.
    
    This is understood to be the result of viscous-scale kinetic fluctuations that most rapidly stretch the magnetic field, giving rise to subviscous magnetic structures that are relatively straight, and become increasingly interspersed with regions of field reversals on smaller scales. \citet{Kazantsev1968} showed that the eigenfunction of the induction equation during the kinematic phase of a subsonic SSD, assuming a balance between field stretching and diffusion, leads to a $\ell^{-3/2}$ self-similar scaling of magnetic energy that concentrates energy at the resistive scale. \citet{schekochihin2004simulations} further predicted that the scale on which the magnetic field becomes dominated by dissipation is determined by the balance of the viscous-scale stretching rate and magnetic dissipation rate (more on this in \autoref{sec:results:dissipation:keta}), which yields $\ell_\tp \sim \ell_\eta \sim \ell_\nu \,\Pranm^{-1/2}$. In \citetalias{kriel2022fundamental} we confirmed this predicted scaling using DNSs of subsonic SSDs with resolved physical dissipation, and also measured power-law exponents of the magnetic energy that was consistent with $\ell^{-3/2}$.

    Our subsonic simulations (\eg top row of \autoref{fig:field_slices}) once again align with this picture. Here we see that as $\Reyh$ increases and $\ell_\nu \sim \Reyh^{-3/4}$ shifts to smaller scales, the flow becomes increasingly more turbulent, supporting smaller scale kinetic fluctuations. These turbulent motions give rise to magnetic structures that are significantly smaller scale than the kinetic motions, via the process we described above. For constant $\Reyh$ subsonic plasmas, one also expects to see smaller scale magnetic structures emerge, since $\ell_\tp \sim \Reym^{1/2}$, however, here our choice to highlight simulations where $\Reym = 3000$ is fixed, is to demonstrate the effect of viscosity, which will become important in \autoref{sec:results:peak}.
    
    Now, in the supersonic regime (bottom row of \autoref{fig:field_slices}), we observe the magnetic morphology converge on a completely different configuration in the transition from viscous to turbulent plasmas. While both the subsonic and supersonic plasmas shown in the viscous regime (left column) have similar large-scale magnetic structures, the highly turbulent plasmas (right column) produce distinctively different energy distribution patterns. The turbulent, subsonic plasma (top-right panel) produces significantly smaller-scale magnetic energy structures compared with the turbulent, supersonic plasma (bottom-right panel), where magnetic energy becomes primarily concentrated in elongated, coherent, shocked regions of gas, as illustrated by the tightly-packed density contours that coincide with the boundaries of plasma regions where the magnetic energy is strongest.
    
    Embedded within these shocked structures, we still observe sub-structures that appear to have field correlations similar to the subsonic plasma. However, in addition to the smallest scale field structures associated with magnetic dissipation, we now see the emergence of another, larger, scale in the magnetic energy distribution, which appears to be associated with shocks. This scale appears to vary across the supersonic simulations, despite all the supersonic runs being forced on the same scale and with the same strength ($k_\tturb = 2$ and $\Mach = 5$). This suggests that the location of the second scale depends not just on turbulent driving, but also on the visco-resistive properties of the plasma.
    
    While there exist a number of systematic studies of the volume-averaged SSD properties (\eg growth rate and saturated energy ratio) across this transition in the supersonic regime \citep[see for example][]{federrath2011mach, chirakkara2021efficient}, a direct, systematic study of the underlying magnetic field properties has yet to be performed, and therefore we focus on this aspect for the remainder of this study.

    \subsection{Measuring Characteristic Scales from Energy Spectra}
    \label{sec:results:tools}
    
    To understand the differences in magnetic field characteristic scales that we saw visually in the previous section, we now quantify three key characteristic scales: the size of viscous fluctuations ($\ell_\nu$), which most dominantly drive SSD amplification during the kinematic phase of subsonic SSDs; the resistive scale ($\ell_\eta$), where Ohmic dissipation becomes significant; and the scale where magnetic energy is concentrated ($\ell_\tp$). By establishing a relationship between these scales and plasma parameters (\eg $\Mach$, $\Reyh$, and $\Pranm$), we aim to compare and contrast subsonic and supersonic plasma SSDs, thereby shedding light on how amplification differs in supersonic SSDs compared with their subsonic counterparts.
    
    Since the flows we are studying are homogeneous and isotropic, we study characteristic scales in Fourier space, where length-scales and wavenumbers are directly related ($k \equiv 2\pi / \ell$). In \citetalias{kriel2022fundamental}, we measured characteristic wavenumbers from our simulations by fitting semi-analytical models for the kinetic ($E_\tkin(k)$) and magnetic ($E_\tmag(k)$) energy spectra. These spectra were derived from 1D shell-integrated power spectra calculated from our simulations in the usual way of summing the total power in discrete, radial shells in k-space. More explicitly, the energy spectrum of field $\vect{\psi}$ is computed as
    \begin{align}
        E_\psi(k, t)
            = \sum_{k^* < \|\vect{k}\| < k^* + \Delta k^*}
                4\pi k^2 \widetilde{\psi}(\vect{k}, t) \, \widetilde{\psi}^*(\vect{k}, t)
    \end{align}
    where we choose to bin in integer $k$-bins ($k^* \in \mathds{Z}^+ : k^* \ne 0$) separated by $\Delta k^* = 1$, and
    \begin{align}
        \widetilde{\psi}(\vect{k}, t)
            = \frac{1}{(2 \pi \ell_\tbox)^{3/2}} \int_\vol \psi(\vect{\ell}, t) \exp\rbrac{-i \vect{k}\cdot\vect{\ell}} \,\dd{^3 \ell}
            \label{eqn:kin_fourier}
    \end{align}
    is the Fourier transform of $\psi(\vect{\ell}, t)$, with $\widetilde{\psi}^*(\vect{k}, t)$ its complex conjugate.
    
    We also attempted this approach for the present study, but found that the existing functional models for both energy spectra fail to reliably measure characteristic scales in our supersonic simulations, especially in lower-resolution runs, which were necessary to perform our resolution study. Two main issues arose: (1) due to the limited inertial range for our low-$\Reyh$ simulations, we could not effectively constrain $k_\nu$; (2) supersonic simulations had a broader energy spectrum than the subsonic simulations, which was not well-fit by the functional form we had used in \citetalias{kriel2022fundamental} (\ie the $E_{\tmag}(k)$ model originally derived by \citet{kulsrud1992spectrum} for a magnetic field coupled to a delta-correlated in time velocity field during the kinematic phase of a subsonic, $\Pranm \gg 1$ SSD). Furthermore, at very high-$\Reyh$, as shown in \citet{Beattie2024_10k_MHD}, if there are break scales in the spectrum it is not obvious what the dissipation scales should be if derived directly from a fit.
    
    Prompted by these challenges, we develop new and simpler, spectral model-free methods for measuring $k_\nu$, $k_\eta$, and $k_\tp$, based on the underlying turbulence and fluid physics theory, which we apply to all our simulations, and demonstrate yields robust results for all plasma and flow regimes.
    
    \subsubsection{Characteristic Viscous Wavenumber}
    \label{sec:tools:k_nu}

    \begin{figure}
        \includegraphics[width=\linewidth]{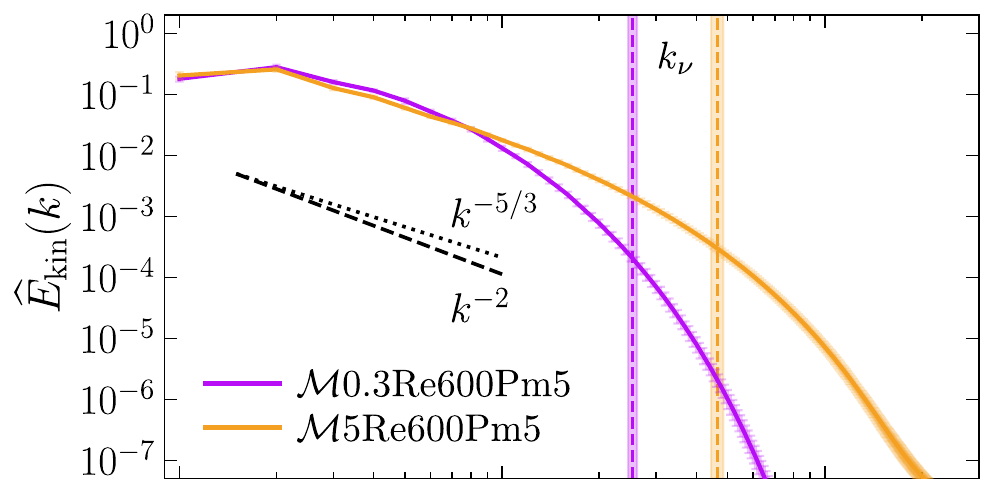}
        \includegraphics[width=\linewidth]{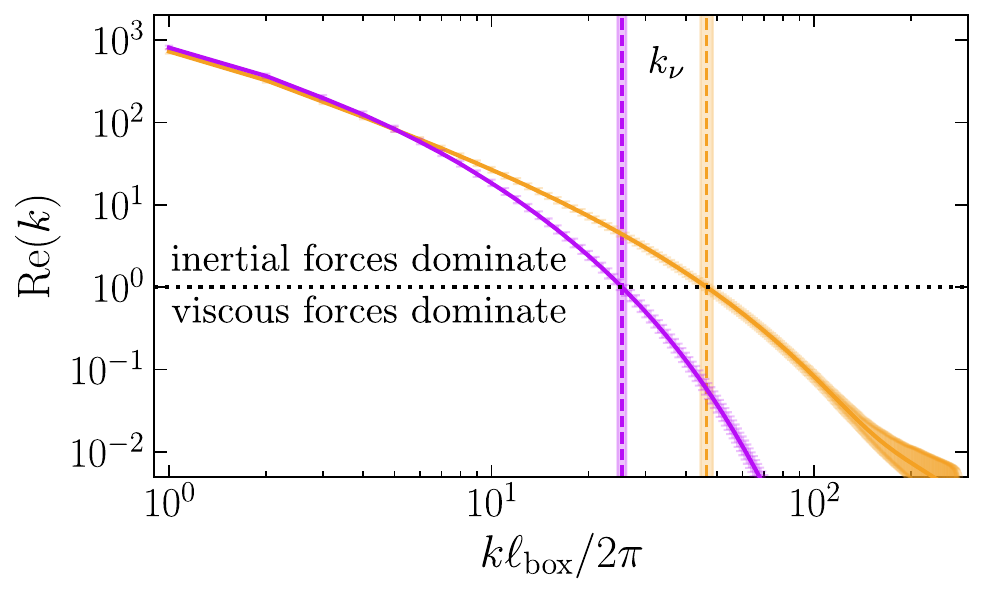}
        \caption{Normalised and time-averaged (over the kinematic phase) kinetic energy spectrum ($\widehat{E}_\tkin(k)$; top panel) constructed with $\vect{\psi} = \vect{u} / \sqrt{2}$ in \autoref{eqn:kin_fourier}, for the \SimName{0.3}{600}{5} (purple) and \SimName{0.3}{600}{5} (yellow) simulations run at $N_\tres = 576$. The solid lines shows the $50^{\rm th}$ percentile of the $\widehat{E}_\tkin(k)$ spectrum over all snapshots during this phase, and the bands show the $16^{\rm th}$ to $84^{\rm th}$ percentile variance of $\widehat{E}_\tkin(k)$. We also plot the wavenumber-dependent hydrodynamic Reynolds number ($\Reyh(k)$ computed via \autoref{eqn:reynolds_spectrum}; bottom panel). From $\Reyh(k)$ we measure a characteristic kinetic energy dissipation (viscous) wavenumber ($k_\nu: \Reyh(k_\nu) = 1$ via \autoref{eqn:knu_measure}), which we annotate with vertical lines for both simulations in the top and bottom panels. For reference, we also plot $\widehat{E}_\tkin(k) \sim k^{-5/3}$ and $\widehat{E}_\tkin(k) \sim k^{-2}$ in the top panel, which corresponds with the expected inertial range scaling of \citet{kolmogorov1941dissipation} and \citet{burgers1948_turbulence_model} kinetic energy spectra, respectively.}
        \label{fig:kin_spectra}
    \end{figure}

    \begin{figure}
        \centering
        \includegraphics[width=\linewidth]{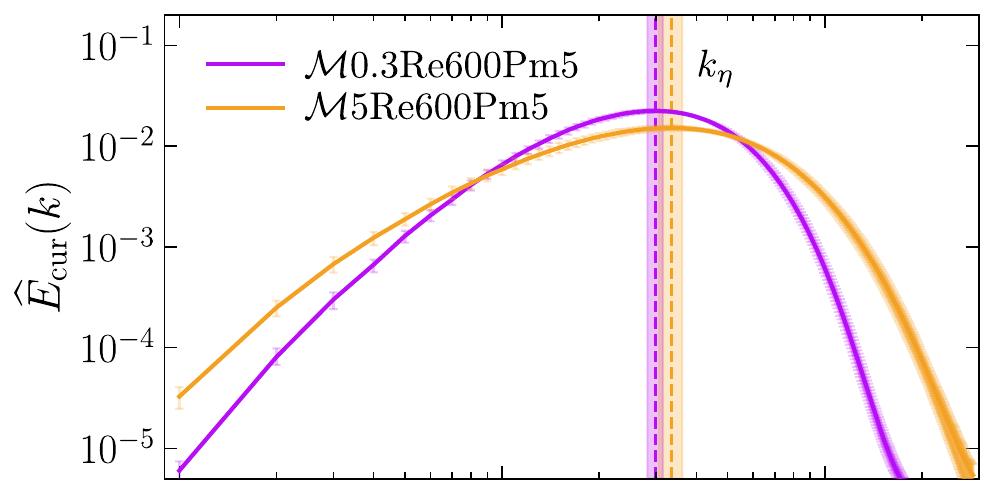}
        \includegraphics[width=\linewidth]{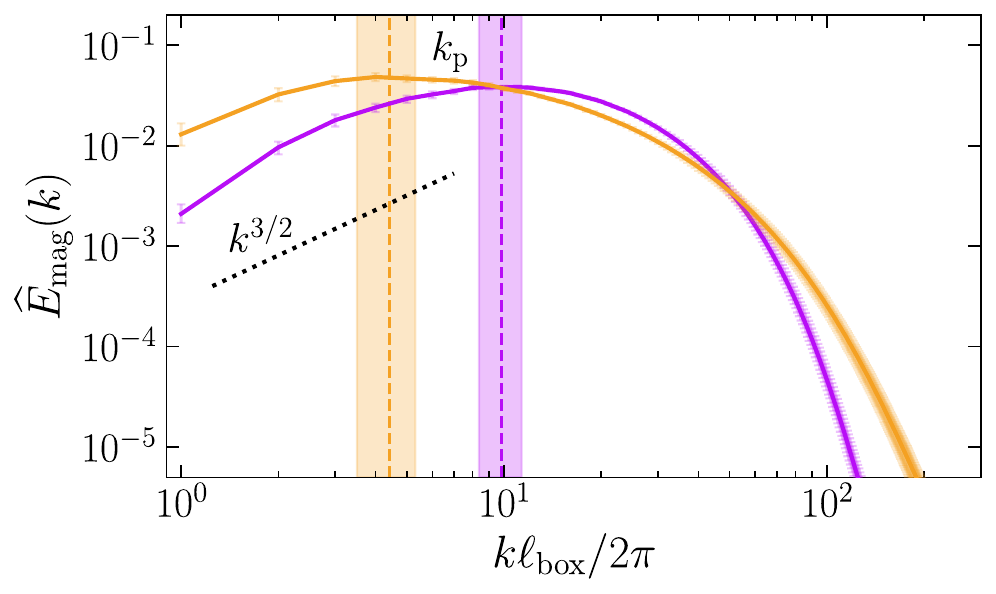}
        \caption{As in \autoref{fig:kin_spectra}, but for the current density power spectra ($\widehat{E}_\tcur(k)$; top panel) and magnetic energy spectrum ($\widehat{E}_\tmag(k)$; bottom panel) for \SimName{0.3}{600}{5} and \SimName{5}{600}{5}. We annotate the measured characteristic magnetic dissipation (resistive) wavenumber in the top panel, ($k_\eta$; \autoref{eqn:keta_measure}), and the magnetic peak wavenumber ($k_\tp$; \autoref{eqn:kp_measure}), in the bottom panel. We also annotate $\widehat{E}_\tmag(k) \sim k^{3/2}$ for reference, which is the expected scaling of $E_\tmag(k)$ in the subviscous range during the kinematic phase of a subsonic SSD \citep{Kazantsev1968, schekochihin2002model}.}
        \label{fig:cur_mag_spectra}
    \end{figure}

    We define the viscous wavenumber, $k_\nu$, directly from the definition in Kolmogorov turbulence, \ie the wavenumber where the scale-dependent hydrodynamic Reynolds number equals one, $\Reyh(k_\nu) = 1$. This scale marks the transition from inertial forces dominating in the turbulent cascade, $k_\tturb < k < k_\nu$, to dissipation dominating at \mbox{$k_\nu < k$}. Hence, to measure $k_\nu$, we construct the wavenumber-dependent hydrodynamic Reynolds number,
    \begin{align}
        \Reyh(k, t)
            = \frac{u_\tturb(k,t)}{\nu k}
            , \label{eqn:reynolds_spectrum}
    \end{align}
    which follows from the turbulent velocity as a function of $k$,
    \begin{align}
        u_\tturb (k, t)
            = \rbrac{
                \frac{2}{\rho_0} \int_{k}^{\infty} E_\tkin(k^\prime, t) \,\dd{k^\prime}
            }^{1/2}
            . \label{eqn:velocity_k}
    \end{align}
    In practice we solve for $k_\nu$ from our simulations using a root finding method,
    \begin{align}
        k_\nu(t)
            = \texttt{argmin}_k\Big[
                \mathcal{I}\Big\{ \Big| \Reyh(k, t) - 1 \Big| \Big\}
            \Big]
            , \label{eqn:knu_measure}
    \end{align}
    where $|\dots|$ is the absolute operator, $\mathcal{I}[\dots]$ is a piecewise cubic polynomial (spline) interpolation operator, as performed in \citet{beattie2022growth}, and $\texttt{argmin}_k[h(k)]$ returns the argument $k$ which minimises the function $h(k)$. For each of our simulations, we evaluate \autoref{eqn:knu_measure} at each time-snapshot during the kinematic phase.

    While \autoref{eqn:knu_measure} provides a simple way of extracting the viscous wavenumber from $E_\tkin(k)$, there remains the question of which field, $\vect{\psi}$, to Fourier transform in the supersonic regime. When $\vect{\psi} = \vect{u}/\sqrt{2}$ in \autoref{eqn:kin_fourier}, then $E_\tkin(k)$ is the velocity power spectrum, which leads \autoref{eqn:reynolds_spectrum} to carry the same units as the usual definition of the hydrodynamic Reynolds number. For incompressible flows, this approach is directly proportional to the kinetic energy spectrum, because the density field is constant. However, this velocity power spectrum definition for $E_\tkin(k)$ ignores the covariance between the density field and the square velocity field, \ie $\ave{\rho u^2}_\vol / 2$ (see \autoref{note:naming_convesions}), as well as fluctuations in the density field, which for compressible, isothermal plasmas are a factor of $\Mach$ larger than velocity fluctuations (this factor follows from a unit analysis of the ideal-hydrodynamic momentum equation in steady state). \citet{beattie2020filaments} showed that the density spectrum becomes dominated by high-$k$ modes that can lead to large pressure gradients, which mediate the exchange of kinetic and internal energy \citep[see for example][]{federrath2010comparing, federrath2013universality, schmidt2019kinetic, grete2021matter, grete2023matter}. Therefore, we also check whether density fluctuations affect our measurements of $k_\nu$, by also considering the definition for $E_\tkin$ which carries the units of kinetic energy, namely with $\vect{\psi} = \vect{u} \sqrt{\rho/2}$ \citep[see, e.g., ][]{federrath2010comparing, grete2021matter, grete2023matter}.
    
    Here, in the main text, we focus on $k_\nu$ derived from \autoref{eqn:reynolds_spectrum} constructed with $\vect{\psi} = \vect{u}/\sqrt{2}$, which we demonstrate in \autoref{fig:kin_spectra} for the same two representative simulations shown in \autoref{fig:time_evolution}, namely \SimName{0.3}{600}{5} and \SimName{5}{600}{5}. Then, in \aref{app:knu:density_weighted}, we demonstrate that regardless of the choice of definition for $E_\tkin(k)$, whether based on $\Reyh(k)$ having dimensionless units (\ie constructed with $\vect{\psi} = \vect{u} / \sqrt{2}$), or choosing $\vect{\psi} = \vect{u} \sqrt{\rho/2}$ such that $E_\tkin(k)$ carries units of kinetic energy, we recover the same scaling behaviour for $k_\nu$, and thus the choice of definition for $E_\tkin$ does not affect any of the conclusions presented in our study.
    
    \subsubsection{Characteristic Resistive Wavenumber}
    \label{sec:tools:k_eta}
    
    While our approach of defining $k_\nu$ in terms of the wavenumber-dependent hydrodynamic Reynolds number performs well, we find that an analogous approach to defining the resistive wavenumber in terms of the spectrum of $\Reym$ yields results that are inconsistent with those derived from early methods \citep{kulsrud1992spectrum, kriel2022fundamental, brandenburg2023dissipative}; see \aref{app:krm} for details. For this reason, we adopt a different approach. Recognising that, since we employ Ohmic dissipation in the induction equation (\autoref{mhd:induction}), the magnetic dissipation rate at any point in space is exactly equal to $\eta j^2$. On this basis we define the resistive wavenumber as the wavenumber of maximum $j^2(k)$, corresponding to the maximum magnetic dissipation (since $\eta$ is a constant). Explicitly, we define $k_\eta$ as the value of $k$ corresponding to the maximum of the 1D shell-integrated current density spectrum (which has units of current density squared), $E_\tcur(k)$, which is defined similarly to $E_\tkin(k)$ in \autoref{sec:tools:k_nu}, but for the field $\vect{\psi} = \nabla\times\vect{b} / (4\pi)$.
    
    In practice we implement this as
    \begin{align}
        k_\eta(t)
            = \texttt{argmax}_k\Big[
                \mathcal{I}\Big\{ E_\tcur(k, t) \Big\}
            \Big]
            , \label{eqn:keta_measure}
    \end{align}
    where $\mathcal{I}\{\dots\}$ is defined as in \autoref{eqn:knu_measure}, and $\texttt{argmax}_k[h(k)]$ returns the argument $k$ which maximises $h(k)$. As with $k_\nu$, we compute this quantity for every snapshot during the kinematic phase. We illustrate this procedure in the top panel of \autoref{fig:cur_mag_spectra}, where in analogy with the top panel of \autoref{fig:kin_spectra}, we plot the normalised and time-averaged (over the kinematic phase) $E_\tcur$ for our two representative simulations, with the corresponding measured resistive wavenumbers $k_\eta$ indicated with vertical bands.

   \subsubsection{Peak Magnetic Wavenumber}
   \label{sec:tools:k_p}
   
    Finally, we define the magnetic peak wavenumber as the maximum of $E_\tmag(k)$, defined similarly to the current density power spectra but with $\vect{\psi} = \vect{b} / {8\pi}$. Explicitly,
    \begin{align}
        k_\tp(t)
            = \texttt{argmax}_k\Big[
                \mathcal{I}\Big\{ E_\tmag(k, t) \Big\}
            \Big]
            . \label{eqn:kp_measure}
    \end{align}
    We illustrate $E_\tmag(k)$ and $k_\tp$ for our two representative simulations in the bottom panel of \autoref{fig:cur_mag_spectra}. In \aref{app:kcor} we point out that, while the magnetic correlation wavenumber is directly proportional to peak wavenumber during the kinematic phase of a subsonic (both viscous and turbulent) SSD \citep{schekochihin2004simulations, galishnikova2022tearing, beattie2022growth}, this scaling breaks down for supersonic, turbulent SSDs.

    \subsection{Convergence of Measured Wavenumbers}
    \label{sec:results:nres}
    
    \begin{figure}
        \centering
        \includegraphics[width=\linewidth]{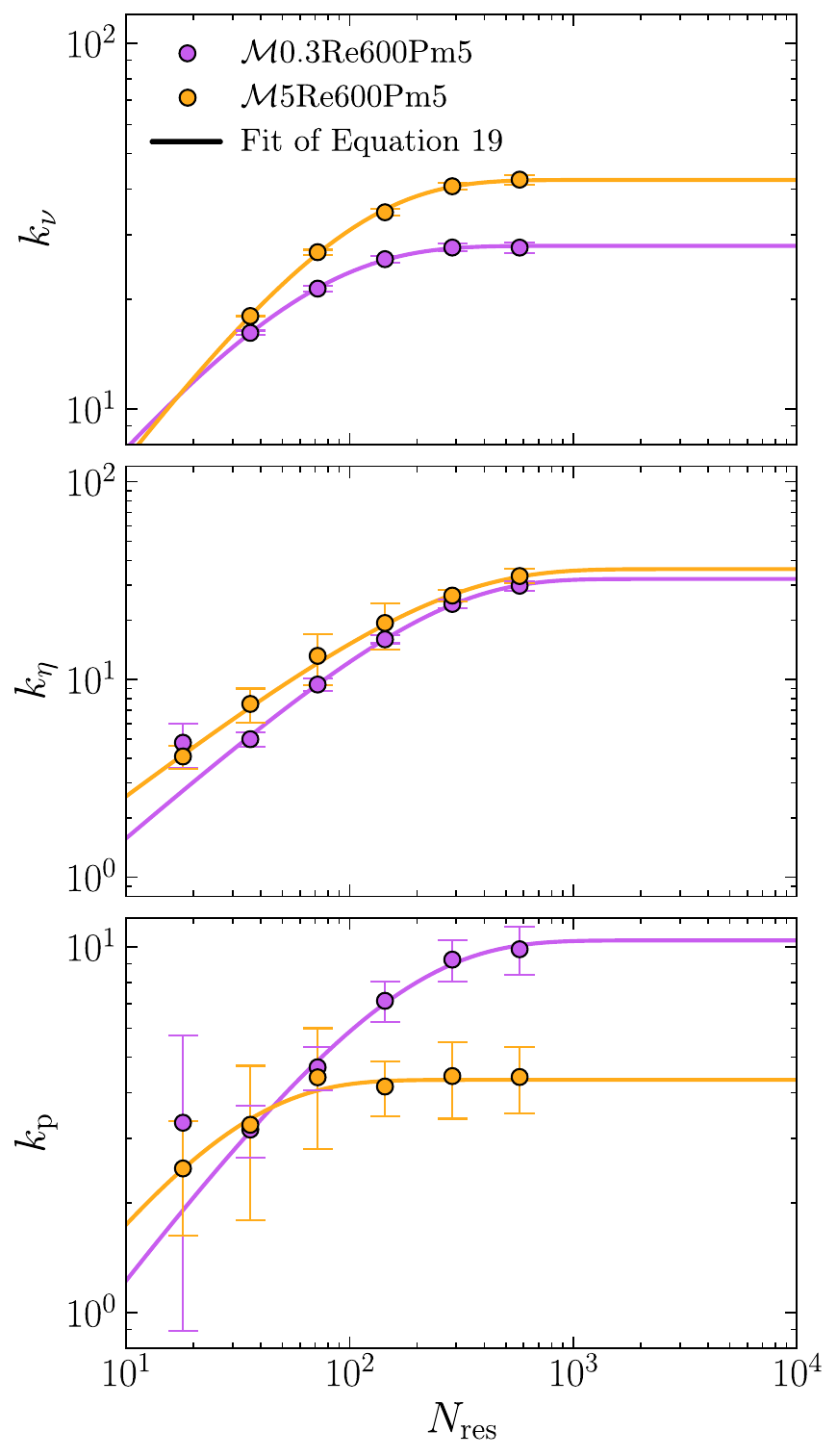}
        \caption{Measured viscous ($k_\nu$; top panel), resistive ($k_\eta$; middle panel), and magnetic peak ($k_\tp$; bottom panel) wavenumbers plotted against linear grid resolution for \SimName{0.3}{600}{5} (magenta) and \SimName{5}{600}{5} (yellow) as circles, indicating the 1-sigma standard deviation in the temporal variation of each scale with error-bars. We overlay a best-fit of our convergence model (\autoref{eqn:nres}), to each set of wavenumbers to measure characteristic wavenumbers that have converged with resolution. We report the converged $k_\nu$, $k_\eta$, and $k_\tp$ for each of our simulations in columns (10), (11), and (12) of \autoref{table:summary}, respectively.}
        \label{fig:nres_scales}
    \end{figure}

    \begin{figure*}
        \centering
        \includegraphics[width=0.49\linewidth]{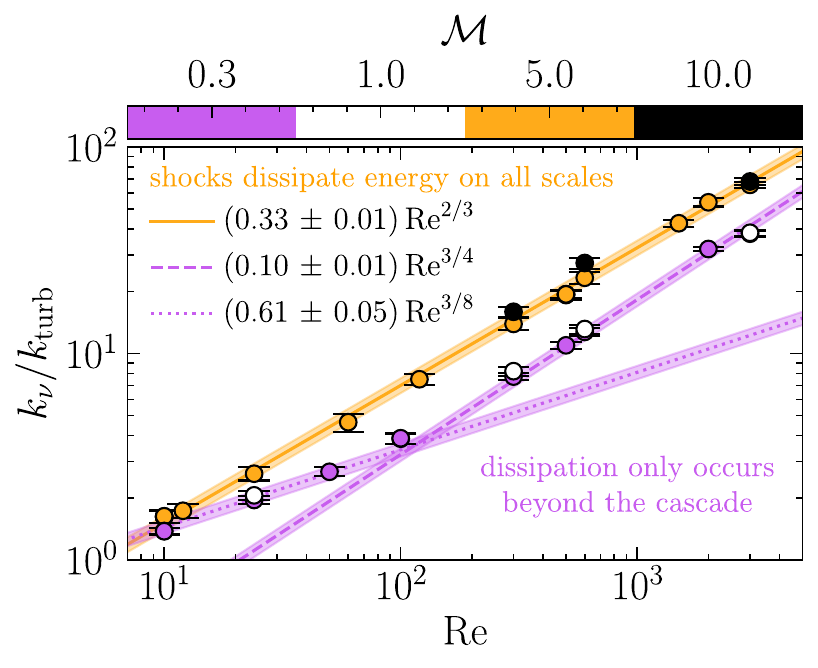}
        \includegraphics[width=0.49\linewidth]{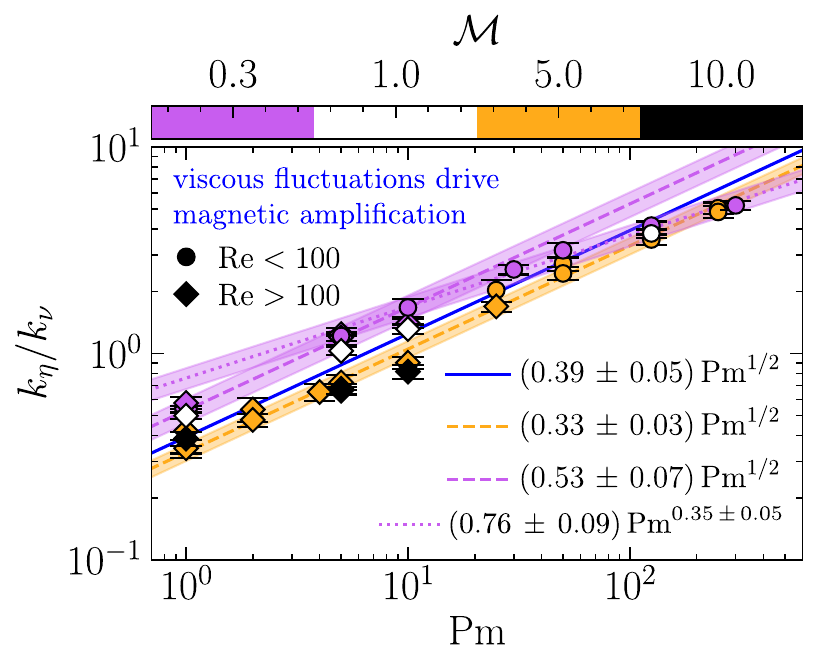}
        \caption{In the left panel we plot the scale separation between the converged kinetic dissipation wavenumber ($k_\nu$) and the external forcing wavenumber ($k_\tturb = 2$) as a function of the hydrodynamic Reynolds number ($\Reyh$) for each simulation setup in \autoref{table:summary}. In the right panel we also compare the scale separation between the converged magnetic and kinetic dissipation wavenumbers ($k_\eta / k_\nu$) against the magnetic Prandtl number ($\Pranm$). Points are coloured by $\Mach$, where $\Mach = 0.3$ simulations are coloured magenta, $\Mach = 1$ are coloured white, $\Mach = 5$ are coloured yellow, and $\Mach = 10$ are coloured black. To guide the eye in the left panel, we fit theoretical scalings of $k_\nu$ in different flow regimes: $k_\nu \sim \Reyh^{2/3}$ to supersonic ($\Mach > 1$) simulations (solid yellow line; expected for \citealt{burgers1948_turbulence_model} turbulence), $k_\nu \sim \Reyh^{3/4}$ to the turbulent ($\Reyh \geq \Reyh_\tcrit \approx 100$) subset of trans- and subsonic ($\Mach \leq 1$) simulations (dashed magenta line; expected for \citealt{kolmogorov1941dissipation} turbulence), and $k_\nu \sim \Reyh^{3/8}$ to viscous ($\Reyh < \Reyh_\tcrit$), trans- and subsonic velocity fields (dotted magenta line; see \citetalias{kriel2022fundamental}). In the right panel we plot viscous simulations ($\Reyh < \Reyh_\tcrit$) as circles, and turbulent simulations ($\Reyh \geq \Reyh_\tcrit$) as diamonds. We also fit a $k_\eta/k_\nu \sim \Pranm^{1/2}$ scaling to the full dataset (solid blue line), the supersonic simulations (dashed yellow line), and the turbulent subset of the trans- and subsonic simulations (dashed magenta line). Finally, we also fit a general power-law model to the viscous subset of trans- and subsonic simulations, and recover a fit (dotted magenta line) that is consistent with our findings in \citetalias{kriel2022fundamental}, where the range of subviscous scales has a shallower dependence on $\Pranm$ than in the turbulent regime. We report all fitted parameters and their 1-sigma uncertainties in their respective labels.}
        \label{fig:dissipation_scaling}
    \end{figure*}
    
    As previously discussed in \autoref{sec:ICs:nres}, we ensure that all the characteristic wavenumbers we measure from our different simulation setups are converged with respect to the grid resolution $N_\tres$. We do this by running each simulation setup in \autoref{table:summary} across a wide range of $N_\tres$, measuring our three wavenumbers of interest, $k_{\rm scale} \in \{k_\nu, k_\eta, k_\tp\}$ in the kinematic phase, and then fitting a generalised logistic model of the functional form
    \begin{equation}
        k_{\rm scale}(N_\tres)
            = k_{\rm scale}(\infty) \rbrac{
                    1 - \exp\cbrac{ -\rbrac{\frac{N_\tres}{N_{\rm res, crit}}}^R }
                }
            , \label{eqn:nres}
    \end{equation}
    to the resolution-dependent wavenumbers. The parameters in our fits are $k_{\rm scale}(\infty)$, which represents the converged ($N_\tres \to \infty$) value of each wavenumber $\{k_\nu, k_\eta, k_\tp\}$, the rate of convergence $R$, and the critical resolution $N_{\rm res, crit}$ at which characterises the resolution where convergence begins.
    
    As discussed in \autoref{sec:ICs:nres}, we start by running each simulation configuration across $N_\tres = 18 - 288$, separated by factors of 2, and then we fit those data for $N_\mathrm{res,crit}$. If we find that our best-fit value of $N_\mathrm{res,crit}$ is larger than 288, or that our data does not yield a fit for $N_\mathrm{res,crit}$ with reasonable uncertainty, we rerun that setup at the higher resolution $N_\mathrm{res} = 576$. We then repeat the convergence test, doubling the resolution until we obtain a well-constrained value of $N_\mathrm{res,crit}$, such that our highest-resolution run for each setup satisfies $\texttt{max}[\{N_\tres\}] > N_\mathrm{res, crit}$.
    
    In \autoref{fig:nres_scales} we plot the resolution ($N_\mathrm{res}$) dependent $k_\nu$, $k_\eta$, and $k_\tp$, and our best fits to these wavenumbers, for our two representative simulations. Broadly, for both simulations, we find evidence of convergence at $N_\tres \approx 576$, but since $k_\nu$, $k_\eta$, and $k_\tp$ may all exist on different scales, one would expect the convergence properties of each wavenumber to be different, since small-scale (high-$k$) structures are expected to require higher resolution to converge than larger-scale (low-$k$) structures. This is what we find. For \SimName{0.3}{600}{5} we find $k_\nu$ begins to converge at $N_{\rm res, crit} = 43.0 \pm 2.2$, while $k_\eta$ and $k_\tp$ begin converging at $N_{\rm res, crit} = 190.4 \pm 20.2$ and $N_{\rm res, crit} = 117.2 \pm 40.0$, respectively. This is expected, since we are operating in the $\Pranm \geq 1$ regime where magnetic structures are smaller-scale than velocity structures, \viz \mbox{$k_\nu < k_\eta \sim k_\tp$}, and therefore require a higher grid resolution to resolve. By contrast, for \SimName{5}{600}{5}, we find that $k_\nu$ shows convergence at $N_{\rm res, crit} = 79.4 \pm 4.6$, whereas $k_\eta$ requires $N_{\rm res, crit} = 173.5 \pm 40.7$ and $k_\tp$ requires $N_{\rm res, crit} = 20.6 \pm 18.3$. Comparing the subsonic and supersonic cases, it is noteworthy that $k_\eta$ shows similar convergence behaviour, but that $k_\tp$ converges at significantly lower grid resolution, consistent with the visual differences in size-scale visible in \autoref{fig:field_slices}.
    
    We report the converged values $k_{\rm scale}(\infty)$ (extrapolated to infinite resolution) for all our simulations in columns 10--12 of \autoref{table:summary}, and use this in all of our analysis that follows, but for compactness from this point on (and in the header of \autoref{table:summary}) we drop the notation $(\infty)$. We also report the measured $N_{\rm res, crit}$ and $R$ for all three derived characteristic wavenumbers, from each of our simulations, in \autoref{table:nres}. These fits are based on using all data up to the highest resolution we have run for each simulation configuration.

    \subsection{Where Does Kinetic and Magnetic Energy Dissipate?}
    \label{sec:results:dissipation}
    
    Now that we have obtained converged dissipation wavenumbers, we are prepared to explore how these scales depend on the dimensionless plasma parameters $\Mach$, $\Reyh$ and $\Pranm$. In the kinematic phase of the dynamo, where magnetic fields are subdominant on all scales, we expect the separation between $k_\tturb$ and $k_\nu$ (\ie the inertial range) to depend only on the velocity field properties (\ie $\Reyh$), and the separation between $k_\nu$ and $k_\eta$ (\ie the sub-viscous range) to be a function of only $\Pranm$. The exact relationship between these dissipation scales and the principal plasma parameters should change between the subsonic and supersonic regimes, an effect we explore in \autoref{fig:dissipation_scaling}. Here we plot $k_\nu / k_\tturb$ against $\Reyh$ in the left panel and $k_\eta / k_\nu$ against $\Pranm$ in the right panel, in both cases colour-coding the simulations by $\Mach$.

    \subsubsection{Viscous Scaling (Inertial Range)}
    \label{sec:results:dissipation:knu}
    
    We first consider the scaling behaviour of $k_\nu$, and its dependence on $\Reyh$ (left hand panel of \autoref{fig:dissipation_scaling}). Here, for our turbulent ($\Reyh \gtrsim \Reyh_\tcrit$), subsonic ($\Mach = 0.3$; plotted in yellow) and transsonic ($\Mach = 1$; plotted in white) simulations, we recover $k_\nu / k_\tturb \sim \Reyh^{3/4}$ as expected for \citet{kolmogorov1941dissipation} turbulence where $E_{\tkin}(k) \sim k^{-5/3}$. This is the same scaling behaviour we had previously demonstrated in \citetalias{kriel2022fundamental} using our previous methods, and now, here, we confirm that our new method (\autoref{eqn:knu_measure}) recovers the same scaling in the same flow regime. This scaling has also been extensively demonstrated in both numerical simulations \citep{yeung1997universality, schumacher2007sub, schumacher2014small} and laboratory experiments \citep{barenblatt1997scaling}, and reflects the conservation of energy flux through the self-similar, inertial cascade, with dissipation only becoming significant at the viscous scale.
    
    For our viscous ($\Reyh < \Reyh_\tcrit$), trans- and subsonic ($\Mach \leq 1$) simulations, we recover $k_\nu / k_\tturb \sim \Reyh^{3/8} = (\Reyh^{3/4})^{1/2}$. This result again agrees with our previous findings in \citetalias{kriel2022fundamental}, where we had found that viscous, subsonic flows have sub-Gaussian velocity gradients. These gradients reflect a lack of intense, localised dissipation events, characteristic of more stable, laminar-like flow regimes \citep[\eg][]{schumacher2007sub, schumacher2014small}, where energy cascade is less efficient compared with \citet{kolmogorov1941dissipation} turbulence.

    Finally, for our supersonic ($\Mach = 5$ and $10$) simulations we measure $k_\nu / k_\tturb \sim \Reyh^{2/3}$, which corresponds with \citet{burgers1948_turbulence_model} turbulence where $E_{\tkin}(k) \sim k^{-2}$ \citep{schober2012magnetic, Federrath2021_sonic_scale}. This scaling of $k_\nu$ sits intermediate between that expected for \citet{kolmogorov1941dissipation} turbulence and the scaling for a viscous, subsonic velocity field \citepalias{kriel2022fundamental}. It reflects shock-dominated energy transfer, where dissipation occurs on all inertial scales, unlike in \citet{kolmogorov1941dissipation} turbulence, where dissipation only occurs on and below the viscous scale.

    \begin{figure*}
        \centering
        \includegraphics[width=\linewidth]{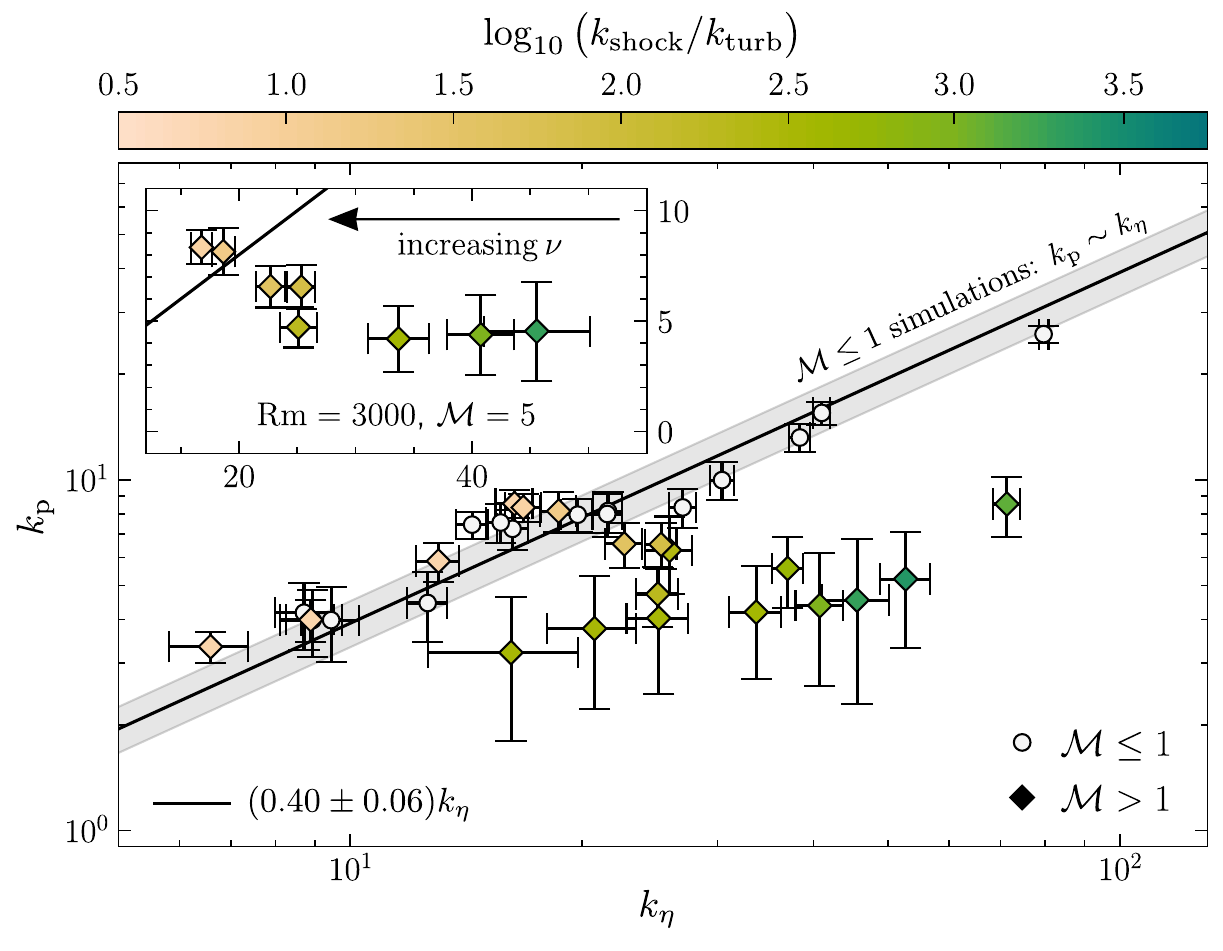}
        \caption{Converged magnetic peak wavenumber ($k_\tp$), plotted against the converged characteristic magnetic dissipation wavenumber ($k_\eta$), for each simulation setup in \autoref{table:summary}. In the main panel, we plot sub- and transsonic simulations with white circles, and in both (main as well as inset) panels, we plot supersonic simulations with diamonds that are coloured based on the reciprocal of our characteristic shock width model ($k_{\tshock}$; see \autoref{eqn:shock_width}). To provide further insights into the supersonic dynamics, we add an inset panel where we focus on a subset of the supersonic simulations. In this inset, we plot the set of $\Mach = 5$ simulations where $\Reym = 3000$ is fixed, and $\Reyh$ is changed via $\nu$. In both panels, we also annotate the same $k_\tp \sim k_\eta$ fit in black, indicating the 1-sigma standard deviation of $k_\tp / k_\eta$ in the $\Mach \leq 1$ subset of points, using a grey coloured band in the main panel.}
        \label{fig:kp_scaling}
    \end{figure*}

    To summarise these three regimes, we find
    \begin{align}
        \frac{k_{\nu}}{k_{\tturb}}
            \sim \left\{\begin{matrix}
                \Reyh^{3/8}, & \Mach \leq 1, & \Reyh < \Reyh_\tcrit, \\
                \Reyh^{3/4}, & \Mach \leq 1, & \Reyh \geq \Reyh_\tcrit, \\
                \Reyh^{2/3}, & \Mach > 1 .
            \end{matrix}\right.
        \label{eqn:knu_scaling}
    \end{align}
    
    The fact that during the kinematic phase of our SSD simulations, we have, in all cases, recovered well-known viscous scaling results for purely hydrodynamic turbulence, is not surprising, since during the kinematic phase \mbox{$E_\tmag(k) \ll E_\tkin(k)$} for all $k$ (see \citealt{beattie2022growth} for the sub-sonic $E_\tmag(k)/E_\tkin(k)$ functions showing this), and therefore the magnetic field exerts a negligible Lorentz force\footnote{\citet{brandenburg2023dissipative}, for example, recognised this simplification and explicitly removed the Lorentz force from the momentum conservation equation, thereby allowing them to indefinitely explore the kinematic SSD phase, as the plasma has no means to become aware of the magnetic field.}. Thus \autoref{mhd:momentum} becomes independent of $\vect{b}$, and approximately hydrodynamical. Once $\vect{b}$ is strong enough (as the SSD process transitions into the linear-growth and saturated phases), we do not expect these dissipation scalings to persist, but measuring these scalings in the other growth regimes is beyond the scope of the current study.

    \subsubsection{Resistive Scaling (Sub-Viscous Range)}
    \label{sec:results:dissipation:keta}
    
    Next, in the right hand panel of \autoref{fig:dissipation_scaling} we consider the subviscous range of scales ($k_\nu < k < k_\eta)$, and determine how the scale-separation between $k_\eta$ and $k_\nu$ depends upon $\Pranm$. This is the most direct way to test whether, even in supersonic flows, the smallest scale kinetic fluctuations are responsible for amplifying (shearing) magnetic fields (see \autoref{sec:ICs:Pm} for details). And in fact, we find evidence that the size of the subviscous range of scales is directly proportional to $\Pranm^{1/2}$ in the turbulent, subsonic regime, as well as in the supersonic flow regime. However, as in \citetalias{kriel2022fundamental}, we find that this scaling does not appear to hold in the viscous, trans- and subsonic flow regimes.

    Now, perhaps this is not completely unexpected, because the theoretical expectation that $k_\eta \sim k_\nu \,\Pranm^{1/2}$ follows from arguments that do not make any assumption about the underlying flow properties of the velocity field (\ie if the flow is subsonic or supersonic). For example, \citet{schekochihin2002spectra} predicted this scaling by assuming that viscous-scale fluctuations most dominantly amplify the magnetic field, with characteristic shearing (or stretching) rate $u_\nu / \ell_\nu$; where $u_\nu$ is the velocity on scale $\ell_\nu$. Balancing this rate of energy injection (into the magnetic field) with the rate of magnetic (Ohmic) dissipation, $\eta / \ell_\eta^2$, rearranges to give
    \begin{align}
        \ell_\eta
            \sim \rbrac{\frac{\eta \, \ell_\nu}{u_\nu}}^{1/2}
            = \rbrac{\ell_\nu^2 \frac{\eta}{\nu} \frac{\nu}{\ell_\nu u_\nu}}^{1/2}
            \sim \ell_\nu \, \Pranm^{-1/2}
            ; \label{eqn:keta:result}
    \end{align}
    where $\ell_\nu \sim \nu / u_\nu$ follows from a straightforward unit analysis.

    Here, the key assumption is the existence of a characteristic viscous scale, where shearing occurs fastest, while the fluctuations still carry a significant amount of energy. For simple plasma flows in the kinematic SSD phase, where there is a single turbulent cascade, this corresponds with the smallest scale in the turbulent cascade. At smaller scales, fluctuations may induce shearing on faster timescales, but their energy dissipates to the point where they contribute negligibly to magnetic amplification. The nature of the energy transfer process that leads to the viscous scale is also not important, as long as the concept of a viscous scale -- below which the kinetic energy cascade dissolves -- remains meaningful. Moreover, the relation $\ell_\nu \sim \nu / u_\nu$, derived from dimensional considerations, represents a general balance between viscous forces and inertial effects. Therefore, while the model of \citet{schekochihin2002spectra} draws upon concepts inspired by \citet{kolmogorov1941dissipation}-like turbulence, its applicability also extends beyond, as we have shown, for example, to the supersonic regime.
    
    On this basis, we conclude that \mbox{$k_\eta \sim k_\nu\, \Pranm^{1/2}$} is a universal scaling in the kinematic phase of turbulent ($\Reyh \geq \Reyh_\tcrit$), $\Pranm \geq 1$ SSDs, invariant to the presence of shocks (\ie $\Mach$). This result supports the idea that the smallest (viscous) fluctuations are always responsible for converting kinetic into magnetic energy, regardless of whether the kinetic energy cascade follows a \citet{burgers1948_turbulence_model}-like or \citet{kolmogorov1941dissipation}-like behaviour. However, we find that the scale separation between $k_\nu$ and $k_\eta$ is generally larger for subsonic compared with supersonic flows (evidenced by larger fitted coefficients in the right panel of \autoref{fig:dissipation_scaling}). This aligns with previous works which have shown that turbulent, subsonic flows are more efficient at driving SSD growth than their supersonic counterparts, even for plasmas with the same $\Pranm$ \citep{schober2012magnetic}.
    
    \subsection{What Sets the Scale Where Magnetic Energy Peaks?}
    \label{sec:results:peak}

    In \autoref{fig:kp_scaling} we plot the wavenumber where the magnetic energy spectrum peaks ($k_\tp$), and compare it with the characteristic resistive wavenumber ($k_\eta$), for all of our simulations. As in \autoref{fig:field_slices}, we again notice an interesting dichotomy between the subsonic and supersonic simulations, with the scaling of $k_\tp$ derived from the $\Mach \leq 1$ (sub- and transsonic) behaving differently to the $\Mach > 1$ (supersonic) simulations. To make this difference between these two flow regimes clear, we introduce a new colouring criteria for all our simulations, one where we colour all the $\Mach \leq 1$ simulations white, and colour the $\Mach > 1$ simulations based on a colour map that we will discuss and motivate below.

    Before we move to the supersonic simulations, we again verify that our new methods for measuring $k_\tp$ and $k_\eta$ are robust and reliable. To demonstrate this, we highlight that we find $k_\tp = (0.40 \pm 0.06)\, k_\eta$, which recovers $k_\tp \sim k_\eta$ for our $\Mach \leq 1$ subset of SSD simulations (white points). This is a well known theoretical result \citep[\eg][]{schekochihin2002spectra}, which was confirmed in previous numerical simulations (\citetalias{kriel2022fundamental}; \citealt{brandenburg2023dissipative}), and tells us that in the kinematic phase of (even approximately) incompressible SSDs, magnetic energy becomes concentrated at the smallest scales allowed by magnetic dissipation \citep[\eg][]{schekochihin2002spectra, Xu2016_dynamo, kriel2022fundamental, brandenburg2023dissipative}. Now, notice that the $0.40 \pm 0.06$ constant of proportionality in this relation is dependent upon the models used to measure $k_\tp$ and $k_\eta$, and therefore expectantly, it is different from the $1.2 \pm 0.2$ found in \citetalias{kriel2022fundamental} (see \autoref{sec:results:tools} for a discussion on our choice of new models for $k_\eta$ and $k_\tp$).

    The behaviour of $k_\tp$ for supersonic SSDs during the kinematic phase is dramatically different from the subsonic scaling, though. While a portion of our $\Mach > 1$ simulations follow the same $k_\tp \sim k_\eta$ scaling as the subsonic cases, the rest deviate significantly to smaller wavenumbers, \viz $k_\tp < k_\eta$. To develop an intuition for why this happens, we briefly turn our attention back to the bottom row panels in \autoref{fig:field_slices}, which show runs \SimName{5}{24}{125}, \SimName{5}{600}{5}, and \SimName{5}{3000}{1}, respectively. In \SimName{5}{3000}{1}, for example, magnetic energy (red) seems to be preferentially concentrated inside of shocked regions of gas, where there are large jumps in the density field (black contours), that have previously been shown to be coherent up to $k_\tturb$ (and even beyond, depending upon how strong the magnetic field is;  \citealt{beattie2020filaments,Beattie2021_multi_shock_model}), even though they fill very little of the volume \citep[e.g.,][]{Hopkins2013_s_pdf, Robertson2018_dense_regions, Mocz2019_s_pdf, Beattie2022_spdf}. Based on $k_\tp$, shocks do not seem to be present in \SimName{5}{24}{125}, where, even though the velocity dispersion is large ($u_\tturb \gtrsim 5 c_\ts$), the kinetic energy diffusion coefficient, $\nu$, is large enough to dissipate supersonic velocities before they form shocks. A likely criteria for this effect is that the shock lifetime, $t_\tshock \sim \ell/u_\tturb$ (which is a fraction of the sound crossing time across the shocked region, $\ell/c_s$; \citealt{Robertson2018_dense_regions}), is shorter than the diffusion timescale, $t_\nu \sim \ell_\nu^2 / \nu$. When this condition is met, shocks, which naturally occur at scales larger than the viscous fluctuations, are able to create and sustain large-scale structure in the magnetic field faster than the viscosity diffuses them away, and generates subviscous scale magnetic structure.

    \begin{figure}
        \centering
        \includegraphics[width=0.75\linewidth]{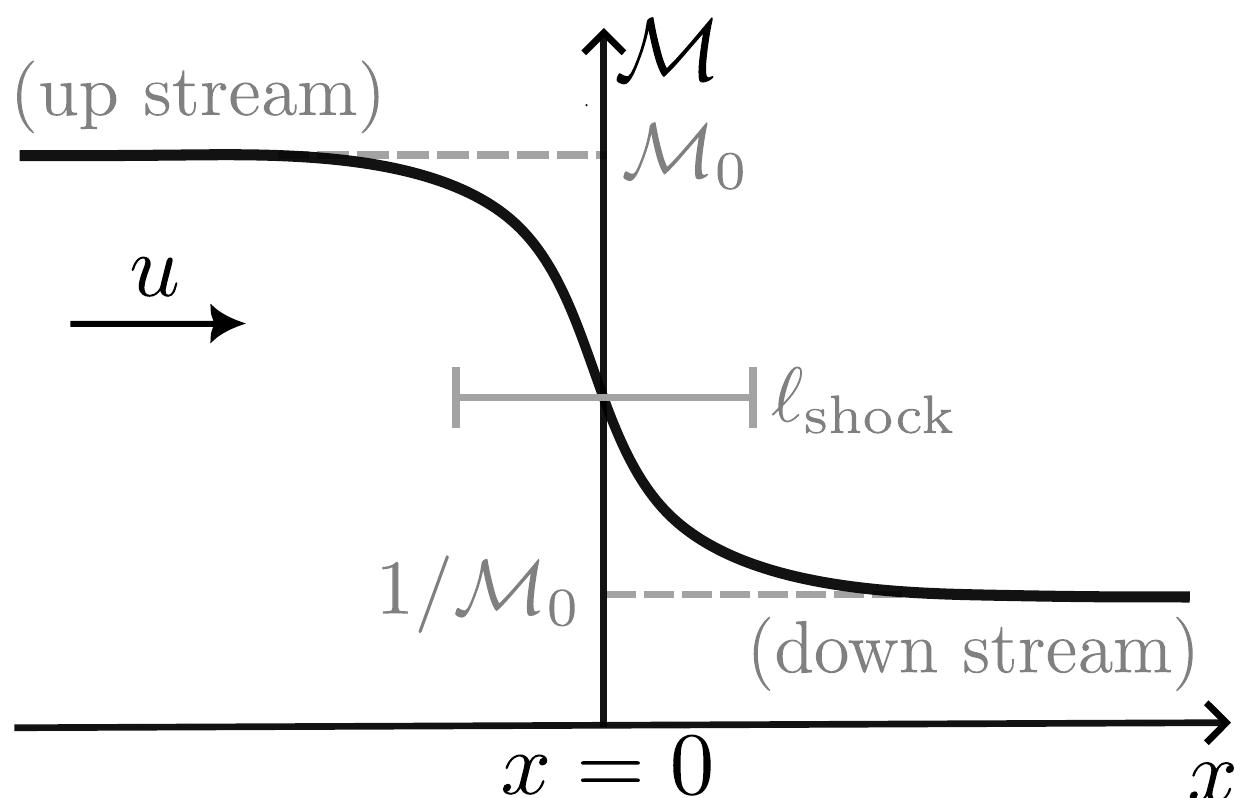}
        \caption{Schematic of a 1D velocity shock, in the reference frame of the shock: the up stream material ($x < 0$), with up stream sonic Mach number, $\Mach_0$, approaches the shock located at $x = 0$ with a velocity $u$.}
        \label{fig:schematic:1D_shock}
    \end{figure}

    \subsubsection{The Role of Isotropic Hydrodynamic Shocks}
    
    Our supersonic simulations appear to support a hypothesis that, when shocks are present, $\ell_\tp$ approaches a value much larger than $\ell_\eta$, which depends somehow on the typical width of shocks, $\ell_\tshock$. To test this conjecture, we estimate $\ell_\tshock$ from the quasi-equilibrium state of the momentum equation, omitting the magnetic terms because the Lorentz force is unimportant in the kinematic phase, even in shocked regions ($E_\tmag(k)/E_\tkin(k) \ll 1,$ for all $k$; \citealt{beattie2022growth}), and excluding external forcing, but not setting viscosity or the sound speed to zero (\eg not going to the pressureless, Burgers' equation limit, $\nabla P/\Mach^2 \to 0$ as $\Mach \to \infty$), namely,
    \begin{align}
        \nabla\cdot\rbrac{
                \rho\vect{u}\otimes\vect{u}
                + c_\ts^2 \rho\tensor{\imatrix}
                - 2\nu\rho\tensor{S}
            } &= 0 . \label{eqn:momntum:steady_state}
    \end{align}
    Since shocks are generated isotropically in our supersonic turbulent simulations, we simplify \autoref{eqn:momntum:steady_state} by considering a single characteristic shock travelling in 1D (adopting the usual convention that $x < 0$ and $x > 0$ are the up and down stream directions, respectively; see \autoref{fig:schematic:1D_shock} for a schematic of this setup). It follows that
    \begin{align}
        \driv{}{x}\sbrac{
            \rho u^2 + c_\ts^2 \rho - \frac{4\nu \rho}{3} \driv{u}{x}
        } = 0 .
        \label{eq:momentum_flux}
    \end{align}
    The quantity in square brackets represents the momentum flux, which remains constant (conserved) across the shock. Since the velocity gradient vanishes far up stream of the shock, \ie \mbox{$\dd u/\dd x \to 0$} as \mbox{$x \to -\infty$}, the momentum flux there must be $\rho_0 u_0^2 + c_s^2 \rho_0$, where subscript $0$ indicates quantities in the far up stream region. Conservation of mass flux further implies that $\rho u = \rho_0 u_0 = \tconst$, which allows us to write position-dependent quantities (without subscripts), in terms of the upstream values. Making use of these relationships, we integrate \autoref{eq:momentum_flux} and solve for
    \begin{align}
        \driv{\Mach}{x}
            = \frac{3 c_\ts}{4 \nu} \rbrac{
                \Mach^2 - \rfrac{\Mach_0^2 + 1}{\Mach_0} \Mach + 1
            } ,
            \label{eqn:shock_width:ode}
    \end{align}
    which captures how $\Mach$ varies across the shocked region.

    \begin{figure}
        \centering
        \includegraphics[width=\linewidth]{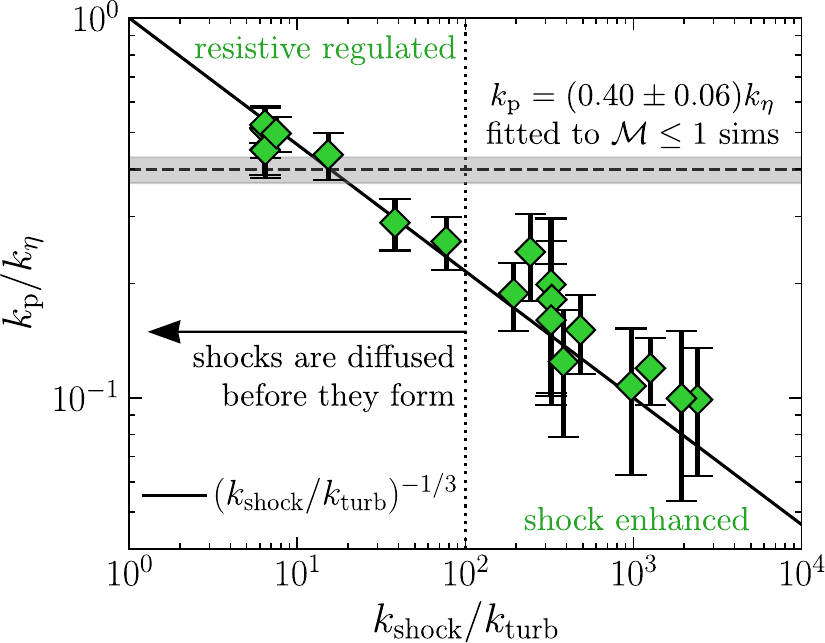}
        \caption{For each of our $\Mach > 1$ simulations we compare $k_\tp/k_\eta$ with the reciprocal of our model for the shock aspect ratio ($k_\tshock/k_\tturb$ via \autoref{eqn:shock_width}) plotted as green diamonds. We annotate the scaling relation $k_\tp/k_\eta = (k_\tshock/k_\tturb)^{-1/3}$ with a solid line, and the measured average $k_\tp/k_\eta = 0.40 \pm 0.06$ from the $\Mach \leq 1$ simulations with a dashed line. We also indicate the 1-sigma standard deviation in the $\Mach \leq 1$ subset of $k_\tp/k_\eta$ points with a grey band, and annotate $k_\tshock/k_\tturb = 10^2$ with a vertical dotted line. We find that the scale-separation between $k_\tp$ and $k_\eta$ for the supersonic transitions from incompressible ($k_\tshock/k_\tturb < 10^2$) to compressible ($k_\tshock/k_\tturb > 10^2$) scaling behaviour when the aspect ratio of filamentary shocks becomes large (captured by \autoref{eqn:shock_width} when $\Mach \gtrsim 1$ and $\Reyh \gtrsim \Reyh_\tcrit \approx 100$; $k_\tshock/k_\tturb \sim 10^2$). This is because, in the compressible regime, supersonic flows support and become dominated by shocks, where flux-frozen magnetic fields become compressed by shocks into larger-scale, filamentary structures (see for example the bottom-right panel in \autoref{fig:field_slices}), which brings $k_\tp$ from $k_\eta$ towards larger scales (smaller wavenumbers, $k$) by an amount proportional to the reciprocal shock aspect ratio, $k_\tshock/k_\tturb$ (see \autoref{eqn:shock_width}). Conversely, in the incompressible, supersonic regime, where even though $\Mach > 1$, the flow is so viscous ($\Reyh < \Reyh_\tcrit$), that shocks are dissipated before they form ($\nu > u_\tturb \ell_\tturb$), and therefore compressed structures do not form, and magnetic energy remains organised in a ``folded field'' configuration (see \autoref{sec:results:kappa}).}
        \label{fig:kshock_scaling}
    \end{figure}
    
    \autoref{eqn:shock_width:ode} has the form of a Riccati equation (nonlinear, quadratic differential equation of first-order) with constant coefficients, and has the solution that $\dd \Mach/\dd x = 0$ when $\Mach = \Mach_0$ or $1/\Mach_0$; the former solution represents the up stream region, and the latter the down stream region. Since $x$ does not explicitly appear on the right-hand side of \autoref{eqn:shock_width:ode}, we are free to choose our coordinate system such that $\Mach = 1$ at $x=0$ (as we have done in \autoref{fig:schematic:1D_shock}). From this, we estimate the characteristic shock width as
    \begin{align}
        \frac{\ell_\tshock}{\ell_\tturb}
            &\sim \frac{1}{\ell_\tturb} \left|
                    \frac{\Mach}{\dd\Mach/\dd x}
                \right|_{x=0}
            \sim \frac{\nu}{u_0 \ell_\tturb} \frac{\Mach_0^2}{(\Mach_0 - 1)^2}
            .
    \end{align}
    Now, in supersonic turbulence, one expects to see a population of shocks that take on a wide distribution of $\ell_\tshock$ \citep[\eg][]{smith2000distribution, brunt2002interstellar, donzis2012shock, lesaffre2013low, squire2017distribution, Park2019_pops_MHD_shocks, Beattie2020_ansotropic_shocks, Beattie2022_spdf}, where each of the different $\ell_\tshock$ are determined by the $\Mach_0$ that is up stream from it. That being said, the distribution of shock widths is regulated by the turbulent properties on $\ell_\tturb$ \citep{Park2019_pops_MHD_shocks}, where on average, \mbox{$\Mach_0 \approx \Mach$}, and \mbox{$u_0 \approx u_\tturb$}. Therefore, we model the characteristic aspect ratio of a shock in our isotropic turbulent simulations as
    \begin{align}
        \frac{k_\tturb}{k_\tshock}
            = \frac{\ell_{\tshock}}{\ell_\tturb}
            \sim \frac{\Mach^2}{\Reyh\, (\Mach - 1)^2}
            . \label{eqn:shock_width}
    \end{align}
    
    In the inset axis of \autoref{fig:kp_scaling} we test whether this shock aspect ratio model can explain the difference between $k_\tp$ and $k_\eta$ (note that we colour points by the reciprocal of \autoref{eqn:shock_width}, to emphasise that in our experiments, the numerator, $k_\tshock$, is changing, and $k_\tturb = 2$ remains fixed in the denominator). Here we plot the full set of $\Mach = 5$ and $\Reym = 3000$ simulations. Within this collection of runs, $\Reyh$ is varied via changing $\nu$ (\ie $u_\tturb$ and $k_\tturb$ are fixed), and we find that the most turbulent simulation in this set, \SimName{5}{3000}{1}, lies farthest from the $k_\tp \sim k_\eta$ relation, while the most viscous simulation, \SimName{5}{24}{125}, lies on the relation. Between these two limits, the inverse shock width (\autoref{eqn:shock_width}) scales directly proportional to the deviation of each point from the $k_\tp \sim k_\eta$ relation. 
    
    We demonstrate this more explicitly in \autoref{fig:kshock_scaling}, which shows $k_\tp/k_\eta$ as a function of $k_\tshock/k_\tturb$, and here it is clear that all the $\Mach > 1$ simulations (green diamond points) are well-fit by the empirical relationship
    \begin{align}
        \frac{k_\tp}{k_\eta}
            = \rfrac{k_\tshock}{k_\tturb}^{-1/3}
            . \label{eqn:kp_scaling}
    \end{align}
    We interpret this to mean that shocks compress flux-frozen magnetic fields into filamentary structures, bringing $k_\tp$ from $k_\eta$ towards larger scales by an amount determined by the aspect ratio of typical shocks, \ie the ratio of shock length ($\ell_\tturb$), compared with shock width ($\ell_\tshock$).
    
    Now, while compressions of flux-frozen magnetic fields do not yield irreversible growth in the magnetic field (see, for example, the top panel of Figure 7 in \citealt{beattie2023bulk}; because growth due to compression is soon negated by dilation), we do see that the compressions introduce qualitatively different and lasting imprints on the field amplitude structure (see, for example, the bottom row panels in \autoref{fig:field_slices}). Because the forcing modes in all of our simulations are $\ell_\tturb = \ell_\tbox/2$, the aspect ratio of the shocks, \ie $\ell_\tshock/\ell_\tturb$, can only change via the shock width, which is mediated by the properties of the medium (\ie $\nu$) and flow ($\Reyh$ and $\Mach$). The exact nature in which the aspect ratio of shocks, and thereafter, the scale on which magnetic energy becomes concentrated, depends on fundamental plasma numbers, is well captured by \autoref{eqn:shock_width} and \autoref{eqn:kp_scaling}, respectively\footnote{We hypothesise that the exponent $1/3$ in \autoref{eqn:kp_scaling} is likely associated with shocks concentrating magnetic energy into filamentary structures \citep{das2024magnetic}. Therefore, the slope of the $\ell_\tp/\ell_\eta$ scaling reflects how magnetic energy is distributed within the plasma, and would likely differ if the shocked regions were not filamentary, but, for example, sheet-like instead.}.
    
    Moreover, in \autoref{fig:kshock_scaling} we once again see evidence for a critical hydrodynamic Reynolds number $\Reyh_\tcrit \approx 100$ that separates the $\Mach > 1$ flows that are able to support shocks ($\Reyh \geq \Reyh_\tcrit$), from those flows where strong viscosity prevents shocks from forming ($\Reyh < \Reyh_\tcrit$). This becomes clear in the high $\Mach$ limit, where \mbox{$k_\tshock/k_\tturb \to \Reyh$}, and thus $\Reyh = \Reyh_\tcrit \approx 100$ corresponds with $k_\tshock/k_\tturb \approx 100$; we highlight this value with a dotted vertical line in \autoref{fig:kshock_scaling}. It is clear from the figure that this line roughly identifies where the $\Mach > 1$ simulations transition from $k_\tp/k_\eta \approx \tconst$ to $k_\tp/k_\eta = (k_\tshock/k_\tturb)^{-1/3}$. Indeed, one should notice that the seven red points that lie to the left of the vertical line in \autoref{fig:kshock_scaling} correspond with the seven that lie along the $k_\tp \sim k_\eta$ relation in \autoref{fig:kp_scaling}, where in all seven cases $\Reyh < 100$.
    
    \begin{figure}
        \centering
        \includegraphics[width=\linewidth]{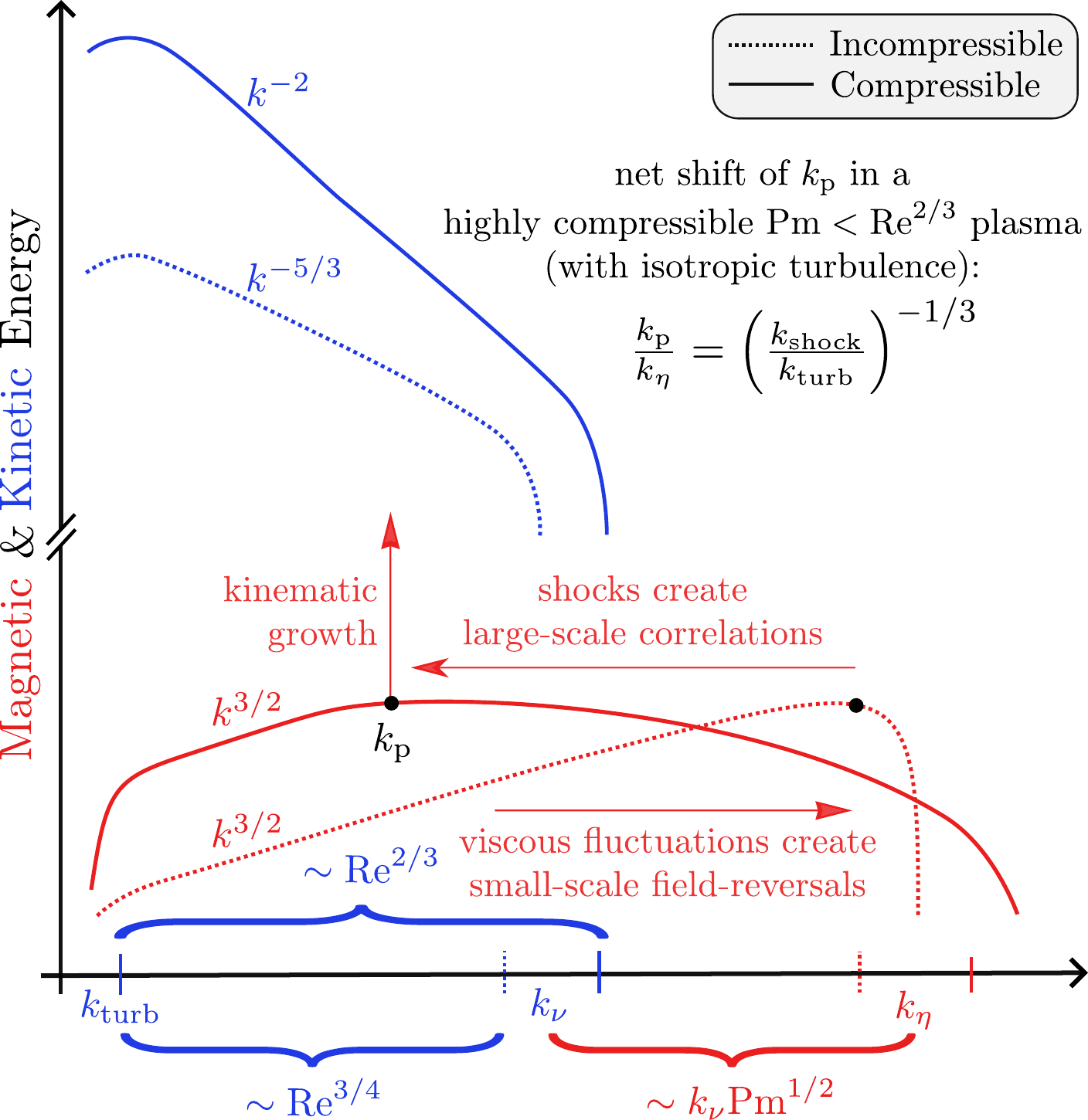}
        \caption{Schematic of the important scaling properties of the kinetic (blue) and magnetic (red) energy spectra during the kinematic phase of a turbulent SSD. We summarise our main findings in \autoref{fig:dissipation_scaling} and \ref{fig:kshock_scaling} for incompressible (dotted line) and highly compressible ($\Mach \to \infty$; solid line) plasmas with the same dimensionless plasma parameters $\Reyh \geq \Reyh_\tcrit$ and $1 \leq \Pranm < \Reyh^{2/3}$. Here we highlight a compressible plasma where the subviscous range is smaller than the net shift in $k_\tp$ ($\Pranm < \Reyh^{2/3}$), and thus the magnetic tug-of-war between shocks and viscous-scale fluctuations shifts $k_\tp$ from $k_\eta$ to beyond $k_\nu$.}
        \label{fig:schematic:energy_spectrum}
    \end{figure}

    \subsubsection{A Hierarchy of MHD Scales}
    
    We conclude that \autoref{fig:kp_scaling} and \autoref{fig:kshock_scaling} can be explained simply by two distinct flow regimes (see \autoref{fig:schematic:energy_spectrum}). In the first, \textit{incompressible} regime, the development of shocks is not supported by the plasma, whether it is because the velocity dispersion is too small ($\Mach \leq 1$) or because the viscosity is large enough to dissipate the supersonic velocities before shocks are able to form (\ie $\Mach > 1$ and $\Reyh < \Reyh_\tcrit$). This results in the well-known hierarchy
    \begin{align}
       \ell_\tturb > \ell_\nu > \ell_\eta \sim \ell_\tp
       . \label{eqn:kp:subsonic}
    \end{align}
    In the second, \textit{compressible} (shock-dominated; $\Mach > 1$ and $\Reyh \geq \Reyh_\tcrit$) flow regime, shocks concentrate magnetic energy in large-scale, filamentary structures. As a result, the peak magnetic field scale shifts from the resistive scale to $\ell_\tp \sim (\ell_\tturb/\ell_\tshock)^{1/3} \ell_\eta \gg \ell_\eta$, approaching $\ell_\tp \sim \Reyh^{1/3} \ell_\eta$ as \mbox{$\Mach \to \infty$}.
    
    A key question then becomes whether magnetic structures can grow larger than the viscous scale, since scales smaller than $\ell_\nu$ (the subviscous range) are generally on relatively small scales (but not always, e.g., the intracluster medium, with low $\Reyh \sim 10^2 \iff k_\nu^{-1} \sim 3\;{\rm kpc}$ and extremely high $\Pranm \gtrsim 10^{29}$; \citealt{Schekochihin2006_ICM,StOnge2020_weakly_collisional_dynamo}). Our analysis shows that this can occur, but only when the subviscous range is not too large. To be precise, in the $\Mach \to \infty$ limit, our results (\autoref{eqn:kp_scaling}) indicate that the scale separation between the peak magnetic energy and viscous scales is $\ell_\tp/\ell_\nu \sim \Reyh^{1/3} \Pranm^{1/2}$, and therefore $\ell_\tp > \ell_\nu$ when $\Pranm < \Reyh^{2/3}$. This yields the following hierarchy of scales
    \begin{align}
        \ell_\tturb > \ell_\tp > \ell_\tshock > \ell_\nu > \ell_\eta
        , \label{eqn:kp:supersonic}
    \end{align}
    which we contrast in \autoref{fig:schematic:energy_spectrum} with the hierarchy of scales produced by an incompressible SSD. On the other hand, if $\Pranm > \Reyh^{2/3}$ then $\ell_\tp$ still shifts from $\ell_\eta$ toward larger scales, but it remains contained in the subviscous range, producing
    \begin{align}
        \ell_\tturb > \ell_\nu > \ell_\tp > \ell_\tshock > \ell_\eta
        .
    \end{align}
    
    Here we have shown that shocks in compressible (\mbox{$\Reyh \geq \Reyh_\tcrit$} and $\Mach > 1$) flow regimes are able to move the magnetic peak scale out of the subviscous range when $1 \leq \Pranm < \Reyh^{2/3}$. Physically, this condition can be understood as a statement that shock lifetimes are longer than the characteristic timescale associated with kinetic fluctuations at the viscous scale (follows from comparing $\ell_\nu/\ell_\eta$ given by \autoref{eqn:keta:result} and $\ell_\tp/\ell_\eta$ given by \autoref{eqn:kp_scaling}). If this condition is met, then large scale correlations in the magnetic field are produced through flux freezing in shocks faster than by kinetic amplification on the viscous scale, the mechanism that dominates in all incompressible SSDs. This distinction has important implications for the transition to the saturated regime, which we discuss in \autoref{sec:discussion:saturation}.

    \subsection{Underlying Field Structure}
    \label{sec:results:kappa}
    
    \begin{figure*}
        \centering
        \includegraphics[width=\linewidth]{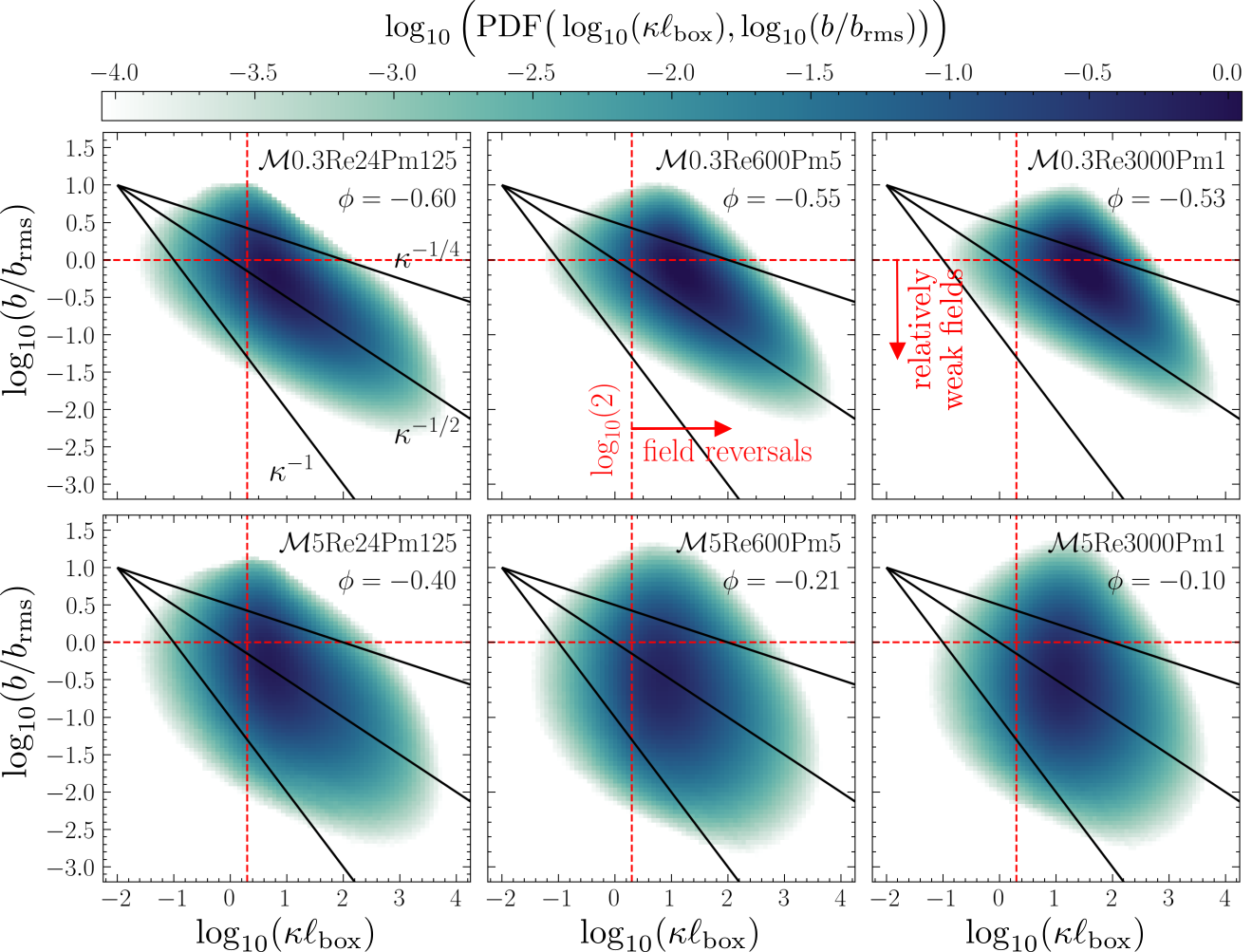}
        \caption{Joint distributions of the relative magnetic field strength ($b/b_\trms$), and the magnetic field-line curvature ($\kappa$; measured in units of inverse box-length, $\ell_{\tbox}$), for the same six simulations as in \autoref{fig:field_slices}. In all panels we report the Pearson correlation coefficient $\phi$ (given by \autoref{eq:pearson_correlation}) between $\log_{10}(b/b_\trms)$ and $\log_{10}(\kappa/\kappa_\trms)$. We also annotate a $b \sim \kappa^{-1/2}$ scaling, which is associated with a curvature relation where the magnetic tension is exactly balanced by the turbulent stretching (the symmetric rate of shear component in $\nabla\otimes\vect{u}$; \citealt{schekochihin2001structure, schekochihin2004simulations}); see \autoref{sec:results:kappa} for a discussion. We also show $b \sim \kappa^{-1/4}$ and $b \sim \kappa^{-1}$ to guide the eye. The red, horizontal, dashed lines indicates $b = b_\mathrm{rms}$, while the red, vertical, dashed lines indicate $\kappa \ell_\tbox = 2$, which marks the transition between fields that can ($\kappa \ell_\tbox \geq 2$) and cannot ($\kappa \ell_\tbox < 2$) reverse within the box-domain. The $b\sim \kappa^{-1/2}$ scaling holds well (on average) for the subsonic plasmas (top row), but breaks down for the supersonic plasmas (bottom two right panels), which is most likely due to the presence of shocks being able to compress and grow the field without necessarily modifying the curvature.}
        \label{fig:curvature}
    \end{figure*}

    \begin{figure*}
        \centering
        \includegraphics[width=\linewidth]{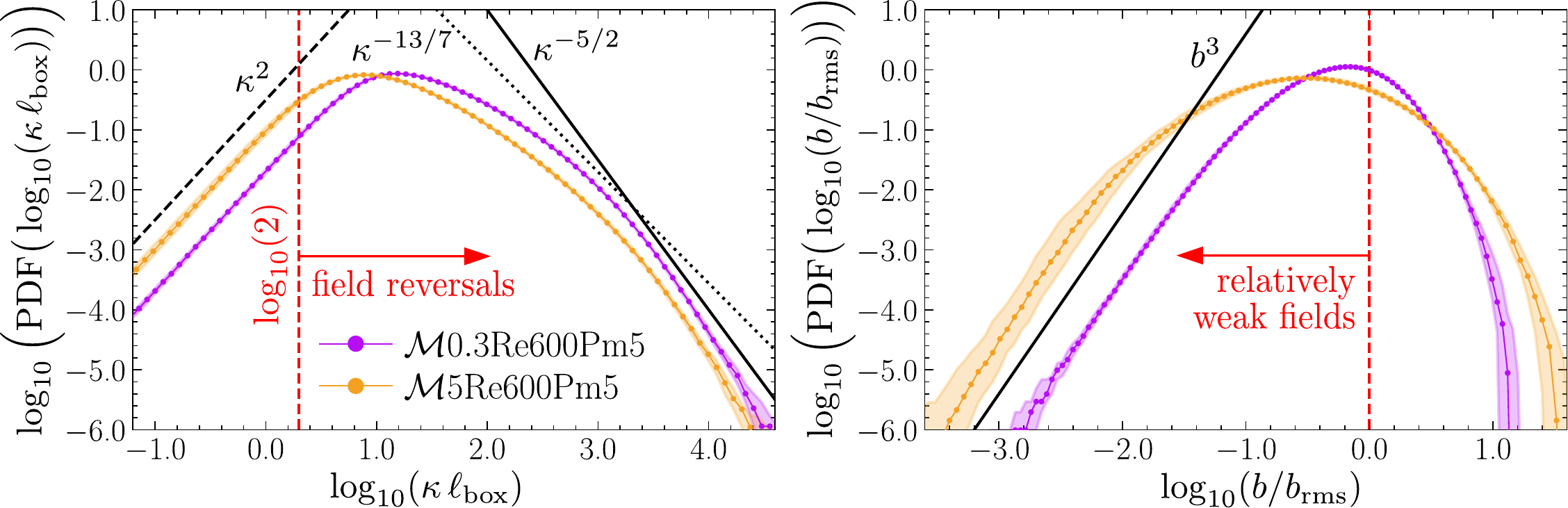}
        \caption{Volume-weighted PDFs of the magnetic field-line curvature ($\kappa$, measured in units of $\ell_\tbox$; left panel), and the normalised magnetic field amplitude ($b/b_\trms$; right panel), for \SimName{0.3}{600}{5} and \SimName{5}{600}{5}. We annotate $\kappa^2$ (dotted), $\kappa^{-13/7}$ (solid; \citealt{schekochihin2002small}), and $\kappa^{-5/2}$ (dash-dotted; \citealt{yang2019role} and \citealt{pfrommer2022simulating}) in the left panel, and $b^{3}$ in the right panel. We only find a negligible difference between the PDFs of $\log_{10}(\kappa \ell_{\tbox})$ for the $\Mach < 1$ (purple) and $\Mach > 1$ (yellow) simulations, but there is significantly more low and high $b/b_\trms$ probability densities in the $\Mach > 1$ simulation. Hence we conclude that the weakening correlation between $\kappa$ and $b$ (shown in \autoref{fig:curvature}) is caused from shock compressions that grow and shrink $b$ without influencing $\kappa$.}
        \label{fig:pdfs}
    \end{figure*}

    While we have shown that shocks reorganise and concentrate magnetic energy into large-scale, filamentary structures, here we want to show that, in doing so, they also completely change the underlying magnetic field-line properties. To quantify field geometry, we compute the curvature of magnetic field-lines,
    \begin{equation}
        \kappa
            = \nbrac{ \rbrac{\tbasis\cdot\nabla}\tbasis }
            , \label{eqn:kappa}
    \end{equation}
    at every point in a simulation domain, where $\tbasis = \vect{b}/\|\vect{b}\|$ is the tangent vector to the field\footnote{Note that in regions where the field changes substantially over the scale of a single cell, as it does in shocked regions, numerical evaluation of $\kappa$ requires some care in choosing a stencil that maintains exact orthogonality of the tangent-normal-binormal basis vectors. See \aref{app:kappa} for a discussion of our method, and \citet{schekochihin2004simulations} for a brief acknowledgement of this issue.}, and curvature points in the $\nbasis = (\tbasis\cdot\nabla)\tbasis / \kappa$ direction. Note that the radius of curvature of the field in units of box length is then $\ell_\tbox/\kappa$.

    \subsubsection{Relationship Between Magnetic Field Strength and Field-Line Curvature}

    In \autoref{fig:curvature} we plot the time-averaged, joint distributions of the relative magnetic field strength, $b/b_\trms$ (normalising out the dynamo growth in $b$), and the field-line curvature magnitude, $\kappa \ell_\tbox$, for the same six simulations as in \autoref{fig:field_slices}, namely \SimName{0.3}{24}{125}, \SimName{0.3}{600}{5}, and \SimName{0.3}{3000}{1} in the top row, and \SimName{5}{24}{125}, \SimName{5}{600}{5}, and \SimName{5}{3000}{1} in the bottom row, respectively. Focusing first, only on the three subsonic simulations (top row of \autoref{fig:curvature}), we find that the $(\kappa,b)$ distributions in both the viscous (left panel) and turbulent SSDs (middle and right panels) appear to have the same overall shape, where the weakest magnetic fields have the largest curvature, \eg have the most field reversals, and conversely the strongest fields are the straightest. Here we find evidence of a scaling consistent with $b \sim \kappa^{-1/2}$, which was first demonstrated by \citet{schekochihin2004simulations} (as opposed to $b \sim \kappa^{-1}$ as previously suggested by \citealt{brandenburg1995size} and \citealt{schekochihin2002small}) for incompressible SSDs, and more recently by \citet{kempski2023cosmic} using $\Mach < 1$ SSD and weak mean-field simulations to study cosmic ray propagation.
    
    \citet{schekochihin2004simulations} associates the $b \sim \kappa^{-1/2}$ anti-correlation with a ``folded field'' geometry\footnote{A term coined by \citet{schekochihin2004simulations} to describe the configuration of magnetic fields in the incompressible flow regime.}, where the incompressible components of the velocity gradient tensor \citep[see][]{beattie2023bulk} and the magnetic tension are balanced. This $1/2$ exponent is a unique solution that produces a steady-state $(\kappa, b)$ configuration in the co-moving frame of the fluid, based on the cancellation of the $\widehat{\vect{b}}\otimes\widehat{\vect{b}}:\nabla\vect{u}$ term in the co-moving, evolution equation for the correlation between the magnetic field amplitude and magnitude of the magnetic field-line curvature (see Equation 25 in \citealt{schekochihin2004simulations}).
    
    In contrast, the supersonic SSDs (bottom row of \autoref{fig:curvature}), show a significantly weaker relationship between the field strength and field-line curvature, with the relative statistical independence of $b$ and $\kappa$ increasing as we evolve from the most viscous plasmas (bottom-left panel) to the most turbulent (bottom-right panel). In the turbulent, supersonic flow regime, it is clear that both strong ($b > b_\trms$) and weak ($b < b_\trms$) field-lines can maintain a straight configuration ($\kappa \ell_\tbox < 2$), although we still see a dearth of low curvature ($\kappa \ell_\tbox < 2$) for strong magnetic fields ($b > b_\trms$; \ie the upper left quadrant of the plot is empty). We quantify this growing independence between $b$ and $\kappa$ for each of our simulations in \autoref{fig:curvature} by computing the Pearson correlation coefficient between $\log_{10}(b/b_{\trms})$ and $\log_{10}(\kappa)$,
    \begin{align}\label{eq:pearson_correlation}
        \phi = \frac{
                \mathrm{cov}\big[ \log_{10}(b/b_{\trms}) \log_{10}(\kappa\ell_{\tbox}) \big]
            }{
                \sqrt{
                    \mathrm{var}\big[ \log_{10}(b/b_{\trms}) \big]
                    \, \mathrm{var}\big[ \log_{10}(\kappa\ell_{\tbox}) \big]
                }
            }
            ,
    \end{align}
    where $\mathrm{cov}[\hdots]$ and $\mathrm{var}[\hdots]$ are the covariance and variance operators, respectively. We annotate these values in each panel of \autoref{fig:curvature}.
    
    The numerical results confirm our qualitative, visual impression: $b$ and $\kappa$ are strongly anti-correlated (and in the $\log$-$\log$ domain, this translates to being linearly anti-correlated, as one expects for the \citealt{schekochihin2004simulations}-type models) in both the subsonic, viscous flow regime (upper left panel, $\phi = -0.60$) and subsonic, turbulent flow regimes (upper middle and right panels; $\phi = -0.55$ and $\phi = -0.53$, respectively). The anti-correlation remains, but becomes weaker for supersonic flows that are too viscous to support shocks (lower left panel, $\phi = -0.40$), and then the anti-correlation almost completely disappears once strong shocks become ubiquitous (lower middle and right panels; $\phi = -0.21$ and $\phi = -0.10$, respectively).

    \subsubsection{Sensitivity to Compressibility}

    We find that the field which drives this anti-correlation in \autoref{fig:curvature} is \textit{not} the distribution of $\log_{10}(\kappa)$, but the distribution of $\log_{10}(b)$. We demonstrate this in \autoref{fig:pdfs}, where we plot the time-averaged (over the kinematic phase) PDF of $\log_{10}(\kappa)$ (left panel) and $\log_{10}(b)$ (right panel) for our two representative simulations (again, \SimName{0.3}{600}{5} and \SimName{0.3}{600}{5}). Notice that while the $\log_{10}(\kappa)$ distributions look very similar\footnote{We annotate a $\sim \kappa^2$ scaling for the $\log_{10}(\kappa)$ PDF at low values of curvature, and for large values of curvature we contrast \citet{schekochihin2002small}'s \mbox{$\sim \kappa^{-13/7}$} model derived from the Fokker-Planck equation for the one-point PDF of $\kappa$, with the \mbox{$\sim \kappa^{-5/2}$} scaling found both in decaying, incompressible 3D MHD simulations \citep{yang2019role} and also throughout most regions (\ie in the galaxy centre, disc, and halo) of merging galaxy simulations \citep{pfrommer2022simulating}. The $\sim \kappa^{-5/2}$ scaling agrees with theoretical modelling of magnetic field fluctuations as quasi-Gaussian distributions (each spatial component is independent of one another, but still satisfying $\nabla\cdot\vect{b} = 0$), with the curvature force, $\|\vect{f}_\mathrm{c}\|$ (see \autoref{app:kappa}), becoming independent of $\kappa$ \citep{yang2019role, pfrommer2022simulating}. In the kinematic phase of both the incompressible and compressible SSD, we find that the $\sim \kappa^{-5/2}$ scaling better represents the observed high-$\kappa$ end of the PDFs derived from our simulations (we also find that this is true in the saturated phase). At low-$\kappa$ we find $\sim \kappa^2$, which is steeper than the $\sim \kappa$ found by \citet{schekochihin2002small} using incompressible SSD simulations with $k_\tturb = k_\tbox$, and \citet{yang2019role} in their decaying simulations. We suspect that the shallower scaling in their simulations could be a consequence of flow statistics that are not turbulent. Moreover, in the inner (turbulent) portion ($< 1 \ukpc$) of their galaxy merger, where most of the dynamo growth occurs \citep[\eg][]{whittingham2021impact}, \citet{pfrommer2022simulating} found some evidence at moderate values of curvature for a $\sim \kappa$ scaling, however, the scaling at low-$\kappa$ is $\sim \kappa^{3/2}$--$\kappa^3$, which is steeper than $\sim \kappa$ and more consistent with our $\sim \kappa^2$ observed scaling.}, the $\log_{10}(b)$ distribution changes significantly between the two flow regimes, with the magnetic-amplitude distribution becoming much broader (a greater proportion of the field is weaker or stronger than $b_\trms$) in the supersonic compared with the subsonic SSD. This is likely due to shocks growing magnetic energy via compression, $\sim b^2(\nabla\cdot\vect{u})$, from magnetic energy equation \citep{beattie2023bulk}, and flux-freezing, which do not significantly change the underlying field-line structure of the magnetic field. There is, however, a marginal increase in straight fields, with low $\kappa/\kappa_\trms$ in the compressible regime, but for our $\Mach \leq 10$ simulations, this effect is largely negligible, therefore it would be worth checking whether this effect becomes more pronounced in simulations with much larger $\Mach$ flows.

\begin{figure}
        \includegraphics[width=\linewidth]{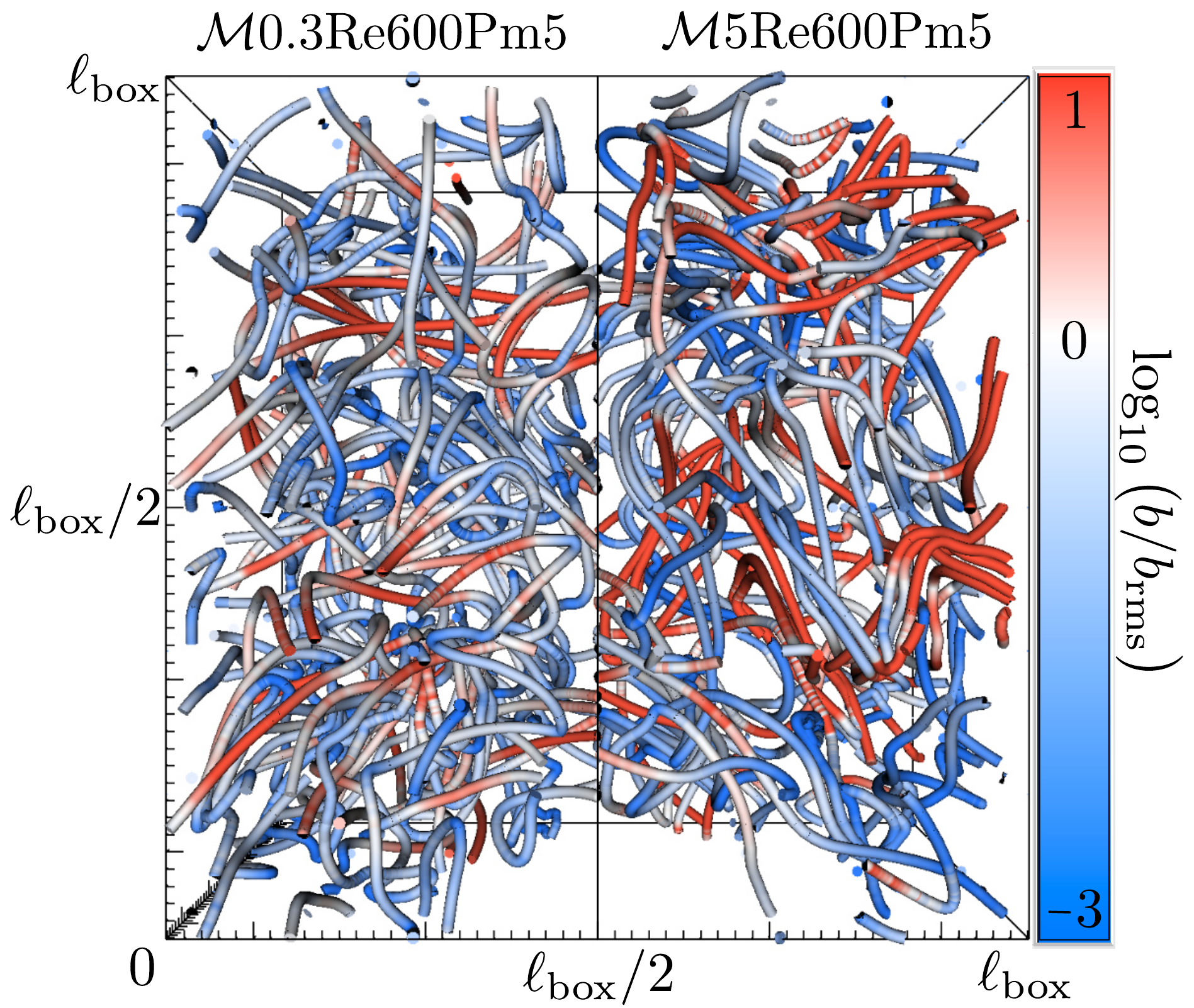}
        \caption{Magnetic field streamlines for \SimName{0.3}{600}{5} and \SimName{5}{600}{5} plotted on the left- and right-hand side of the cube, respectively, for a time realisation midway through the kinematic phase of each simulation. Streamlines are coloured based on the normalised magnetic field magnitude, $b / b_\trms$, using a colour-range that spans across the same range of $b / b_\trms$ as plotted in the right panel of \autoref{fig:pdfs}. Notice how the relationship between field magnitude and field-line curvature (see \autoref{fig:curvature}) for \SimName{0.3}{600}{5} is readily apparent, with strong fields (red) being straight and weak fields (blue) bent, whereas for \SimName{5}{600}{5}, this relationship weakens, and some strong fields are bent significantly, and some weak fields are straight.}
        \label{fig:cube}
    \end{figure}

    This picture is supported by \autoref{fig:cube}, where we plot $100$ magnetic field streamlines for our two representative simulations at a time realisation midway through the kinematic phase. \SimName{0.3}{600}{5} is plotted on the left half of the cube, and \SimName{5}{600}{5} is plotted on the right half. The key takeaway is, in line with our findings above, that while magnetic field-lines in both regimes thread the simulation domain with an equally ``chaotic structure'' (the field-line curvature distribution remains largely unchanged between the two flow regimes), the magnitude of the field corresponding with different field-line structures is different in the incompressible, compared with the compressible flow regime. In \SimName{0.3}{600}{5}, for example, all strong ($b > b_\trms$; red) field-lines are straight, and all weak ($b < b_\trms$; blue) field-lines are located in regions with intense curvature. In contrast, for \SimName{5}{600}{5} we see that some curved fields are strong and others are weak, and likewise, some straight fields are weak, although most straight fields are still strong. Phenomenologically, this is expected, since shocks are able to bend all but the strongest magnetic field-lines, and, conversely, even weak fields can be relatively straight if they happen to lie in regions of the flow far from shocks.
    
    \subsubsection{Compressible vs. Incompressible Field Geometry}

    Overall, our quantitative analysis of the field geometry supports our interpretation and \citet{schekochihin2004simulations, schekochihin2002spectra}'s models for the relationship between the magnetic peak energy scale and resistive dissipation scale in the previous section: in the incompressible flow regime (where shocks are absent), the SSD in the kinematic phase produces a ``folded field'' geometry where magnetic energy becomes concentrated at $k_\eta$ -- the smallest scales possible (a state that \citet{schekochihin2004simulations} shows persists into the saturated phase), and $b \sim \kappa^{-1/2}$, but in the compressible flow regime (where shocks are ubiquitous), this correlation between the amplitude and the field geometry disappears. Again, we emphasise that the distribution of magnetic field-line curvature remains essentially the same, regardless of the flow regime, but the magnetic field amplitude distribution changes significantly, reducing the amplitude - correlation covariance and giving rise to strong deviations away from the incompressible $b \sim \kappa^{-1/2}$ relation. It would be an interesting additional avenue of exploration to understand if the correlation is re-established at small enough $\ell$, perhaps $\ell_{\rm s}$, the sonic scale, where the plasma transitions from supersonic to subsonic, and becomes highly-magnetised and fundamentally incompressible (see \eg \citealt{Beattie2024_10k_MHD} for an $\Reyh \sim 10^6$ simulation that resolves this transition). We leave this investigation for future studies. 

\section{Discussion}
\label{sec:discussion}

    Our findings for the scaling behaviour of velocity and magnetic fields generated during the kinematic phase of SSDs are relevant to a wide range of astrophysical systems, but here we highlight galaxy mergers and cosmic ray propagation as two particularly interesting cases. The former case creates an environment that excites a kinematic SSD, and the latter has recently been explored in terms of underlying curvature statistics of the magnetic field. In both of these examples we highlight how it is of key importance to differentiate between the incompressible and compressible SSD regimes (see the end of \autoref{sec:results:peak} for our distinction between these two regimes) for these astrophysical processes, since the underlying fields are completely different in these two regimes, even based on the power spectrum alone. We also motivate the importance of this distinction in regime for the transition of magnetic fields out of the kinematic phase, and into the saturated phase, which is ultimately the phase the Milky Way's ISM is in today.
    
    \subsection{Small-Scale Dynamos in Galaxy Mergers}

    Young galaxies \citep{geach2023polarized} and galaxy mergers are two environments where turbulent SSD amplified magnetic fields play a critical role in shaping the plasma dynamics \citep[\eg][]{pakmor2014magnetic, pakmor2017magnetic, martin2018three, martin2022towards, brzycki2019parameter, whittingham2021impact, whittingham2023impact, pfrommer2022simulating}. Merger events, particularly those involving gas-rich disc galaxies, are highly transformative, with strong magnetic fields influencing angular momentum transport and providing pressure support against collapse \citep{whittingham2021impact}. However, in these environments, a notable but as-yet unexplained property of the amplified magnetic fields is that magnetic energy peaks on scales of $\ell_\tp \gtrsim \ukpc$ during even the kinematic phase of the SSD \citep[\eg][]{rodenbeck2016magnetic, basu2017detection, brzycki2019parameter, whittingham2021impact, pfrommer2022simulating}, which is much larger than predicted by incompressible SSD theory, which limits field growth to smaller scales (\ie magnetic dissipation scale, $\ell_\eta$).

    Since, here, in this study, we have found that compressible SSDs can drive fields to larger scales than incompressible ones, it is natural to ask if considering compressibility resolves this problem. To answer this question, recall that we found that compressibility in the highly turbulent, supersonic plasma flows can shift the peak magnetic scales $\ell_\mathrm{p}$ from $\ell_\eta$ to much larger scales, even larger than the viscous scale $\ell_\nu$. The condition for this to occur is $\Pranm < \Reyh^{2/3}$, where $\Reyh$ and $\Pranm$ varies with plasma density, temperature, and ionisation throughout different phases of the ISM \citep{rincon2019dynamo, ferriere2019plasma, shukurov2021astrophysical, brandenburg2023galactic}. This condition is met in some phases (most notably the cold neutral medium, CNM), but not others, and it is unclear how to extend our condition to a realistic environment where different phases are intermixed and thus there is no single value of $\Reyh$ and $\Pranm$. Nonetheless, even if we are generous and assume that this condition is satisfied, it is clear that compressibility alone cannot explain the large observed scales of magnetic fields in mergers, simply because the ratio $\ell_\tp/\ell_\nu \sim \Reyh^{1/3}\Pranm^{1/2}$ is not large enough -- even for the most favourable case of the CNM, for which \citet{brandenburg2023galactic} estimate $\Reyh\sim 10^{10}$ and $\Pranm \sim 10^{4}$, we only have $\ell_\tp/\ell_\nu \sim 10^5$, whereas $\ell_\tturb/\ell_\nu \sim \Reyh^{2/3} \sim 10^{7}$. Thus, even using the most generous possible characteristic numbers, a compressible SSD would still yield a peak magnetic scale that is $\sim 2$ orders of magnitude smaller than the turbulent driving scale, which is much too small to explain the observed $\sim\ukpc$-scale fields observed in galaxy mergers. Adopting values of $\Reyh$ and $\Pranm$ for other ISM phases (especially the warm and hot phases, which are volume-filling in the disk) only strengthen this conclusion.

    That said, the fact that compressibility \textit{alone} cannot explain the existence of large-scale fields in galaxy merger does not mean that it is irrelevant. While compressible SSDs still leave $\ell_\tturb \gg \ell_\tp$, they nonetheless produce $\ell_\tp$ that are larger than in the incompressible case by a factor of $\Reyh^{1/3}$ (in highlight compressible, $\Mach\to\infty$ plasma flows) -- a factor that, depending on the ISM phase and the large-scale velocity dispersion, ranges from hundreds to tens of thousands. The resulting intermediate-scale fields can then serve as seeds for processes such as shear from galactic rotation or gravitational and buoyancy instabilities \citep[e.g.][]{pakmor2017magnetic, steinwandel2023towards} that drive the field to yet larger scales, significantly reducing the time required for this growth compared to what would be expected if one had to begin from the smaller seed fields produce by an incompressible SSD.

    \subsection{Cosmic Ray Propagation in Compressible Plasmas}

    Cosmic ray propagation is a longstanding problem, the bulk of which is far beyond the scope of this paper. However, there is a class of propagation models for which our conclusions regarding curvature statistics have important implications. For example, \citet{butsky2023galactic} has recently suggested that intermittent magnetic field fluctuations are necessary for regulating low-energy ($\sim {\rm MeV}$ -- ${\rm TeV}$) cosmic ray scattering in the Milky Way. \citet{kempski2023cosmic} and \citet{Lemoine2023_field_reversal_CR_transport} provide a potential source of magnetic field intermittency: intense regions of curved magnetic fields that reverse upon themselves, \ie magnetic field reversals. They argue that if magnetic fields are dominated by their fluctuating field ($\delta b/b_0 \gg 1$; as is the case for our simulations because there is no $\vect{b}_0$) then there is an energy dependent diffusion process associated with cosmic ray particles scattering off of magnetic fields due to the field's reversal being on scales that are resonant with the gyroradius of the cosmic ray.
    
    The \citet{kempski2023cosmic} model contains a resonance criterion derived from the assumption that $b \sim \kappa^{-1/2}$, which we have shown holds during the kinematic phase of incompressible SSDs but breaks down in compressible (\ie supersonic, turbulent) SSDs\footnote{It is not immediately clear whether this breakdown is relevant for cosmic ray propagation in galaxies like the Milky Way, where magnetic fields are maintained in a saturated dynamo state. While one might consider the possibility that $b \sim \kappa^{-1/2}$ breaks down for supersonic SSDs during the kinematic phase but holds in the saturated phase, this seems unlikely. In fact, \citet{schekochihin2004simulations} showed that for incompressible SSDs, the relationship between magnetic field-line curvature and field strength remains largely unchanged from the kinematic to the saturated phase, with the main difference being simply that high values of $b/b_\trms$ are suppressed as $b_\trms$ converges towards the saturated value. In a forthcoming paper we extend this result to the compressible regime, meaning that our conclusions about the implications of a breakdown in the $b\sim\kappa^{-1/2}$ relation for cosmic ray propagation apply to both high-redshift galaxies and present-day galaxies like the Milky Way.}. Instead, \autoref{sec:results:kappa} shows that magnetic field strength and curvature tend towards being independent in the compressible regime. That is not to say that the magnetic field does not also support reversals in the compressible regime; in probability density, it has just as many as the incompressible regime (see \autoref{fig:pdfs}). However the magnetic field amplitude is no longer constrained to a curvature relation, and hence the gyroradii of the CRs are free to vary more independently of the curvature than would be true for the incompressible case. Therefore these models will need to be modified for the $\Mach > 1$ turbulent regime, where parts of the volume will have no resonances. The caveat being, we do not know how these relations scale as a function of length scale, and it could be that the required correlations emerge on small enough scales and the resonances are restored.

    \subsection{The Transition out of Kinematic Growth}
    \label{sec:discussion:saturation}

    Finally, we would like to briefly motivate the importance of performing an analysis similar to the present study for the dynamo transition through the nonlinear and into the saturated phase. This would not only reveal the hierarchy of scales ultimately produced and sustained by SSDs in steady state, but would also determine what mechanism mediates the magnetic backreaction onto the velocity field as the magnetic energy saturates. This exercise carries direct relevance for galaxy formation, where repeated mergers induce SSD action that evolve through the kinematic phase, which we have studied here, and converge on a remnant ISM where the post-merger galaxy is maintained in the saturated phase of the SSD. It is therefore important for building an understanding of the magnetic fields that provide the backdrop for all post-merger interstellar processes.

    For an incompressible SSD, \citet{schekochihin2001structure, schekochihin2002small, schekochihin2004simulations} first predicted, and \citetalias{kriel2022fundamental} confirmed with DNSs, that during the kinematic phase where $E_\tmag(\ell) \ll E_\tkin(\ell)$ for all $\ell$, magnetic fields naturally become organised into a folded field geometry with magnetic energy concentrated at the smallest scales allowed by magnetic dissipation \citep[see also \eg][]{brandenburg2023dissipative}, namely $\ell_\tp \sim \ell_\eta$. In this phase magnetic fields are most rapidly grown through stretching motions \citep{beattie2023bulk} induced by viscous fluctuations, since these fluctuations have the shortest turnover time (see \autoref{sec:results:dissipation:keta}). These fluctuations, however, cannot remain the most efficient driver of SSD growth indefinitely. Once the magnetic field reaches an amplitude where $\vect{u}\cdot\nabla\vect{u} \sim \vect{b}\cdot\nabla\vect{b}$, it suppresses hydrodynamic motions on that scale. Following \citet{schekochihin2002model}, this happens when $E_\tmag \sim E_\tkin(\ell_\ts)$, where in the kinematic phase, the stretching scale is $\ell_\ts \sim \ell_\nu$. Once stretching on $\ell_\nu$ becomes suppressed, SSD action is expected to transition into a reduced growth regime. For incompressible SSDs, there is a theoretical consensus that magnetic energy growth transitions to linear-in-time \citep{cho2009growth, beresnyak2012universal, Xu2016_dynamo, seta2020saturation}, characterised by $E_\tmag \sim \varepsilon t$ (see \autoref{fig:time_evolution}). 
    
    The situation is less clear for compressible SSDs, where there is an ongoing debate about whether the growth mechanism is universal, or if SSDs in the presence of \citet{burgers1948_turbulence_model}-like turbulence transition into quadratic-in-time regime, $E_\tmag \sim t^2$, as proposed by \citet{Schleicher2013_quadratic_growth}. Regardless of the answer to this question, the mechanism behind the backreaction is thought to be generally similar to the incompressible case: $\ell_\ts$ shifts from $\ell_\nu$ to the next smallest unsuppressed turbulent fluctuations. Since these fluctuations are larger than the viscous-scale fluctuations, they have longer turnover times, and thus generate magnetic flux at a slower rate \citep{Schober2015_dynamo_saturation}. Magnetic amplification continues to suppress progressively larger scales \textit{ad infinitum}, until the dynamo saturates at the driving scale, with $\ell_\tp > \ell_\nu$. This has very clearly been shown in the time-dependent spectral ratios in \citet{beattie2022growth} and \citet{beattie2023bulk}. Various phenomenological models that describe the details of this process have been proposed, but they have never been tested at a fundamental level using DNSs, and it is unclear exactly which details of the incompressible picture are extensible to the compressible regime. Thus, in the third part of this series, we will explore the statistics of the nonlinear and saturated phase using the comprehensive dataset we have built in this study, as well as in \citetalias{kriel2022fundamental}. Where it is important to reveal the mechanism, and thereby the timescale, of the backreaction, as this process also provides an avenue for generating ``large-scale'' magnetic fields, in addition to LSDs (see the Introduction and the previous section for a discussion), and thus play an important role in the context of explaining $\sim \ukpc$ fields observed in young and merger-disturbed galaxies.

\section{Summary \& Conclusions}
\label{sec:conclusion}

    In this study we explore how both the energy spectrum (see \autoref{sec:results:peak}) and field-line curvature (see \autoref{sec:results:kappa}) of magnetic fields produced during the exponential-growing (kinematic) phase of small-scale dynamos (SSDs) depend upon the plasma flow regime (from subsonic $\Mach < 1$ to supersonic $\Mach > 1$, and viscous $\Reyh < \Reyh_\tcrit \approx 100$ to turbulent $\Reyh \geq \Reyh_\tcrit$ flows). To do so, we have used DNSs where we explicitly control the velocity field magnitude, as well as the dissipation rates of kinetic and magnetic energy to explore a wide range of $\Mach$, hydrodynamic Reynolds number, $\Reyh$, and magnetic Prandtl number, $\Pranm$. This has allowed us to extend our understanding of $\Pranm \geq 1$ SSD-amplified magnetic fields into the previously poorly-understood regime of supersonic SSDs. In particular, we have identified new relationships between the important characteristic length scales in a SSD -- the outer scale of turbulence $\ell_\tturb$, the kinetic energy dissipation scale $\ell_\nu$, the magnetic energy dissipation scale $\ell_\eta$, and the peak magnetic energy scale $\ell_\tp$ -- as a function of the fundamental plasma numbers: $\Mach$, $\Reyh$, and $\Pranm$. We complement this with a study of the statistics of magnetic field-line curvature, which support our findings.
    
    We list our key results below:
    \begin{itemize}        
        \item During the kinematic phase of SSDs, the kinetic energy dissipation scale $\ell_\nu$ varies with $\Reyh$ and $\Mach$ as expected for hydrodynamic flows (see the left panel in \autoref{fig:dissipation_scaling}). That is, we find three regimes: $\ell_\nu \sim \ell_\tturb \,\Reyh^{-3/4}$ for subsonic, turbulent flows (where $\Mach \leq 1$ and $\Reyh \geq \Reyh_\tcrit$), which corresponds with \citet{kolmogorov1941dissipation}-like scaling, $\ell_\nu \sim \ell_\tturb \,\Reyh^{-2/3}$ for supersonic ($\Mach > 1$) flows, which corresponds with \citet{burgers1948_turbulence_model}-like scaling, and finally $\ell_\nu \sim \ell_\tturb \,\Reyh^{-3/8}$ for viscous, subsonic flows (where $\Mach \leq 1$ and $\Reyh < \Reyh_\tcrit$ first shown in \citetalias{kriel2022fundamental}). \\

        \item The magnetic dissipation scale $\ell_\eta$ is related to the kinetic dissipation scale as $\ell_\eta \sim \ell_\nu \,\Pranm^{-1/2}$ in all $\Pranm \geq 1$ turbulent plasma flow regimes. Since this scaling relation, first proposed by \citet{schekochihin2002spectra}, follows from assuming that $\ell_\eta$ is located on the scale where the viscous shearing rate is balanced by Ohmic dissipation, we conclude that the viscous (smallest-scale) kinetic fluctuations are always the most efficient at amplifying magnetic energy during the kinematic phase of $\Pranm \geq 1$ SSDs, invariant to the $\Mach$ of the plasma. \\

        \item The magnetic peak scale $\ell_\tp$ and the associated statistics of the magnetic field-line curvature $\kappa$ produced by SSDs are very different for \textit{incompressible} (\ie either $\Mach \leq 1$, or $\Mach > 1$ and $\Reyh < \Reyh_\tcrit$) and \textit{compressible} (\ie $\Mach > 1$ and $\Reyh \geq \Reyh_\tcrit$) flows. The incompressible case behaves as predicted by the ``folded field'' model of \citet{schekochihin2002spectra, schekochihin2004simulations}, where the field-line curvature $\kappa$ and the field magnitude $b$ are strongly anti-correlated (see \autoref{fig:curvature}) as $b \sim \kappa^{-1/2}$, and the magnetic energy becomes concentrated on the smallest scales allowed by magnetic dissipation, which results in the hierarchy of characteristic length scales $\ell_\tturb > \ell_\nu > \ell_\eta \sim \ell_\tp$. By contrast, in the compressible regime, supersonic turbulence naturally gives rise to shocks with characteristic shock width $\ell_\tshock \sim \Mach^2 / (\Reyh\, (\Mach-1)^2)$, which we derive in \autoref{sec:results:peak}. These shocks create magnetic structures via flux-freezing and compression, which change the distribution of $b$ amplitudes, essentially destroying the anti-correlation between $b$ and $\kappa$, but notably, does not change the distribution of field-line curvature. These (isotropic) shocks also concentrate magnetic fields on a scale $\ell_\tp \sim (\ell_\tturb/\ell_\tshock)^{1/3} \ell_\eta \gg \ell_\eta$, where, in the $\Mach\to\infty$ limit, this produces $\ell_\tp \sim \Reyh^{1/3} \ell_\eta$. In plasmas where $\Pranm < \Reyh^{2/3}$, the subviscous range is smaller than the net shift in $\ell_\tp$ to larger scales (from $\ell_\eta$), which gives rise to a hierarchy $\ell_\tturb > \ell_\tp > \ell_\tshock > \ell_\nu > \ell_\eta$. Moreover, in the compressible regime, magnetic field-line curvature and magnitude tend towards independence, $b \sim \kappa^0$. \\

        \item We discuss a longstanding problem about the generation of large-scale magnetic fields in the context of galaxy mergers (but potentially can be more broadly applied to other young galaxies, such as the recent observation of a young starburst galaxy with a large-scale field in \citealt{geach2023polarized}). Since the turbulence in these galaxies is likely supersonic due to high gas densities that favour rapid cooling, the SSDs within them are likely compressible rather than incompressible. We show that this is not by itself sufficient to explain the large-scale fields observed in these galaxies, because, while compressibility increases the characteristic field size by many orders of magnitude compared to the incompressible case (at least for some combinations of plasma parameters), this increase still leaves fields concentrated on scales much smaller than those observed. Nonetheless, we argue that the increase in field size scale enabled by compressibility have the potential to help accelerate other, larger-scale dynamo mechanisms that produce large-scale fields. \\

        \item We discuss the implications our results have on models of cosmic ray scattering based on resonances between the size-scale of magnetic field reversals and the gyro-radius of the cosmic ray, \citep[e.g.,][]{kempski2023cosmic}. We suggest that for supersonic plasmas, which may be a significant portion of the interstellar medium of the Milky Way, these resonances may not work, since the magnetic field amplitude ($\sim$ gyro-radius) and underlying field curvature become independent from one another, at least in the sense that the global correlation reduces between the curvature and field amplitude.
    \end{itemize}

\section*{Acknowledgements}
    We thank the reviewer for their feedback, as well as Amitava Bhattacharjee, Axel Brandenburg, Frederick Gent, Pierre Lesaffre, Philipp Kempski, Mordecai-Mark Mac Low, Sergio Martin Alvarez, Christoph Pfrommer, Bart Ripperda, Alexander Schekochihin, Anvar Shukurov, Juan Diego Soler, and Enrique V\'azquez-Semadeni, along with Naomi M. McClure-Griffiths' research group, especially Yik Ki (Jackie) Ma, Hilay Shah, and Lindsey Oberhelman, for helpful discussions which have greatly improved the quality of this study.
    
    N.~K. acknowledges financial support from the Australian Government via the Australian Government Research Training Program Fee-Offset Scholarship and the Research School of Astronomy \& Astrophysics for the Joan Duffield Research Award. J.~R.~B.~acknowledges financial support from the Australian National University, via the Deakin PhD and Dean's Higher Degree Research (theoretical physics) Scholarships, the Australian Government via the Australian Government Research Training Program Fee-Offset Scholarship, and the Australian Capital Territory Government funded Fulbright scholarship. C.~F.~acknowledges funding provided by the Australian Research Council (Future Fellowship FT180100495 and Discovery Project DP230102280), and the Australia-Germany Joint Research Cooperation Scheme (UA-DAAD). C.~F.~further acknowledges high-performance computing resources provided by the Leibniz Rechenzentrum and the Gauss Centre for Supercomputing (grants~pr32lo, pr48pi and GCS Large-scale project~10391), the Australian National Computational Infrastructure (grant~ek9) and the Pawsey Supercomputing Centre (project~pawsey0810) in the framework of the National Computational Merit Allocation Scheme and the ANU Merit Allocation Scheme. M.~R.~K acknowledges support from the Australian Research Council through its Discovery Projects and Laureate Fellowships funding schemes, awards DP230101055 and FL220100020 and from the Australian National Computational Merit Allocation Scheme awards at the National Computational Infrastructure and the Pawsey Supercomputing Centre (award jh2). J.~K.~J.~H. acknowledges funding via the ANU Chancellor’s International Scholarship, the Space Plasma, Astronomy \& Astrophysics Higher Degree Research Award, and the Boswell Technologies Endowment Fund.

    The authors acknowledge valuable discussions that took place during the Nordita workshop, ``Towards a Comprehensive Model of the Galactic Magnetic Field,'' which was supported by NordForsk and the Royal Astronomical Society, as well as the Interstellar Institute's program ``II6,'' hosted at the Institut Pascal at Paris-Saclay University, which nourished the development of the ideas in this work.
    
    The simulation software, \textsc{flash}, was in part developed by the Flash Centre for Computational Science at the Department of Physics and Astronomy of the University of Rochester. As discussed in \autoref{sec:ICs:Mach}, we produce a turbulent forcing field, $\vect{f}$, via \textsc{TurbGen} developed by \citet{federrath2010comparing}. We also relied on the following programming languages/packages to analyse our simulation data and produce visualisations: \textsc{C++} \citep{stroustrup2013cpp}, \textsc{python} along with \textsc{numpy} \citep{oliphant2006guide, van2011numpy, harris2020array}, \textsc{matplotlib} \citep{hunter2007matplotlib, bisong2019matplotlib}, and \textsc{cython} \citep{behnel2010cython, smith2015cython}, as well as the \textsc{WebPlotDigitizer} app \citep{Rohatgi2022} and \textsc{visit} \citep{childs2012visit}.

\section*{Data Availability}

    A summary dataset, which includes the data underlying \autoref{table:summary} and \autoref{table:nres}, is publicly available from MNRAS. Access to data beyond these summary statistics will be shared upon reasonable request to the corresponding author.

\bibliographystyle{Header/mnras.bst}
\bibliography{Header/refs.bib}

\appendix
\section{Density-Weighted Hydrodynamic Reynolds Number}
\label{app:knu:density_weighted}

    \begin{figure}
        \centering
        \includegraphics[width=\linewidth]{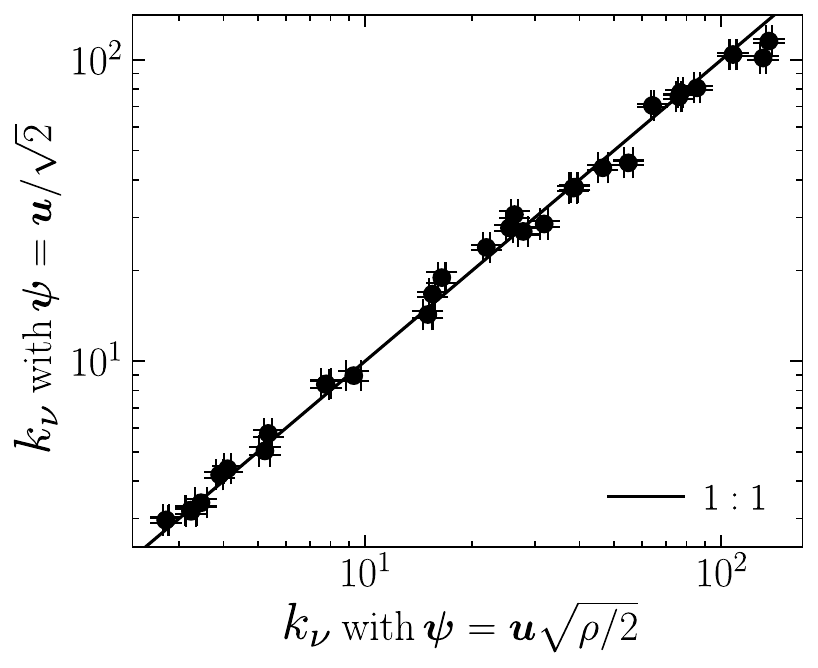}
        \caption{For each simulation setup in \autoref{table:summary}, we compare the converged (see \autoref{sec:results:nres} for details) characteristic viscous wavenumber, $k_\nu$, measured from \autoref{eqn:knu_measure} constructed with $\vect{\psi} = \vect{u} / \sqrt{2}$ (y-axis; studied in the main text), and $k_\nu$ constructed with $\vect{\psi} = \vect{u} \sqrt{\rho/2}$ (x-axis). We find that our measurement of $k_\nu$ is insensitive to the inclusion of density ($\rho$) variations in the plasma.}
        \label{fig:hydro_reynolds_compared}
    \end{figure}

    In the main text (see \autoref{sec:tools:k_nu}) we measured characteristic viscous wavenumbers, $k_\nu$, from kinetic energy spectra, $E_\tkin(k)$, constructed only from the velocity field (that is, $\vect{\psi} = \vect{u}/\sqrt{2}$ in \autoref{eqn:kin_fourier}). An alternative, popular definition for compressible flows \citep[see for example][]{federrath2010comparing, federrath2013universality, schmidt2019kinetic, grete2021matter, grete2023matter}, which accounts for density fluctuations, is $E_\tkin(k)$ constructed from \autoref{eqn:kin_fourier} with $\vect{\psi} = \vect{u} \sqrt{\rho/2}$. This definition is based on the idea that $E_\tkin$ should have units of kinetic energy, and such that $\psi^2$ needs to be a positive definite quantity \citep{kida1990energy}.
    
    In \autoref{fig:hydro_reynolds_compared} we compare converged (with numerical resolution; see \autoref{sec:results:nres}) $k_\nu$ measured from both definitions for $E_\tkin(k)$, for all our simulations, plotting $k_\nu$ derived from the $\vect{\psi} = \vect{u} \sqrt{\rho/2}$ spectrum on the x-axis, and $\vect{\psi} = \vect{u} / \sqrt{2}$ on the y-axis. We show that $k_\nu$ measured from the different definitions for $E_\tkin(k)$ scales 1:1, and therefore our measurements for $k_\nu$ in the kinematic phase of the SSD are robust and insensitive to density fluctuations, even for our $\Mach > 1$ simulations.

\section{Resistive Scale Measured from the Magnetic Reynolds Spectrum}
\label{app:krm}

    \begin{figure}
        \centering
        \includegraphics[width=\linewidth]{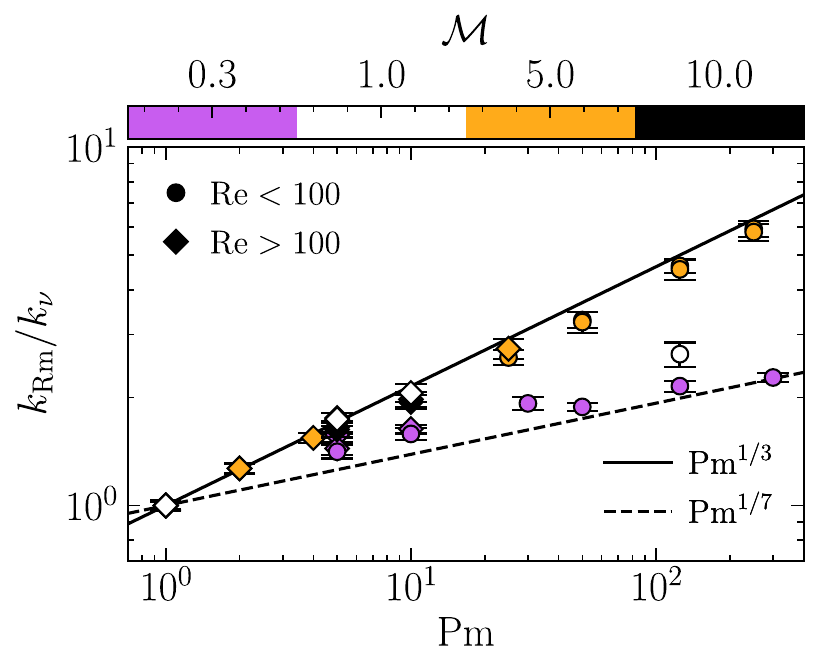}
        \caption{For each simulation in \autoref{table:summary}, we plot the separation between the wavenumber where the magnetic Reynolds number is one ($k_{\rm Rm} : \Reym(k) = 1$), and the viscous wavenumber ($k_\nu$; using the definition in the main text, see \autoref{sec:tools:k_nu}), via $k_{\rm Rm}/k_\nu$, and compare this scale separation with the magnetic Prandtl number. We plot simulations with $\Reyh < \Reyh_\tcrit \approx 100$ as circles, and $\Reyh \geq \Reyh_\tcrit$ as diamonds, and colour points by the sonic Mach number. We also annotate two power laws $\Pranm^{1/3}$ (solid line) and $\Pranm^{1/7}$ (dashed line) to highlight the different empirical scaling of $k_{\rm Rm}/k_\nu$ for subsonic and supersonic SSDs. Since $k_{\rm Rm}$ does not return a scaling consistent with $\Pranm^{1/2}$, as we previously found in \citetalias{kriel2022fundamental} for $\Mach < 1$ flows, we conclude that this measurement does not effectively probe magnetic dissipation, unlike $k_\nu : \Reyh(k) = 1$, which accurately probes the dissipation of the kinetic flow.}
        \label{fig:Rm_scaling}
    \end{figure}
    
    Given the success of our definition for the viscous scale (\autoref{eqn:knu_measure}), one may be tempted to define the resistive scale analogously. That is
    \begin{align}
        k_{\rm Rm}(t)
            = \texttt{argmin}_k\Big[
                \mathcal{I}\Big\{ \Big| \Reym(k, t) - 1 \Big| \Big\}
            \Big]
            , \label{eqn:Rm_measure}
    \end{align}
    where $\Reym(k, t) \equiv u_\tturb(k, t) / \eta k$. However, in \autoref{fig:Rm_scaling} we show that this wavenumber, $k_{\rm Rm}$, does not correspond with $k_\eta$ derived from \autoref{eqn:keta_measure}. In this figure we plot the separation between the converged (with numerical resolution; see \autoref{sec:results:nres}), time averaged (over the kinematic phase) $k_{\rm Rm}$ and $k_\nu$, for each of our simulations in \autoref{table:summary}, and show that this separation does not scale like $\Pranm^{1/2}$, which we showed in \autoref{fig:dissipation_scaling} is the case irrespective of the flow regime.

    We interpret this to mean that, whilst a unit analysis of the magnetic Reynolds number gives $\Reym \sim u_\tturb \ell_\tturb / \eta$, more directly, $\Reym$ (\autoref{dfn:Reym}) controls the relative importance of the induction term, $\nabla\times(\vect{u}\times\vect{b})$, compared with magnetic dissipation, $\eta\nabla\times\vect{j}$. However, each of these terms operate on different characteristic scales, (scales associated with $\ell_\nu$ for the induction, as we showed in \autoref{fig:dissipation_scaling}, and scales associated with $\ell_\eta$ for the dissipation). These details are not captured by only considering the influence of $u_\tturb \ell_\tturb / \eta$, \ie only considering the velocity structures for building the scale dependence (as we did with \autoref{eqn:reynolds_spectrum}). In fact, we know from \autoref{sec:results:dissipation} that $k_\eta$ should be largely independent of the kinetic cascade statistics, which \autoref{eqn:Rm_measure} is completely dependent upon (along with the material properties of the magnetic field).

\section{Magnetic correlation scale}
\label{app:kcor}

    \begin{figure}
        \centering
        \includegraphics[width=\linewidth]{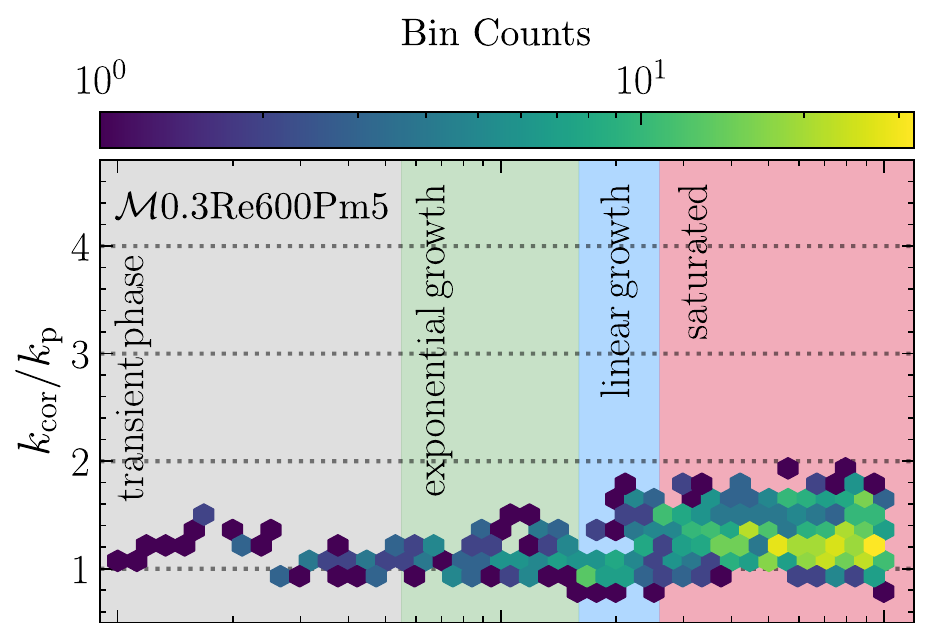}
        \includegraphics[width=\linewidth]{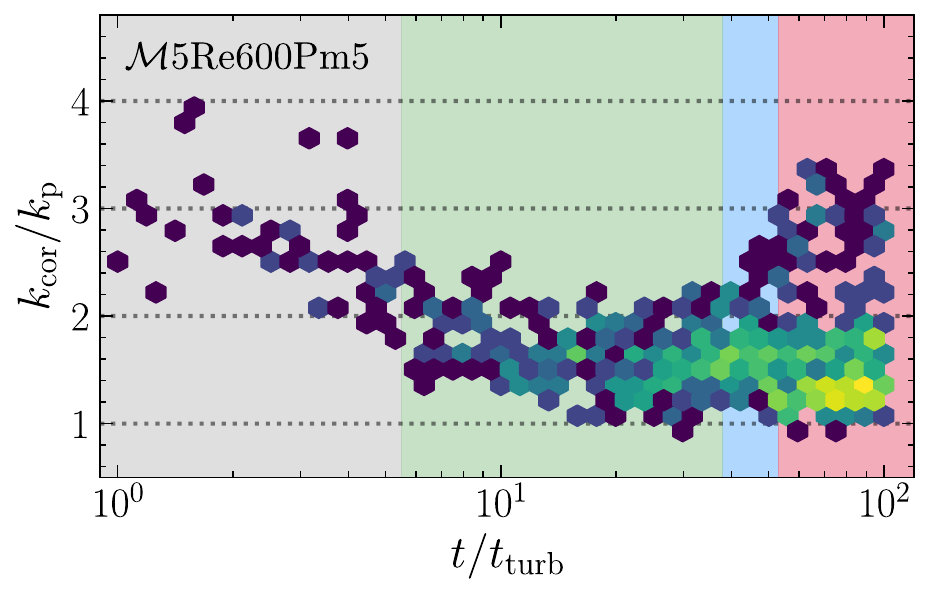}
        \caption{Time evolution of the magnetic correlation scale ($k_\tcor$; from \autoref{eqn:cor_scale}), compared with the magnetic peak scale ($k_\tp$; from \autoref{eqn:kp_measure}), for \SimName{0.3}{600}{5} and \SimName{5}{600}{5} in the top and bottom panels, respectively. We highlight the transient phase of the simulation in grey, followed by the exponential growth (kinematic), linear-growth, and saturated SSD phases in green, blue, and red, respectively.}
        \label{fig:mag_scales}
    \end{figure}

    \citet{schekochihin2004simulations} and \citet{galishnikova2022tearing} assume that the magnetic correlation (coherence) wavenumber,
    \begin{align}
        k_\tcor(t)
            &= \frac{
                    \int_{0}^{\infty} E_\tmag(k, t) \,\dd{k}
                }{
                    \int_{0}^{\infty} k^{-1} E_\tmag(k, t) \,\dd{k}
                }
            , \label{eqn:cor_scale}
    \end{align}
    is proportional to the magnetic peak wavenumber, $k_\tp$, which \citet{beattie2022growth} showed seemed to be qualitatively true not only in the kinematic phase, but also in the saturated phase. It is not clear whether this assumption would be valid for supersonic SSD amplified fields, where the magnetic energy spectrum is broadened by shocks when $\Mach > 1$ (see the bottom panel in \autoref{fig:cur_mag_spectra}). To test this, we compute $k_\tcor$ and $k_\tp$ for our two representative simulations, \SimName{0.3}{600}{5} and \SimName{5}{600}{5} at $N_\tres = 576$. In \autoref{fig:mag_scales} we compare the time evolution of these two scales by plotting $k_\tcor/k_\tp$ computed throughout the simulation for both setups. We colour individual hexagon-bins based on the number of occurrences, and shade the time range corresponding with the transient, exponential growth (kinematic), linear-growth, and saturated phases grey, green, blue, and red, respectively.

    As expected for \SimName{0.3}{600}{5}, we find evidence that during the kinematic phase $k_\tcor \sim k_\tp$, however, this relationship appears to weaken in the saturated state. For \SimName{5}{600}{5} we find that $k_\tcor > k_\tp$ at all times, where initially $k_\tcor \gg k_\tp$ due to shocks compressing magnetic energy into long, thin (\ie filamentary) structures, which broaden the magnetic energy distribution. The typical length of these shocked regions contributes energy to low-$k$ modes, while their typical width (see our model \autoref{eqn:shock_width}) contributes energy to high-$k$ modes. As the magnetic energy gradually amplifies and approaches saturation, the magnetic field becomes strong enough to suppress deformation on the smallest scales. Consequently, the ratio $k_\tcor / k_\tp$ tends toward $\gtrsim 1$.

\section{Fit Parameters Derived from Measuring Scale Convergence}

    In \autoref{sec:results:nres} we highlighted numerical convergence for two of our representative simulations, namely \SimName{0.3}{600}{5} and \SimName{5}{600}{5}. We also performed the same convergence study for all of our other simulation setups in \autoref{table:summary}, so here, for each of our simulation setups, we list the two fitted parameters: (1) critical numerical resolution, $N_{\tres, \tcrit}$, required for wavenumber convergence, and (2) the rate of convergence, $R$, for the viscous (column 2), resistive (column 3), and magnetic peak wavenumbers (column 4).

    For all of our viscous simulations, where $\Reyh < \Reyh_\tcrit \approx 100$, we find that our measurement for $k_\nu$ (and $k_\tp$ in the case of \SimName{5}{10}{25}) converged at even our lowest resolution runs (\ie as expected, we do not need much resolution to resolve low-$\Reyh$ dynamics). In these cases, we report the lowest resolution run we performed as $N_{\tres, \tcrit}$, and report no value for $R$, in \autoref{table:nres}.
    
    \begin{table*}
    \setlength{\tabcolsep}{7pt}
    \caption{Convergence properties of characteristic MHD wavenumbers.}
    \label{table:nres}
    \begin{center}
    \begin{tabular}{l cc l cc l cc}
        \hline\hline
        \multicolumn{1}{c}{Sim. ID}
            & \multicolumn{2}{c}{$k_\nu$}
            && \multicolumn{2}{c}{$k_\eta$}
            && \multicolumn{2}{c}{$k_\tp$} \\
        \cline{2-3} \cline{5-6} \cline{8-9} \\[-1.5ex]
            &  $N_{\rm res, crit}$ & $R$
            && $N_{\rm res, crit}$ & $R$
            && $N_{\rm res, crit}$ & $R$ \\
        \multicolumn{1}{c}{(1)}
            & \multicolumn{2}{c}{(2)}
            && \multicolumn{2}{c}{(3)}
            && \multicolumn{2}{c}{(4)} \\
\hline
\hline
\multicolumn{9}{c}{$\Mach = 0.3$} \\
\hline
$\mathcal{M}$0.3Re500Pm1
	& $34.7 \pm 1.9$ & $0.8 \pm 0.1$
	&& $67.8 \pm 8.7$ & $1.0 \pm 0.2$
	&& $31.6 \pm 13.4$ & $0.9 \pm 0.8$ \\
$\mathcal{M}$0.3Re100Pm5
	& $18.0$ & --
	&& $54.4 \pm 10.4$ & $1.0 \pm 0.2$
	&& $31.4 \pm 14.7$ & $0.9 \pm 1.0$ \\
$\mathcal{M}$0.3Re50Pm10
	& $18.0$ & --
	&& $49.9 \pm 7.5$ & $1.1 \pm 0.3$
	&& $32.5 \pm 12.0$ & $1.2 \pm 1.0$ \\
$\mathcal{M}$0.3Re10Pm50
	& $18.0$ & --
	&& $54.5 \pm 12.6$ & $0.8 \pm 0.2$
	&& $25.8 \pm 15.3$ & $0.8 \pm 0.8$ \\
$\mathcal{M}$0.3Re3000Pm1
	& $116.8 \pm 5.0$ & $1.0 \pm 0.1$
	&& $229.2 \pm 19.0$ & $1.1 \pm 0.1$
	&& $169.7 \pm 32.4$ & $1.0 \pm 0.1$ \\
$\mathcal{M}$0.3Re600Pm5
	& $43.0 \pm 2.2$ & $0.8 \pm 0.1$
	&& $190.4 \pm 20.2$ & $1.0 \pm 0.1$
	&& $117.2 \pm 40.1$ & $0.8 \pm 0.2$ \\
$\mathcal{M}$0.3Re300Pm10
	& $23.9 \pm 0.9$ & $0.9 \pm 0.1$
	&& $120.3 \pm 15.0$ & $1.1 \pm 0.1$
	&& $84.8 \pm 26.6$ & $1.0 \pm 0.3$ \\
$\mathcal{M}$0.3Re100Pm30
	& $18.0$ & --
	&& $124.3 \pm 21.0$ & $1.0 \pm 0.1$
	&& $89.2 \pm 35.2$ & $0.9 \pm 0.3$ \\
$\mathcal{M}$0.3Re24Pm125
	& $18.0$ & --
	&& $102.4 \pm 12.9$ & $1.0 \pm 0.1$
	&& $77.3 \pm 37.5$ & $0.8 \pm 0.3$ \\
$\mathcal{M}$0.3Re10Pm300
	& $18.0$ & --
	&& $102.4 \pm 21.6$ & $0.9 \pm 0.1$
	&& $82.2 \pm 27.1$ & $0.9 \pm 0.2$ \\
$\mathcal{M}$0.3Re2000Pm5
	& $130.3 \pm 4.3$ & $0.8 \pm 0.1$
	&& $534.9 \pm 33.1$ & $1.0 \pm 0.1$
	&& $317.4 \pm 51.9$ & $0.9 \pm 0.1$ \\
\hline
\multicolumn{9}{c}{$\Mach = 1$} \\
\hline
$\mathcal{M}$1Re3000Pm1
	& $116.4 \pm 4.5$ & $1.1 \pm 0.1$
	&& $205.2 \pm 19.8$ & $1.0 \pm 0.1$
	&& $148.7 \pm 36.5$ & $0.9 \pm 0.2$ \\
$\mathcal{M}$1Re600Pm5
	& $48.5 \pm 3.5$ & $0.7 \pm 0.1$
	&& $159.8 \pm 19.0$ & $0.9 \pm 0.1$
	&& $90.8 \pm 36.2$ & $0.7 \pm 0.2$ \\
$\mathcal{M}$1Re300Pm10
	& $18.0$ & --
	&& $117.4 \pm 17.0$ & $1.0 \pm 0.1$
	&& $103.2 \pm 81.7$ & $0.7 \pm 0.3$ \\
$\mathcal{M}$1Re24Pm125
	& $18.0$ & --
	&& $96.7 \pm 16.1$ & $0.9 \pm 0.1$
	&& $95.4 \pm 111.2$ & $0.6 \pm 0.3$ \\
\hline
\multicolumn{9}{c}{$\Mach = 5$} \\
\hline
$\mathcal{M}$5Re10Pm25
	& $18.0$ & --
	&& $35.2 \pm 7.8$ & $0.8 \pm 0.4$
	&& $18.0$ & -- \\
$\mathcal{M}$5Re10Pm50
	& $18.0$ & --
	&& $56.5 \pm 12.1$ & $0.8 \pm 0.2$
	&& $26.1 \pm 9.8$ & $0.8 \pm 0.5$ \\
$\mathcal{M}$5Re10Pm125
	& $18.0$ & --
	&& $101.5 \pm 24.1$ & $0.8 \pm 0.1$
	&& $63.9 \pm 38.9$ & $0.6 \pm 0.2$ \\
$\mathcal{M}$5Re10Pm250
	& $18.0$ & --
	&& $144.2 \pm 36.8$ & $0.8 \pm 0.1$
	&& $151.9 \pm 124.8$ & $0.6 \pm 0.2$ \\
$\mathcal{M}$5Re500Pm1
	& $61.9 \pm 4.0$ & $0.8 \pm 0.1$
	&& $64.6 \pm 35.3$ & $1.0 \pm 0.4$
	&& $14.1 \pm 27.5$ & $1.0 \pm 6.9$ \\
$\mathcal{M}$5Re500Pm2
	& $62.7 \pm 4.2$ & $0.8 \pm 0.1$
	&& $84.1 \pm 32.8$ & $1.0 \pm 0.3$
	&& $11.2 \pm 20.5$ & $0.6 \pm 1.8$ \\
$\mathcal{M}$5Re500Pm4
	& $62.1 \pm 3.8$ & $0.8 \pm 0.1$
	&& $100.3 \pm 28.7$ & $1.0 \pm 0.2$
	&& $22.0 \pm 14.8$ & $0.7 \pm 1.2$ \\
$\mathcal{M}$5Re3000Pm1
	& $146.7 \pm 3.9$ & $1.2 \pm 0.1$
	&& $200.1 \pm 60.5$ & $1.0 \pm 0.1$
	&& $20.1 \pm 18.1$ & $0.7 \pm 1.6$ \\
$\mathcal{M}$5Re1500Pm2
	& $143.6 \pm 8.0$ & $0.9 \pm 0.1$
	&& $199.8 \pm 45.1$ & $1.0 \pm 0.1$
	&& $21.6 \pm 13.1$ & $0.8 \pm 1.6$ \\
$\mathcal{M}$5Re600Pm5
	& $79.4 \pm 4.6$ & $0.8 \pm 0.1$
	&& $173.5 \pm 40.7$ & $0.9 \pm 0.1$
	&& $20.6 \pm 18.3$ & $0.6 \pm 0.8$ \\
$\mathcal{M}$5Re300Pm10
	& $43.1 \pm 3.5$ & $0.7 \pm 0.1$
	&& $120.1 \pm 26.5$ & $0.9 \pm 0.1$
	&& $31.1 \pm 35.0$ & $0.5 \pm 0.5$ \\
$\mathcal{M}$5Re120Pm25
	& $20.3 \pm 1.2$ & $0.7 \pm 0.1$
	&& $157.0 \pm 29.2$ & $0.8 \pm 0.1$
	&& $84.3 \pm 112.4$ & $0.5 \pm 0.3$ \\
$\mathcal{M}$5Re60Pm50
	& $18.0$ & --
	&& $159.8 \pm 36.9$ & $0.8 \pm 0.1$
	&& $75.7 \pm 99.4$ & $0.5 \pm 0.3$ \\
$\mathcal{M}$5Re24Pm125
	& $18.0$ & --
	&& $143.7 \pm 33.1$ & $0.8 \pm 0.1$
	&& $134.4 \pm 169.0$ & $0.5 \pm 0.3$ \\
$\mathcal{M}$5Re12Pm250
	& $18.0$ & --
	&& $137.4 \pm 30.7$ & $0.8 \pm 0.1$
	&& $112.5 \pm 54.6$ & $0.6 \pm 0.2$ \\
$\mathcal{M}$5Re2000Pm5
	& $208.5 \pm 10.1$ & $0.8 \pm 0.1$
	&& $457.0 \pm 68.8$ & $0.9 \pm 0.1$
	&& $146.1 \pm 161.6$ & $0.5 \pm 0.2$ \\
\hline
\multicolumn{9}{c}{$\Mach = 10$} \\
\hline
$\mathcal{M}$10Re3000Pm1
	& $137.8 \pm 4.0$ & $1.2 \pm 0.1$
	&& $228.8 \pm 55.0$ & $1.0 \pm 0.1$
	&& $26.2 \pm 12.2$ & $0.8 \pm 0.6$ \\
$\mathcal{M}$10Re600Pm5
	& $83.7 \pm 5.0$ & $0.8 \pm 0.1$
	&& $198.1 \pm 32.0$ & $0.9 \pm 0.1$
	&& $35.1 \pm 34.6$ & $0.5 \pm 0.4$ \\
$\mathcal{M}$10Re300Pm10
	& $45.7 \pm 3.6$ & $0.7 \pm 0.1$
	&& $125.9 \pm 32.3$ & $0.9 \pm 0.1$
	&& $60.7 \pm 98.6$ & $0.5 \pm 0.4$ \\
\hline
\hline
    \end{tabular}
    \end{center}
    \begin{tablenotes}[para]
         \textit{\textbf{Note:}} All parameters are derived from fits of \autoref{eqn:nres} to the time averaged (over the kinematic phase of the SSD) characteristic wavenumbers derived for each simulation. \textbf{Column (1):} unique simulation ID. \textbf{Column (2):} the characteristic grid resolution, $N_{\tres, \tcrit}$ (scale-height parameter in \autoref{eqn:nres}), where the viscous wavenumber, $k_{\nu}$ (derived from \autoref{eqn:knu_measure} with $\vect{\psi} = \vect{u} / \sqrt{2}$), shows evidence of convergence, and the rate of convergence, $R$. \textbf{Column (3) and (4):} the same as column (2) but for the resistive wavenumber, $k_\eta$ (derived from \autoref{eqn:keta_measure}), and the magnetic peak wavenumber, $k_{\tp}$ (derived from \autoref{eqn:kp_measure}), respectively.
    \end{tablenotes}
\end{table*}

\section{Computing Field-Line Curvature}
\label{app:kappa}

    As mentioned in the main text (and very briefly mentioned in \citet{schekochihin2004simulations}), when computing the curvature $\|\vect{\kappa}\|$ of the magnetic field $\vect{b}$ in the presence of grid-scale structures, it is necessary to take some care to preserve the exact orthogonality of the tangent and normal basis-vectors,
    \begin{align}
        \tbasis
            &= \frac{\vect{b}}{\|\vect{b}\|} , \\
        \nbasis
            &= \frac{\vect{\kappa}}{\|\vect{\kappa}\|}
        , \label{eqn:nbasis}
    \end{align}
    respectively. Here the non-normalised normal vector,
    \begin{align}
        \vect{\kappa}
            = (\tbasis\cdot\nabla)\tbasis
            = \cancelto{0}{\frac{1}{2} \nabla(\|\tbasis\|^2)}
                + (\nabla\times\tbasis) \times \tbasis
            ,
    \end{align}
    points in the direction of maximum curvature, and carries magnitude equal to the inverse-radius of curvature $\|\vect{\kappa}\|^{-1}$.

    In the following subsections we will discuss two different approaches for computing $\|\vect{\kappa}\|$. In \autoref{app:kappa:flawed} we demonstrate how directly calculating $\vect{\kappa}$ from $\tbasis$ produces measurements polluted with significant numerical errors (through a violation of orthogonality between $\tbasis$ and $\nbasis$), whereas the improved expression that is evaluated in this study, and derived in \autoref{app:kappa:stencil}, inherently preserves orthogonality. Finally, in \autoref{app:kappa:improved} we describe how we compute $\|\vect{\kappa}\|$ in practice. Note that when discussing these algorithms (in \autoref{app:kappa:flawed} and \autoref{app:kappa:improved}), we denote the $x$-component of $\vect{b}$ in cell $(i,j,k)$ as $\vect{b}_x^{i,j,k}$, which should not be confused for the index notation used in \autoref{app:kappa:stencil}.

    \subsection{Flawed Algorithm}
    \label{app:kappa:flawed}
    
    The following is a straightforward, but flawed algorithm for computing $\|\vect{\kappa}\|$:
    \begin{enumerate}
        \item Define the tangent vector in every cell:
            \begin{align}
                \tbasis^{i,j,k}
                    = \frac{\vect{b}^{i,j,k}}{\|\vect{b}^{i,j,k}\|} .
            \end{align}
        \item Construct the magnetic gradient tensor by approximating the derivative of each component in each spacial direction, via second-order centered differences, for example the derivative of the $y$-component in the $z$-direction:
            \begin{align}
                (\nabla\otimes\tbasis)^{i,j,k}_{z,y}
                    = \frac{1}{2\Delta x} \rbrac{\widehat{b}^{i,j,k+1}_y - \widehat{b}^{i,j,k-1}_y} .
            \end{align}
        \item Compute the curvature vector in every cell as:
            \begin{align}
                \vect{\kappa}^{i,j,k}
                    = \tbasis^{i,j,k} \cdot (\nabla\otimes\tbasis)^{i,j,k}
                    . \label{eqn:kappa:flawed}
            \end{align}
        \item Finally, compute the curvature-magnitude as:
            \begin{align}
                \kappa^{i,j,k}
                    = \|\vect{\kappa}^{i,j,k}\| .
            \end{align}
    \end{enumerate}
    The difficulty in this algorithm is that there is no guarantee that the vector $\vect{\kappa}^{i,j,k}$ produced by \autoref{eqn:kappa:flawed} will be exactly orthogonal to $\vect{b}^{i,j,k}$; instead, the degree to which orthogonality is maintained depends on the accuracy of the finite difference approximation used to construct $(\nabla\otimes\tbasis)^{i,j,k}$.
    
    Based on numerical experiments, we find that in regions where $\vect{b}^{i,j,k}$ is smooth, this algorithm computes $\nbasis^{i,j,k}$ that is very close to orthogonal to $\tbasis^{i,j,k}$, but in regions where $\vect{b}$ changes directions over length scales of order $\Delta x$ (\eg in shocked regions), these two (basis) vectors deviate away from orthogonality (see \autoref{fig:kappa_error}).

    \subsection{A Stable Approach to Compute Curvature}
    \label{app:kappa:stencil}
    
    As an alternative approach, we expand the expression for $\kappa_i$ in such a way to ensure that any numerical errors introduced by the finite difference approximation are eliminated by construction. Through some algebra-gymnastics, it is straightforward to show that $\kappa_i$ can equivalently be expressed as
    \begin{equation}
        \kappa_i
            = \frac{b_j}{b_k b_k} \frac{\pp b_i}{\pp x_j}
                - \frac{b_i b_m  b_j}{(b_k b_k)^2}\frac{\pp b_m}{\pp x_j} ,
        \label{eqn:nbasis:formula}
    \end{equation}
    or in vector notation
    \begin{align}
        \vect{\kappa}
            &= \frac{1}{\|\vect{b}\|^2} \rbrac{\tensor{\imatrix} - \frac{\vect{b}\otimes\vect{b}}{\|\vect{b}\|^2}} \cdot \big(\vect{b}\cdot\nabla\big)\vect{b}
            ,
    \end{align}
    which, while algebraically equivalent to $(\tbasis\cdot\nabla)\tbasis$, is significantly more accurate when evaluated numerically. To see why, consider the inner product between $\vect{b}$ and $\vect{\kappa}$ (given by \autoref{eqn:nbasis:formula}), \viz
    \begin{equation}
        b_i \kappa_i
            = \frac{b_i b_j}{b_k b_k} \frac{\pp b_i}{\pp x_j}
                - \frac{b_j b_m}{b_k b_k}\frac{\pp b_m}{\pp x_j} = 0
        .
    \end{equation}
    The critical point to notice, is that $\vect{b}\cdot\vect{\kappa}$ vanishes by construction, \textit{regardless} of the finite difference approximation used to compute the $\pp b_i/\pp x_j$ and $\pp b_m/\pp x_j$ tensors. These approximations need not be accurate, they merely need to be the same for both gradient tensors.
    
    This property is not guaranteed for $\vect{\kappa}$ computed via \autoref{eqn:kappa:flawed}, since
    \begin{align}
        b_i \kappa_i
            = \frac{b_i b_j}{(b_k b_k)^{1/2}} \frac{\pp}{\pp x_j}\sbrac{\frac{b_i}{(b_m b_m)^{1/2}}}
    \end{align}
    is only zero if the finite difference approximation for $\pp [b_i / (b_k b_k)^{1/2}]/\pp x_j$ is exact. Of course, in practice this is never the case. Moreover, in our testing, we found that evaluating \autoref{eqn:kappa:flawed} yielded results polluted with significant numerical errors, even when employing a fourth order accurate finite difference stencil, whereas a second order stencil was sufficient to converge on numerically exact results when evaluating \autoref{eqn:nbasis:formula} instead.
    
    \subsection{Improved Algorithm}
    \label{app:kappa:improved}
    
    In practice, the algorithm we adopt is:
    \begin{enumerate}
        \item Compute the tensor field $(\nabla\otimes\vect{b})^{i,j,k}$ as in \autoref{app:kappa:flawed}, for example, for the derivative of the $y$-component in the $z$-direction:
            \begin{align}
                (\nabla\otimes\tbasis)^{i,j,k}_{z,y}
                    = \frac{1}{2\Delta x} \rbrac{\widehat{b}^{i,j,k+1}_y - \widehat{b}^{i,j,k-1}_y} .
            \end{align}
        \item Compute $\vect{\kappa}^{i,j,k}$ as:
            \begin{align}
                \vect{\kappa}^{i,j,k}
                    &= \left\|\vect{b}^{i,j,k}\right\|^{-2} \rbrac{\vect{b}^{i,j,k} \cdot (\nabla\otimes\vect{b})^{i,j,k}} \nonumber\\
                    &\nquad[1] - \left\|\vect{b}^{i,j,k}\right\|^{-4} \rbrac{(\vect{b}^{i,j,k}\otimes\vect{b}^{i,j,k}) : (\nabla\otimes\vect{b})^{i,j,k}} ,
            \end{align}
            where the colon-operator represents the double contraction (double inner product) of tensors, \ie $\tensor{M}:\tensor{N} \equiv M_{ij} N_{ij}$.
        \item Compute the curvature as:
            \begin{align}
                \kappa^{i,j,k}
                    = \|\vect{\kappa}^{i,j,k}\| .
            \end{align}
    \end{enumerate}
    
\begin{figure}
        \centering
        \includegraphics[width=\linewidth]{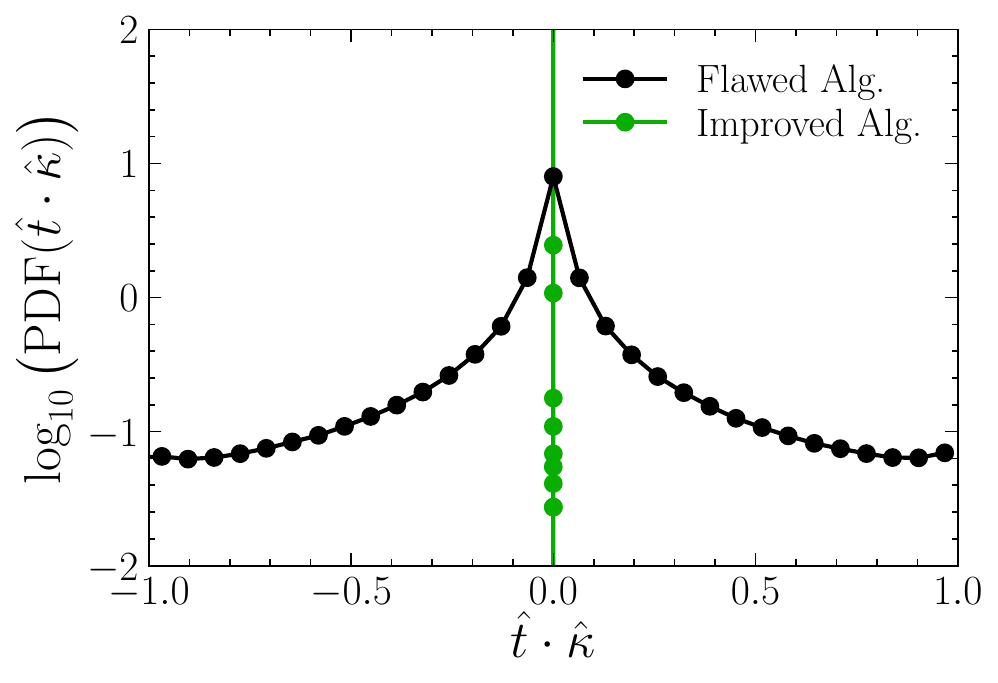}
        \caption{Comparison of how well the orthogonality is maintained between the tangent ($\tbasis$) and normal ($\nbasis$) basis vectors when computed using a straight-forward, but flawed algorithm (plotted in black; see \autoref{app:kappa:flawed}) versus our improved algorithm (plotted in green; see \autoref{app:kappa:improved}). The volume weighted PDFs of $\tbasis\cdot\nbasis$ is shown for the \SimName{0.3}{600}{5} simulation run at $N_\tres = 576$, at a snapshot midway through the kinematic phase. Notice that our improved algorithm maintains orthogonality to machine precision.}
        \label{fig:kappa_error}
    \end{figure}

    In line with our discussion above, we find that this algorithm maintains the orthogonality of $\tbasis$ and $\nbasis$ to machine precision, in every cell, which sits in contrast to the flawed implementation, as demonstrated in \autoref{fig:kappa_error} for \SimName{0.3}{600}{5} at a snapshot during the kinematic phase.

\bsp % typesetting comment
\label{lastpage}
\end{document}